\documentclass[
prd,
twocolumn,
nofootinbib,
floatfix,
showpacs ,amssymb, aps]{revtex4}
\usepackage{amsmath,slashed}

\usepackage{comment}
\usepackage{graphicx}

\usepackage{amsmath,amssymb}
\usepackage{bm}
\usepackage{enumitem}

\usepackage{natbib}
\usepackage{wasysym}

\newcommand{\J}{\mathcal{J}}

\newcommand{\rqm}{{%
	\declareslashed{}{\text{-}}{0.04}{0}{I}\slashed{I}}}

\begin{document}

\title{
Gravitational Waves and Hydromagnetic Instabilities in Rotating Magnetized Neutron Stars}

\author{Paul D. Lasky}
	\affiliation{School of Physics, University of Melbourne, Parkville, VIC 3010, Australia}
	\email{paul.lasky@unimelb.edu.au}
	\altaffiliation{Theoretical Astrophysics, IAAT, Eberhard Karls University of T\"ubingen, T\"ubingen 72076, Germany}
\author{Burkhard Zink and Kostas D. Kokkotas}
	\affiliation{Theoretical Astrophysics, IAAT, Eberhard Karls University of T\"ubingen, T\"ubingen 72076, Germany}

		\begin{abstract}
	We perform nonlinear general relativistic ideal magnetohydrodynamic simulations of poloidal magnetic fields in rotating polytropic neutron stars.  We have three primary goals: i) to understand the nature of magnetohydrodynamic instabilities inherent to poloidal magnetic fields in non-rotating {\it and} rotating neutron stars, ii) to explore the possible space of stable equilibrium configurations and iii) to understand gravitational wave emissions caused by the catastrophic reconfiguration of magnetic fields associated with giant magnetar flares.  Our key physical contributions can be summarized as follows:  i) gravitational waves from $f$-modes caused by magnetar flares are {\it unlikely} to be detected in the current or near-future generation of gravitational waves observatories, ii) gravitational waves from Alfv\'en waves propagating inside the neutron star are more likely candidates, although this interpretation relies on the unknown damping time of these modes, iii) any magnetic field equilibria derived from our simulations are characterized as non-axisymmetric, with approximately $65\%$ of their magnetic energy in the poloidal field, iv) rotation acts to separate the timescales of different instabilities in our system, with the varicose mode playing a more major role due to a delayed kink instability and v) despite the slowing growth rate of the kink mode, it is always present in our simulations, even for models where the rotational period is of the same order as the Alfv\'en timescale.
		\end{abstract}
		
		
		
		
		
		\maketitle

\section{Introduction}
How can we probe magnetic fields in the cores of neutron stars?  Despite their central role in multiple aspects of neutron star physics, this question has eluded sufficient resolution for nigh on five decades.  Two independent sets of observations -- neutron star spin down and thermal emissions -- probe the magnetic field above the {\it surface} of the neutron star.  However, these provide little information about the field lying {\it inside} the crust and core of the star.  It is therefore left to first-principals modelling of neutron star interiors to garner these vital pieces of information.

From a theoretical perspective, knowledge of the strength, topology and dynamics of magnetic fields in the core and crust of neutron stars is crucial to understanding phenomena in garden-variety pulsars including, but not limited to;  boundary conditions feeding magnetospheric models that describe emission and neutron star spin-down \cite{pacini67,pacini68,eastlund68,gunn69}, glitch dynamics \cite{easson79} and gravitational wave emissions from internal fields \cite[e.g.][]{haskell08,gualtieri11,mastrano11} and magnetic ``mountains'' \cite[e.g.][]{payne04,vigelius08}.  Moreover, the class of neutron stars known as the {\it magnetars}, with surface field strengths of $B\gtrsim10^{14}\,{\rm G}$, exhibit more exotic phenomena, whose existence and dynamics crucially depend on the magnetic field.  These include their unusually high surface temperatures \cite[e.g][]{thompson96,heyl98,colpi00} (although see the recent article of \citet{ho11}), the generation of magnetar flares \cite{duncan92,thompson95,thompson96} and their subsequent quasi-periodic oscillations \cite[e.g.][]{strohmayer05,strohmayer06,levin06,levin07,watts07,sotani08a,colaiuda09,cerdaduran09,colaiuda11,gabler11,colaiuda12} and free-precession \cite{melatos99} or lack thereof \cite{glampedakis10}.  These reasons and more provide sufficient impetus to warrant first principals modelling of neutron stars emphasizing the effects of their magnetic fields. 

In two previous short articles we have begun to probe the interior dynamics of neutron stars strong magnetic fields utilising our newly developed three-dimensional, non-linear, general relativistic magnetohydrodynamics code {\sc horizon} \cite{zink11}.  The first of these \cite{lasky11} was a letter investigating a canonical series of non-rotating models with a single equation of state, in a first attempt to understand poloidal magnetic field instabilities in general relativity.  This letter explored the subsequent pseudo-equilibria states, broadening the possible solution space of stable equilibria available in neutron stars.  A following short article \cite{zink12} utilised these models to calculate the gravitational wave emission from the complete reconstruction of the magnetic field due to the magnetic field instability.  This model was viewed as mimicking the behaviour of the interior of a neutron star immediately following a giant magnetar flare, and therefore gave estimates on the gravitational wave detectability of such a situation.  In the present work we follow-up on those two short articles, providing significantly more details on our numerical model, as well as confirming and extending our previous models using both different polytropic equations of state and also introducing rotation into the system.  With this in mind, the present paper has three key, short-term motivations: i) to understand the nature of magnetohydrodynamic (MHD) instabilities inherent to poloidal magnetic fields in both non-rotating and rotating relativistic neutron stars, ii) to substantially extend the full set of MHD equilibria derived as steady-state solutions and iii) to robustly extend the results regarding gravitational wave emission following giant flares to include different polytropic equations of state, and hence include a relationship between the gravitational wave strain, magnetic field, radius and mass of the star.  

The study of axisymmetric MHD equilibria relevant for neutron stars has a rich heritage, established with the early work of \citet{chandrasekhar53,monaghan65,roxburgh66} and \citet{parker66}.  It was soon realised that these idealised analytic fields are prone to various kinds of instabilities including, but not limited to, Tayler ``kink'' instabilities \cite{tayler57,tayler73,wright73,markey73,markey74} and the \citet{flowers77} instability.  In their various guises, these instabilities indicate that purely poloidal and purely toroidal fields are dynamically unstable.  Poloidal fields are unstable in the regions where its field lines are closed within the star, and it is therefore widely believed that threading a toroidal component through this region of the poloidal field acts to stabilise the field \cite{wright73,tayler80}.  Such {\it twisted-torus} configurations have been studied in considerable detail in semi-analytic calculations \cite{ioka01,yoshida06,yoshida06b,haskell08,ciolfi09} and also derived as equilibrium configurations from global MHD simulations \cite{braithwaite06b,braithwaite09}.  However, the stability of such configurations in {\it barotropic} stars has recently been questioned \cite{lander12}.  

In our recent paper beginning with axisymmetric, purely poloidal initial conditions \cite{lasky11}, our evolutions characteristically developed non-axisymmetries during the early phases of the non-linear development of the kink instability.  All of our subsequent equilibria where therefore non-axisymmetric, although they retained certain properties of the twisted-torus configurations.  The development of these non-axisymmetries were consistent with the evolutions of \citet{braithwaite08}, who found that the evolution of fields to non-axisymmetric or axisymmetric states depended on the radius of the initial neutral line.  

Rotation has a marked effect on the development of various MHD instabilities.  From an early stage, \citet{frieman60} understood that rigid-body rotation has an effect on the stability of hydromagnetic equilibria {\it only} when the velocity of the fluid flow is of the same order, or greater than the Alfv\'en velocity.  The first works on the global stability of purely poloidal, rotating fields was that of \citet{geppert06}, who found that stars rotating with sufficiently high rotational velocities have suppressed instabilities.  \citet{lander11b}, on the other hand, utilised their linear code to conclude that some, but not all modes were stabilised by the presence of rotation.  In the present article, we perform the first general relativistic simulations of rotating poloidal fields, concluding that instability timescales are slowed by the rotation, but that the instabilities are not completely suppressed.  We do note, however, that we are only beginning to approach the regime of {\it fast} rotation, whereby the fluid-flow velocity is of the same order as the Alfv\'en velocity.  We leave open the distinct possibility that faster rotation would suppress these instabilities.

Throughout the article we allow Greek indices to range $0\ldots3$ and Latin indices $1\ldots3$.  The paper is set out as follows: in section \ref{numerical} we outline our numerical method, focussing on the equations of general relativistic magnetohydrodynamics in \ref{GRMHD} and details of our specific code implementation in \ref{horizon}.  In section \ref{nonrotate} we look at a non-rotating fiducial model, concentrating on the magnetic field instability in \ref{instability}, gravitational wave emissions in \ref{GWs1}, quasi-equilibrium end-states in section \ref{endstate} and the effect of different magnetic field strengths in \ref{magfieldstrength}.  In section \ref{GWs} we study the relationship between the radius and mass of the star, magnetic field strength and gravitational wave emissions, deriving for the first time an empirical relation between these four variables.  In this section we also look at the detectability of such gravitational waves in current and future gravitational wave detectors such as Advanced LIGO and the proposed Einstein Telescope respectively, concentrating on both the $f$-mode emission and also lower frequency Alfv\'en modes.  In section \ref{rotation} we look at the effect rotation has on the varicose and kink instabilities.  We conclude in section \ref{conclusion}.

\section{Numerical Model}\label{numerical} 
We are studying magnetised neutron stars through the time evolution of the ideal MHD equations in general relativity utilising the {\sc horizon} code \cite{zink11,lasky11,zink12}.  In \citet{zink11} and \citet{lasky11} we have presented brief outlines of our numerical method, however it is worth elaborating on this in more detail.  Throughout this section we assume geometrised units such that $G=c=1$, although we retain specific units for the remainder of the article following this section.

\subsection{General Relativistic Magnetohydrodynamics}\label{GRMHD}
To express the equations of General Relativistic Magnetohydrodynamics (GRMHD) in a form appropriate for numerical integration, we follow closely the formalism outlined in \citet{gammie03}.  To this end, we define a four-vector that is orthogonal to hypersurfaces of constant time, $t$, which has components $n^{\mu}=\frac{1}{\alpha}\left(1,\,-\beta^{i}\right)$.  Here, $\alpha$ is the lapse function and $\beta^{i}$ are the spatial components of the shift vector.  The line element of such a spacetime can then be expressed in coordinates $x^{\mu}=\left(t,x^{i}\right)$ as
\begin{align}
	ds^{2}=-\left(\alpha^{2}-\beta_{a}\beta^{a}\right)dt^{2}+2\beta_{a}dtdx^{a}+\gamma_{ab}dx^{a}dx^{b},
\end{align}
where $\gamma_{\mu\nu}=g_{\mu\nu}+n_{\mu}n_{\nu}$ is the induced three-metric. 
 
We denote the four-velocity comoving with the fluid as $u^{\mu}$, allowing a definition of the three-velocity of the fluid as measured by the observer moving along $n^{\mu}$;
\begin{align}
	v^{i}=\frac{1}{W}{h^{i}_{\alpha}u^{\alpha}},
\end{align}
where $W=-u_{\alpha}n^{\alpha}$ is the relative Lorentz factor between the two observers and $h_{ij}$ is the induced three-metric on the constant $t$ hypersurfaces.

The electromagnetic field is now defined by the anti-symmetric Faraday tensor, $F_{\mu\nu}$, with the four Maxwell equations given by
\begin{align}
	\nabla_{[\mu}F_{\nu\sigma]}=&0,\\
	\nabla^{\alpha}F_{\mu\alpha}=&4\pi \J_{\mu}.
\end{align}
Here, $\J^{\mu}$ is the electric four-current and square brackets denote anti-symmetrization of the indices.  The four-current can be decomposed into components in and orthogonal to the fluid four-velocity which, together with Ohm's law, can be expressed as
\begin{align}
	\J^{\mu}=\varepsilon u^{\mu}+\sigma F^{\mu\alpha}u_{\alpha},
\end{align}
where $\varepsilon$ is the proper charge density and $\sigma$ is the electric conductivity as measured in the comoving frame.

The electric and magnetic fields for any observer can be evaluated by appropriately contracting the Faraday tensor with the observers four velocity.  For example, the electric and magnetic field vectors as observed in the frame orthogonal to hypersurfaces of constant $t$ can be expressed respectively as
\begin{align}
	E^{\mu}=&F^{\mu\alpha}n_{\alpha},\\
	B^{\mu}=&{}^{\star}F^{\mu\alpha}n_{\alpha}=\frac{1}{2}\epsilon^{\mu\alpha\beta\gamma}F_{\alpha\beta}n_{\gamma},
\end{align}
where ${}^{\star}F^{\mu\nu}$ is the Hodge dual of the Faraday tensor and $\epsilon^{\mu\nu\sigma\tau}$ is the Levi-Civita alternating pseudo-tensor.

Moreover, we define the magnetic field observed in the comoving frame of the fluid as 
\begin{align}
	b^{\mu}={}^{\star}F^{\mu\alpha}u_{\alpha}.
\end{align}
Throughout this article we assume our fluid to be perfectly conducting (i.e. we are working in the ideal MHD approximation), implying we take the limit of $\sigma\rightarrow\infty$.  In order to keep the current finite,  this implies $F^{\mu\alpha}u_{\alpha}=0$, implying the electric field observed by the comoving observer vanishes.  This property allows the electric field in any frame to be expressed in terms of the magnetic field and the relevant four-velocities, implying the electric field no longer enters the calculations.  

The stress-energy tensor is expressed in terms of the fluid (assumed herein to be a perfect fluid) plus electromagnetic components, which are respectively given by
\begin{align}
	T^{{\rm fluid}}_{\mu\nu}&=\rho h u_{\mu} u_{\nu}+pg_{\mu\nu},\\
	T^{{\rm EM}}_{\mu\nu}&=\frac{1}{4\pi}\left(F_{\alpha\mu}{F^{\alpha}}_{\nu}-\frac{1}{4}g_{\mu\nu}F_{\alpha\beta}F^{\alpha\beta}\right).
\end{align}
Here, $\rho$ is the energy-density of the fluid, $h=1+\varepsilon+p/\rho$ the specific enthalpy, with $\varepsilon$ being the specific internal energy and $p$ the isotropic pressure.  We assume a polytropic equation of state (EoS) for our neutron star models, such that $p=K\rho^{\Gamma}$, discussing this in more detail below.

With the above stress-energy tensors, the conservation law $\nabla_{\alpha}{T^{\alpha}}_{\mu}=0$, can be expressed in a coordinate basis as
\begin{align}
	\frac{\partial}{\partial x^{0}}\left(\sqrt{-g}{T_{\mu}}^{0}\right)+\frac{\partial}{\partial x^{i}}\left(\sqrt{-g}{T_{\mu}}^{i}\right)=\sqrt{-g}{T_{\alpha}}^{\beta}{\Gamma^{\alpha}}_{\mu\beta},
\end{align}
where ${\Gamma^{\mu}}_{\nu\sigma}$ are the Christoffel symbols.  Moreover, the spatial and temporal components of the induction equation can also be expressed respectively as
\begin{align}
	\frac{\partial}{\partial t}\left(\sqrt{-g}B^{i}\right)+\frac{\partial}{\partial x^{j}}\left[\sqrt{-g}\left(b^{j}u^{i}-b^{i}u^{j}\right)\right]=&0,\\
	\frac{\partial}{\partial x^{i}}\left(\sqrt{-g}B^{i}\right)=&0.
\end{align}

\subsection{The {\sc horizon} Code}\label{horizon}
{\sc Horizon} is a GPU based numerical code borne out of the CPU, general relativistic hydrodynamics code {\sc thor} \cite{zink08,korobkin11}.  As such, {\sc horizon} solves the equations of GRMHD outlined in section \ref{GRMHD}.  Throughout this article we generate initial conditions using the {\sc lorene} spectral code\footnote{http://www.lorene.obspm.fr/}, which produces self-consistent solutions of the coupled Einstein-Maxwell field equations in ideal MHD \cite{bocquet95}.  The {\sc lorene} solver finds only solutions with purely poloidal magnetic fields including rotation, and we therefore restrict our attention to these models in the present article\footnote{It is worth noting that alternative methods exist for generating initial conditions, for example the numerical codes of \citet[][known as {\sc XNS}]{bucciantini11} or \citet{kiuchi08}, which both find axisymmetric equilibria for stars with purely toroidal fields.  Alternatively, one can use the {\sc RNS} code \cite{stergioulas95} for solving the equations of general relativistic hydrodynamic equilibrium, subsequently imposing a magnetic field as a perturbation.  This latter approach, while initially providing a small constraint violation, has the advantage that arbitrary magnetic field topologies can be imposed on the system.}.  

Following the generation of initial conditions, we map the spectral grid to a regular, three-dimensional Cartesian mesh.  Throughout the present article we do not add any artificial perturbation to the system, relying instead on noise created through the mapping process to add a pseudo-random perturbation to the system.  Our Cartesian grid typically contains $120^{3}$ grid points.  We have extensively tested this grid resolution, with some details given below.  We impose a low-density, artificial atmosphere in the region of our spacetime exterior to the star.  That is, for all grid cells for which the density drops below some critical value, we impose $\rho=\rho_{{\rm atm}}$, which for the evolutions presented herein is $\rho_{{\rm atm}}=10^{-8}$ in units where $c=G=M_{\odot}=1$.  We subsequently allow for the full evolution of the magnetic field in this region (see below for details of the evolution).  This is in contrast to many GRMHD numerical studies where, for example, the magnetic field is confined to the interior \cite[e.g.,][]{duez05,shibata05}, the magnetic field is prohibited from evolving in the exterior \cite{cerdaduran08} or where {\it ad hoc} magnetic diffusivity terms are added to the induction equation \cite{giacomazzo07,ciolfi11}.  Our method still does not treat the exterior of the star correctly (one would require significantly more complicated magnetospheric physics such as radiation transport and force-free magnetic fields), however it does allow for a free evolution of the magnetic field at the surface of the star which is important for the dynamical evolution.  

The outer boundary of our star is located approximately $1.4$ times the stellar radius (at the closest point), and there we adopt Dirichlet boundary conditions for the evolution of the magnetic field.  Dirichlet boundary conditions are restrictive in the sense that they do not allow the magnetic field to evolve at the outer boundary.  However, we persist with these conditions as they are found to be the most stable for our numerical simulations.  In particular, we have attempted both Neumann boundary conditions as well as linear extrapolation techniques, both of which eventually introduce numerical instabilities into the atmosphere of the star.  For a limited subset of these simulations we were able to evolve for long enough to track the magnetic field instability (see section \ref{nonrotate}), exhibiting minimal difference in growth timescale and topological behaviour of the instability between different boundary conditions.  We are unable to rigorously test our equilibrium configurations against these boundary conditions due to the numerical instabilites, however we do note that a majority of the magnetic field evolution is driven by the interior dynamics of the star, with the exterior magnetic field evolving very little.  We discuss this in more detail in section \ref{endstate}.  It is also worth noting that we have rigorously tested our evolutions against the location of the outer boundary and found no discernible difference with {\it either} the nature of the instability or the end-state of the magnetic field configuration.  

Throughout the present article we adopt the Cowling approximation, such that the spacetime metric is held fixed throughout the evolution.  This has the obvious benefit of significantly reducing computational costs.  Moreover, throughout the article we shall be looking at reconfigurations of the magnetic field, which generally occurs on small-scales around the stellar interior.  In \citet{lasky11} we discussed how these instabilities act on equipotential surfaces, implying Cowling is a good approximation.  We do note however, that the adoption of the Cowling approximation does not allow us to study the damping of various oscillation modes due to gravitational wave emission.  Particularly for the $f$-mode, this is anticipated to be the main source of damping, acting on timescales of order $0.1\,\mbox{s}$ \cite{detweiler75}.  Other, lower frequency modes however may last significantly longer \cite{mcdermott88}, a point discussed in more detail below.

The initial conditions are given in terms of the primitive variables, and must therefore be converted into the conservative variables utilised in the evolution code.  These relations are algebraic, and can therefore trivially be evaluated in each cell.  The temporal evolution then requires the computation of the cell face flux vectors, which are obtained by solving a local Riemann problem across each cell (see Ref.\cite{zink11} for details).  This method solves for the conserved variables at each time step, however the source terms  require knowledge of the primitive variables at each step.  Transforming the conserved variables to the primitive variables is not, in general, an algebraic process.  To this end we adopt a one-dimensional polytropic recovery scheme \cite{noble06}, employing a Newton-Raphson method to solve the non-linear algebraic equations.  As a final note, we adopt hyperbolic divergence cleaning according to the prescription outlined in \citet{anderson06}, which maintains a divergence free magnetic field throughout the evolution.

\subsection{Shock Tube Code Tests}\label{CodeTests}
Prior to detailing results, we shall spend some time exploring two standard shock tube code tests which give confidence in the numerical accuracy of our results.  These are the \citet{sod78} and \citet{balsara01} tests, which were first presented for the {\sc horizon} code in \citet{zink11}.

The \citet{sod78} test only includes the hydrodynamic portion of the code (i.e. in the absence of magnetic fields).  In particular, a Riemann problem is prepared for an ideal gas with $\Gamma=5/3$ and initial states $\rho_{L}=1$, $P_{L}=1$, $u_{L}^{i}=0$ and $\rho_{R}=0.125$, $P_{R}=0.1$, $u_{R}^{i}=0$.  This problem is prepared on a full $120^{3}$ grid, and simulated up to $t=0.8$.  A reference solution has further been created using the well-tested {\sc thor} code \cite{zink08,korobkin11} with a very fine grid.

\begin{figure}
	\begin{center}
	\includegraphics[angle=0,width=.95\columnwidth]{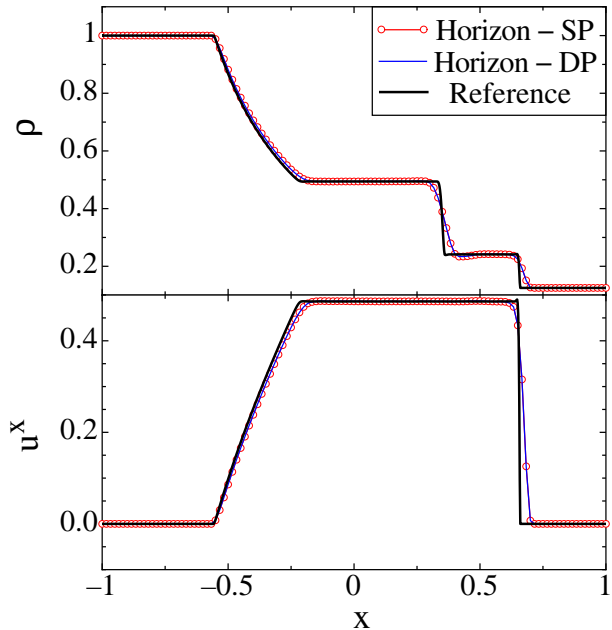}
	\end{center}
	\caption{\label{sod} Density profile (top panel) and $x$ component of the velocity field ($u^{x}$; bottom panel) for the Sod test.  The red line with open circles and the blue thin line are the {\sc horizon} evolutions using single and double precision respectively.  The reference solution (thick black line) was obtained with a very high resolution simulation by the well-tested {\sc thor} code.  These plots show the end of the simulation after $t=0.8$.}
\end{figure}

In figure \ref{sod} we plot the density (top panel) and the $x$ component of the velocity profiles at the end of the evolution for both single and double precision simulations, as well as for the reference solution.  The dissipative nature of the numerical method can be seen with the subtle difference between the {\sc horizon} runs and the finer-grid {\sc thor} simulation.  The largest difference between the single and double precision results are $\left|\delta\rho\right|\approx5\times10^{-5}$ and $\left|\delta u^{x}\right|\approx10^{-4}$, which appear exactly at the location of the shock.  In the smooth parts of the fluid flow, these differences are typically $\left|\delta\rho\right|<10^{-6}$ and $\left|\delta u^{x}\right|<10^{-6}$.  These errors are at or lower than the level of the discretisation error of the numerical scheme.

\begin{figure}
	\begin{center}
	\includegraphics[angle=0,width=.95\columnwidth]{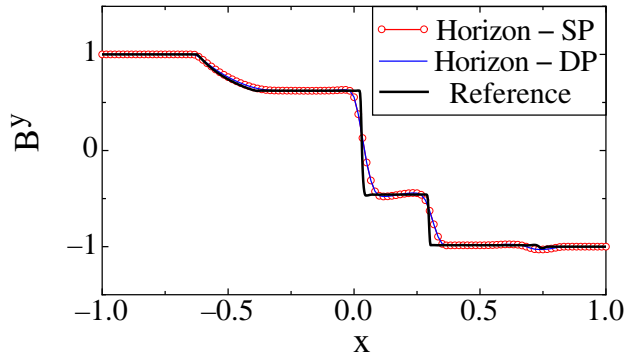}
	\end{center}
	\caption{\label{balsara} Profile of the $y$ component of the magnetic field, $B^{y}$, for the Balsara test after $t=0.8$ showing both single and double precision accuracies.  The reference solution is again obtained using a very high resolution grid in the {\sc thor} code.}
\end{figure}

The \citet{balsara01} test includes the magnetic field, which contains the same initial conditions as for the Sod test, with the additional conditions of $B_{L}^{i}=\left(0.5\,1,\,0\right)$ and $B_{R}^{i}=\left(0.5,\,-1,\,0\right)$.  In figure \ref{balsara} we plot the $y$ component of the magnetic field vector at the end of the evolution for single and double precision and also for the reference solution.  The magnetic field evolution is not seen to be affected by the use of single precision accuracy, with the absolute difference being $\left|\delta B^{y}\right|<10^{-6}$.  Similar values hold for other evolved variables.

Details of tests on rapidly rotating, {\it non}-magnetised neutron stars, as well as tests involving toroidal magnetic fields in non-rotating magnetised stars, can be found in \citet{zink11}.  These include recovering the correct spectral frequencies (as compared with accurate, linear codes) of various fluid modes for rapidly rotating stars.  We omit the details of these tests in the present article as we shall also be recovering the frequencies of various oscillation modes in our fiducial models and, in particular, comparing these with the results from accurate linear codes (see section \ref{modes}).  Moreover, we detail the remainder of our numerical tests performed for our specific fiducial models in the following sections alongside the corresponding results.

\section{Fiducial Simulation}\label{nonrotate}

In this section we review the key results of \citet{lasky11} and \citet{zink12}, providing significantly more details than published in those two works.  In particular, in order to discuss the magnetic field kink and varicose instabilities, as well as the equilibrium configurations presented in \citet{lasky11}, we utilise a fiducial model with polytropic EoS with $K=100$, $\Gamma=2$.  Such a model has gravitational mass of $1.3\,M_{\odot}$ and equatorial radius $R=12.6\,{\rm km}$.  Moreover, this fiducial model has an average internal magnetic field strength (i.e. where $\rho>\rho_{{\rm atm}}$) of $\bar{B}_{15}=13$, where $B_{15}=B/10^{15}\,{\rm G}$.  To make contact with neutron star observations we express the magnetic field in terms of the surface field strength evaluated at the pole of the star, which for this fiducial model is $B_{15}=8.8$.  Finally, the characteristic Alfv\'en timescale, defined according to
\begin{align}
	\tau_{A}=\frac{2R\sqrt{4\pi\left<\rho\right>}}{\left<B\right>},
\end{align}
where $\left<\ldots\right>$ represents a volume weighted average, evaluates to $\tau_{A}=5.0\,{\rm ms}$ for our fiducial models.  

\begin{figure}
	\begin{center}
	\includegraphics[angle=0,width=.95\columnwidth]{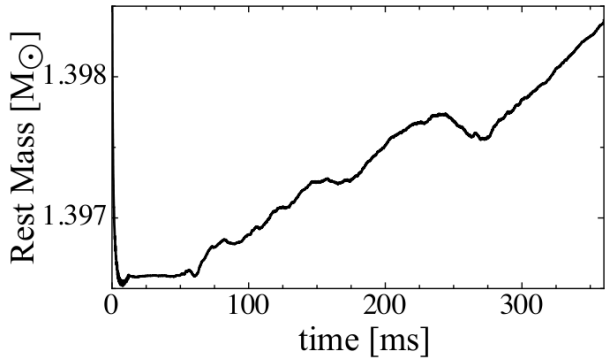}
	\end{center}
	\caption{\label{restmass}Rest mass as a function of time for a non-rotating polytropic model with $K=100$, $\Gamma=2$ and central magnetic field of $1.0\times10^{17}\,{\mbox G}$.  An initial rearrangement of the rest mass is seen in the first milliseconds.  This quantity is conserved on the level of $\sim0.3\%$.}
\end{figure}

In figure \ref{restmass} we plot the evolution of the rest mass for our fiducial simulation.  One can see an initial loss of mass in the first few milliseconds of the simulation.  This corresponds to errors associated with mapping the {\sc lorene} spectral grid to our Cartiesian grid.  This initial mass loss soon reaches a new equilibrium, at which point the mass evolves almost unchanged until approximately $75\,{\rm ms}$ into the simulation.  At this point, which corresponds to the nonlinear saturation of the magnetic field instability (see section \ref{instability}), the mass begins to slowly increase.  After more than $t=360\,{\rm ms}$ we see a total change in the rest mass of less than $0.15\,\%$.  We note that, even for our models with the strongest magnetic fields of more than $10^{17}\,{\rm G}$ in the center of the star, we see less than a $0.3\,\%$ change in the total rest mass over many hundreds of milliseconds.

\begin{figure}
	\begin{center}
	\includegraphics[angle=0,width=.95\columnwidth]{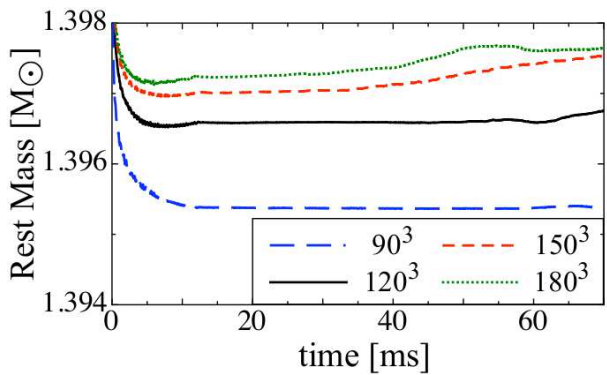}
	\end{center}
	\caption{\label{convrestmass} Evolution of rest mass for model presented in figure \ref{restmass} with $120^{3}$ grid-points at single precision (SP; black thick line), $120^{3}$ grid-points at double precision (DP; red dotted line) and $180^{3}$ grid-points at SP (dashed blue line).}
\end{figure}

In figure \ref{convrestmass} we present evolutions of the rest mass for different resolution simulations, using the same fiducial model presented in figure \ref{restmass}.  In particular we show four simulations with $90^{3}$, $120^{3}$ (our canonical model from figure \ref{restmass}), $150^{3}$ and $180^{3}$ grid-points.  One can see here that the lower resolution simulations have a larger mapping error at the beginning of the simulation, however over the remaining $70\,{\rm ms}$ shown in this plot there is little deviation in the rest mass, independent of resolution.

\subsection{Mode Analysis}\label{modes}
The benefits of understanding pressure related modes in our fiducial model are two-fold.  Firstly, they provide an excellent code check that we are reproducing known results from the literature, and secondly they will aid in our understanding of the gravitational wave emission (see sections \ref{GWs1} and \ref{GWs}).  

\begin{figure}
	\begin{center}
	\includegraphics[angle=0,width=0.95\columnwidth]{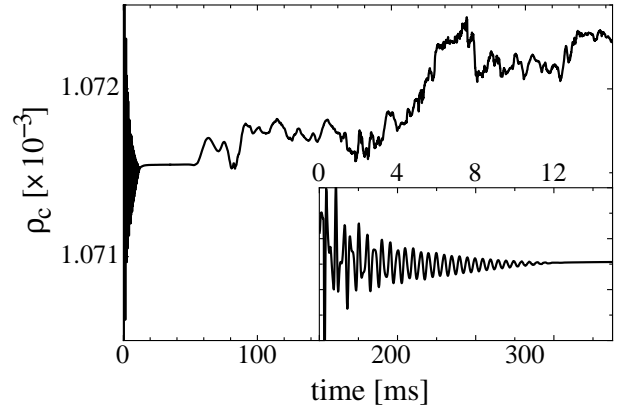}
	\end{center}
	\caption{\label{rhomaxevol} Evolution of the central energy-density, $\rho_{c}$, for our canonical model.  The full evolution lasts over $360\,{\rm ms}$, while the inset in the plot shows only the first $15\,{\rm ms}$ of the evolution.  Note that the central energy-density is given in natural units such that $c=G=M_{\odot}=1$. The large perturbation seen in the initial evolution is due to the mapping between the {\sc lorene} spectral code and our Cartesian grid.}
\end{figure}

In figure \ref{rhomaxevol} we plot the evolution of the central energy-density, $\rho_{c}$, as a function of time.  As mentioned above, our simulations are only perturbed through the mapping process between the {\sc lorene} spectral grid and our Cartesian grid on which the evolutions are performed.  In the initial moments of the simulation, i.e. for approximately the first $15\,{\rm ms}$, this induces a perturbation in the central energy-density that can be seen clearly in the inset of figure \ref{rhomaxevol}.  This initial perturbation is damped by numerical viscosity such that is it no longer visible approximately $15$ to $20\,{\rm ms}$ into the simulation.  The central density then evolves with little variation until approximately $50\,{\rm ms}$, at which point the magnetic field instability is seen to disrupt the density in the middle of the star.  The subsequent evolution of the central density also contains significant motion due to the additional kinetic energy in the star provided by the magnetic field instability.  We discuss this in significantly more detail in the following section.  

\begin{figure}
	\begin{center}
	\includegraphics[angle=0,width=0.95\columnwidth]{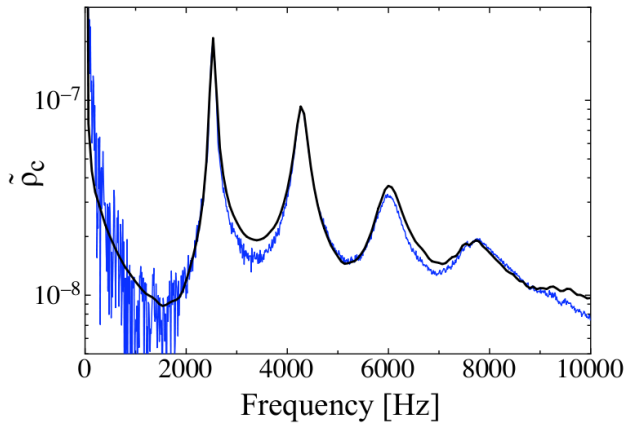}
	\end{center}
	\caption{\label{rhomaxFFT} Fourier transform of the central energy-density, $\tilde{\rho}_{c}$, for our canonical model.  The black line is a Fourier transform of the first $15\,{\rm ms}$ of the evolution, which has the greatest perturbation, and hence cleanest signal for the pressure modes.  The blue line is for the entire evolution lasting more than $350\,{\rm ms}$.  Here we see the fundamental radial $F$-mode at $\sim2530\,{\rm Hz}$ and its overtones. }
\end{figure}

In figure \ref{rhomaxFFT} we show the Fourier transform of the central density, $\tilde{\rho}_{c}$, for the first $15\,{\rm ms}$ of the evolution (black line) and for the entire $360\,{\rm ms}$ (blue line).  The two curves are scaled such that the sizes of the first peak are equivalent.  The black curve provides a significantly cleaner signal than the blue curve due to the initial signal shown in the inset of figure \ref{rhomaxevol}.  The four strong peaks seen at approximately $2530$, $4280$, $5990$ and $7750\,{\rm Hz}$ represent the purely radial, $\ell=0$, $F$-mode and its three lowest overtones respectively.  

We compare this $F$-mode frequency with that found in the literature.  In particular, \citet{gaertig08} utilised a linear numerical code to study stellar oscillations of $\Gamma=2$, $K=100$ polytropes, however with different central densities to those used herein.  They used stellar models with gravitational mass $M=1.4\,{\rm M}_{\odot}$, and equatorial radius $R=14.15\,{\rm km}$, for which the $F$-mode  was found to have a frequency of $2679\,{\rm Hz}$.  Given that the mode frequency scales with the compactness of the star, $M/R$, scaling their frequency to our stellar model gives a corresponding $F$-mode frequency of $2547\,{\rm Hz}$, which can be compared to our derived result of $2530\,{\rm Hz}$.  This small difference in frequency is within our error margin given the calculation, however we further note that the \citet{gaertig08} calculations did not include magnetic fields as was done herein.

\begin{figure}
	\begin{center}
	\includegraphics[angle=0,width=0.95\columnwidth]{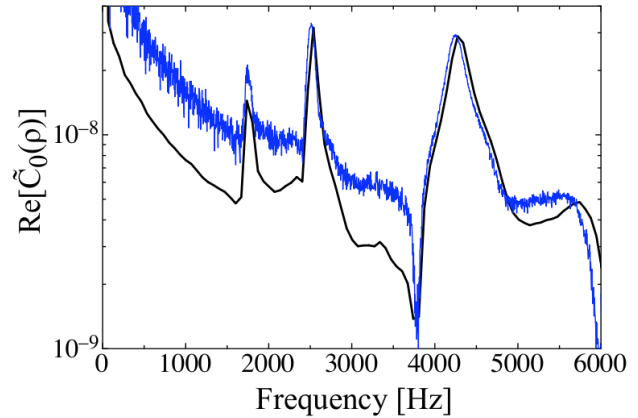}
	\end{center}
	\caption{\label{Crho6FFT} Fourier transform of the real part of $C_{0}\left(\rho\right)$ as defined in equation (\ref{cmB}) and extracted at three-quarters of the stellar radius.  The black line is a Fourier transform of the first $15\,{\rm ms}$ of the evolution, which has the greatest perturbation, and hence cleanest signal for the pressure modes.  The blue line is for the entire evolution lasting more than $360\,{\rm ms}$.  The fundamental $\ell=2$ non-radial $f$-mode is seen at $\sim1750\,{\rm Hz}$, the $F$-mode at $\sim2530\,{\rm Hz}$ and the $\ell=2$, $p_{1}$-mode at $\sim4250\,{\rm Hz}$.}
\end{figure}

As expected, the central energy-density evolution shows no hint of any non-radial pressure modes.  To look at these modes we track various quantities at different radii throughout the star.  In particular, for this, and also for the purpose of tracking the magnetic field instability (section \ref{instability}), we evaluate a Fourier decomposition of various physical quantities on a ring in the equatorial plane.  That is, we compute complex weighted averages \cite[e.g.][]{zink07}
\begin{align}
	C_{m}\left(f\right)=\frac{1}{2\pi}\int_{0}^{2\pi}f\left(\varpi,\phi,z=0\right){\rm e}^{im\phi}d\phi,\label{cmB}
\end{align}
where $\varpi=\sqrt{x^{2}+y^{2}}={\rm const.}$ lies in the initial equatorial plane of the magnetic field.  To track fluid modes we look at the quantity $C_{m}\left(\rho\right)$.  In figure \ref{Crho6FFT} we plot the Fourier transform of the real part of $C_{0}\left(\rho\right)$, extracted at $\varpi=0.75\varpi_{\star}$, where $\varpi_{\star}$ is the equatorial stellar radius.  Again, the black line represents the Fourier transform of only the first $15\,{\rm ms}$ of the evolution, while the blue line is the Fourier transform of the entire evolution.  In this signal we see considerable noise in the lower frequency band of the spectrum, however we also see a considerable number of distinct modes.  In particular, the first mode located at approximately $1750\,{\rm Hz}$ is the fundamental, non-radial $\ell=2$ $f$-mode.  The second peak is again the $F$-mode seen in figure \ref{rhomaxFFT}, and the final mode at $\sim4250\,{\rm Hz}$ is the $p_{1}$-mode.

Taking literature values from \citet{gaertig08} for the $\ell=2$, $f$-mode, we find a re-scaled value of the frequency for their models to be approximately $1797\,{\rm Hz}$, which is again in rough agreement with the $1750\,{\rm Hz}$ frequency found herein.

\subsection{Instability}\label{instability}
\begin{figure}
	\begin{center}
	\includegraphics[angle=0,width=.95\columnwidth]{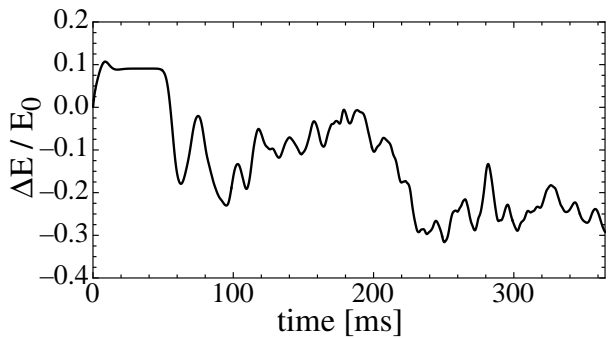}
	\end{center}
	\caption{\label{EnergyCons} Evolution of the change in total magnetic energy, $\Delta E/E_{0}$, for our fiducial model.}
\end{figure}

A broad appreciation of the magnetic field instability can be gained from looking at the time evolution of the change in magnetic energy, plotted in figure \ref{EnergyCons}.  Here, $\Delta E/E_{0}=\left(E-E_{0}\right)/E_{0}$, where $E$ is the total magnetic energy and $E_{0}$ is the magnetic energy at $t=0$.  We see an initial rearrangement of the magnetic field that settles after approximately $t=10\,{\rm ms}=2\,\tau_{A}$.  This is associated with mapping errors present in the rest mass and energy density evolutions, however this takes longer to settle to an equilibrium as the timescale for the magnetic energy is the Alfv\'en crossing time rather than the sound crossing time.  

Following the initial rearrangement, the magnetic energy remains constant until approximately $50{\rm ms}$, at which point a sharp decrease is seen in the energy which is attributed to the kink instability (see below).  This loss of magnetic energy is therefore associated with a conversion to kinetic energy in the system.

This magnetic field instability can be studied in more detail by again computing complex weighted averages of the Fourier decomposition presented in equation (\ref{cmB}).  This time we choose the function $f$ to be the $\phi$ component of the magnetic field, $B_{\phi}$, although we note that other components of the magnetic field, fluid velocity and energy-density also exhibit the instability, however the signal is generally not as clean.

\begin{figure}
	\begin{center}
	\includegraphics[angle=0,width=0.95\columnwidth]{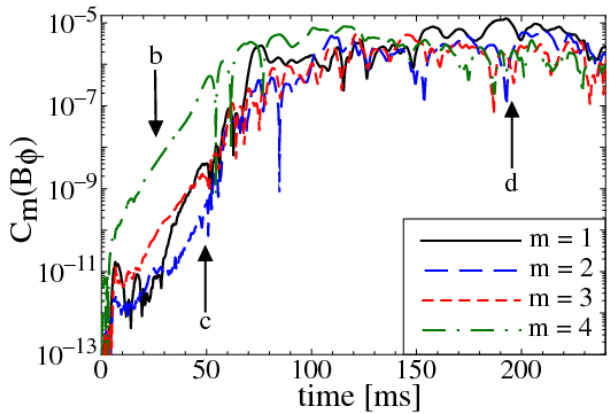}
	\end{center}
	\caption{\label{cmBphi}Evolution of $C_{m}\left(B_{\phi}\right)$ for our fiducial model with average magnetic field strength of $\bar{B}_{15}=13$, which has an Alfv\'en crossing time of $5.0\,{\rm ms}$ and surface polar magnetic field of $B_{15}=8.8$.  The arrows represent the times of the three-dimensional snapshots shown in figure \ref{3D_non_rotate} and the two-dimensional equatorial slices in figure \ref{Slice}.}
\end{figure}

In figure \ref{cmBphi} we plot the evolution of the $m=1,\ldots,4$ modes for $C_{m}\left(B_{\phi}\right)$ for our fiducial model.  These quantities are measured at $\varpi=0.6\varpi_{\star}$, where $\varpi_{\star}$ is the stellar radius.  It is worth noting that the neutral line of these simulations (i.e. the ring around the equatorial plane where ${\bf B}=0$) is located at approximately two-thirds of $\varpi_{\star}$.

\begin{center}
\begin{figure*}
\includegraphics[width=0.49\textwidth]{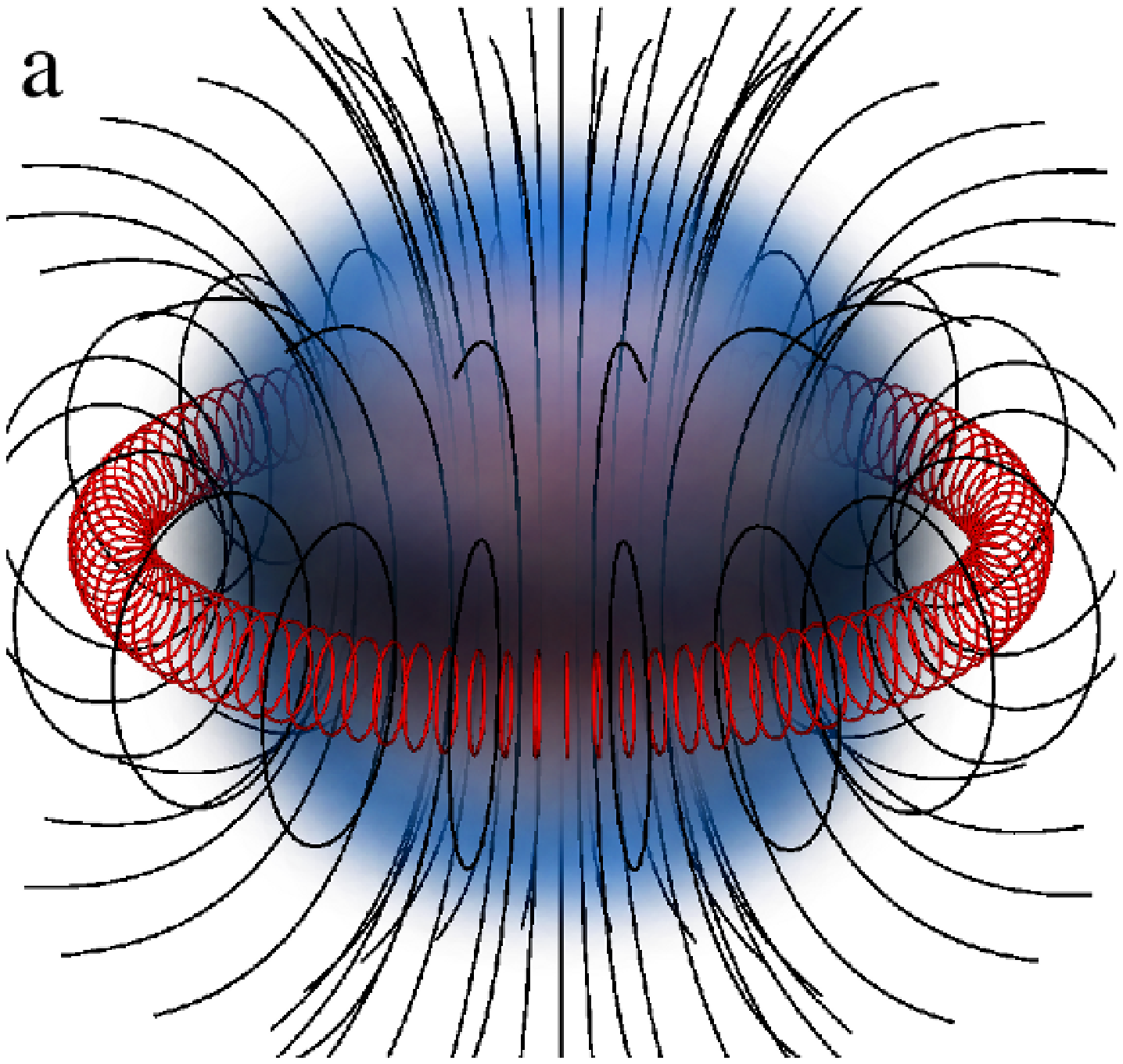}
\includegraphics[width=0.49\textwidth]{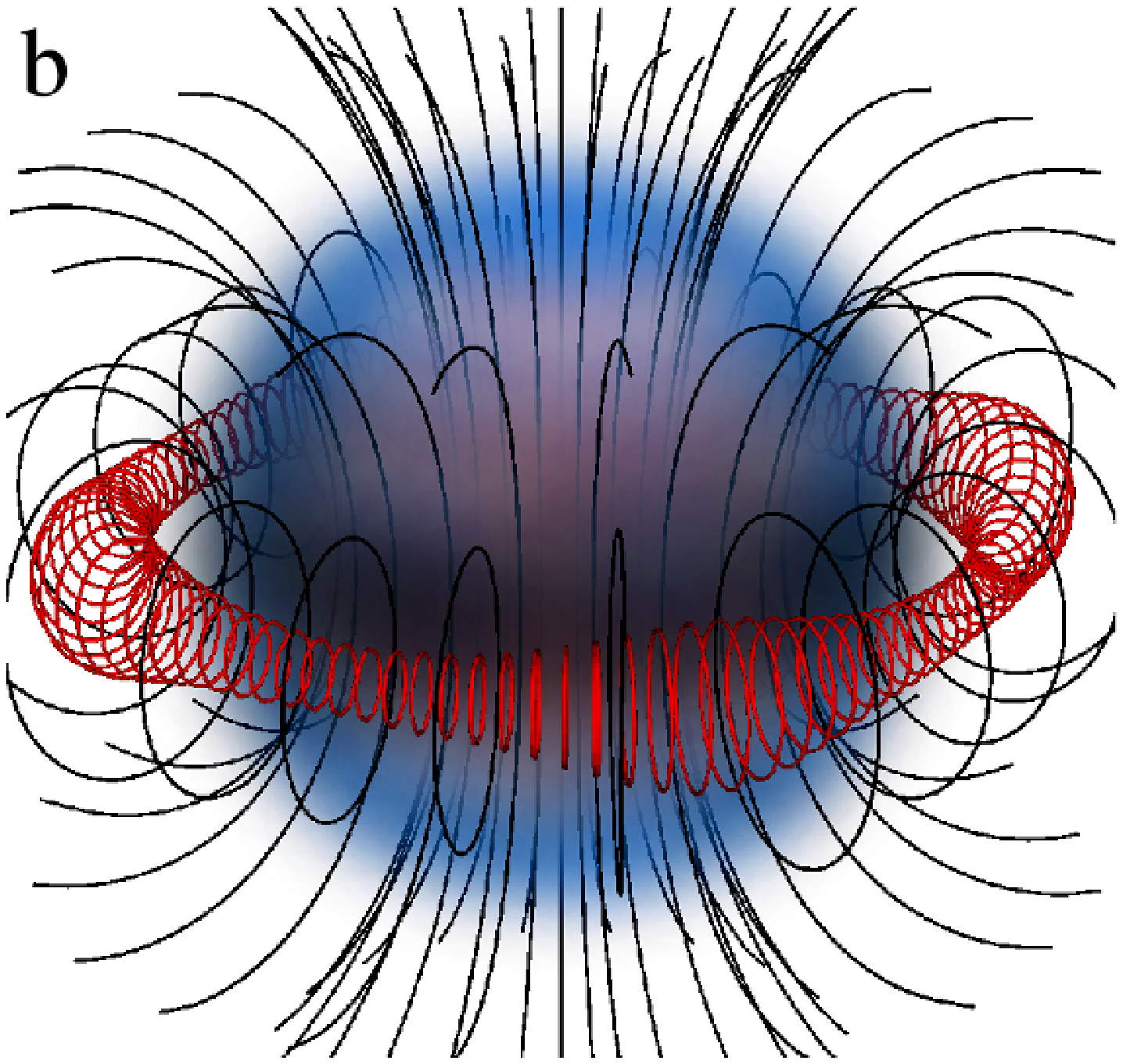}
\includegraphics[width=0.49\textwidth]{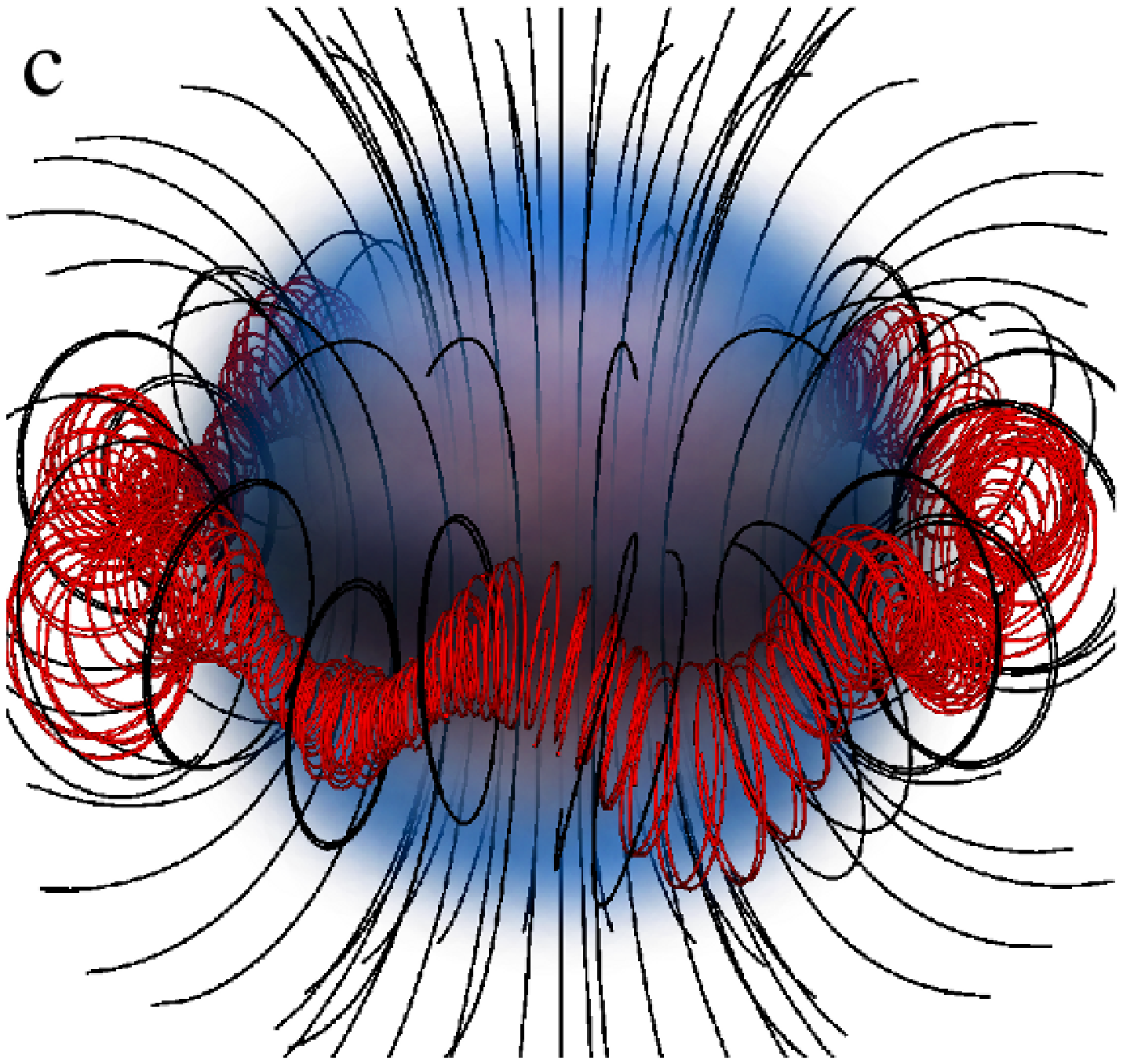}
\includegraphics[width=0.49\textwidth]{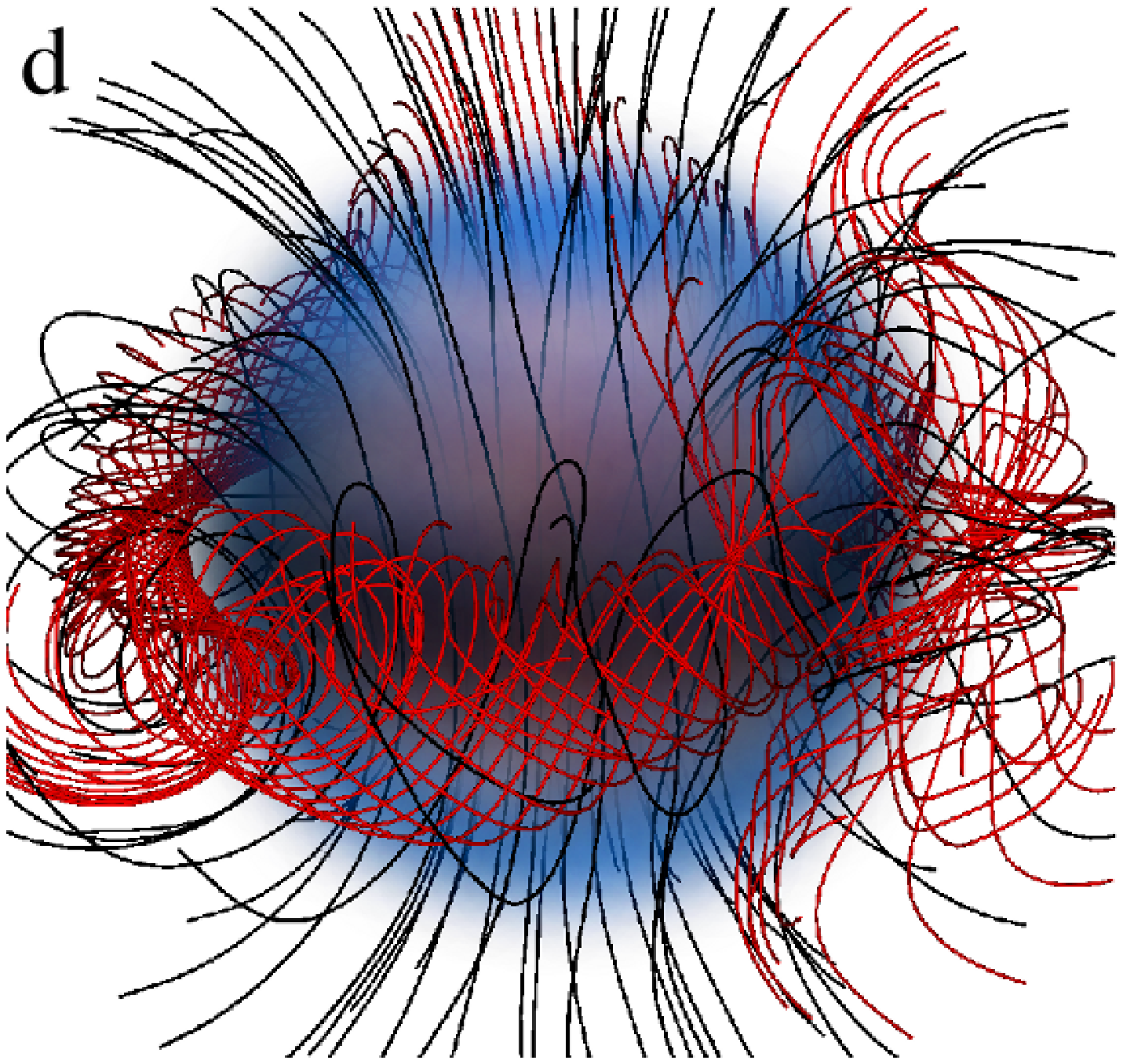}
	\caption{\label{3D_non_rotate}  Time evolution of fiducial model with average magnetic field $\bar{B}_{15}=13$, corresponding to an Alfv\'en wave crossing time of $5\mbox{ms}$.  The figures are a) $t=0\,\mbox{ms}$, b) $t=25\,\mbox{ms}$, c) $t=50\,\mbox{ms}$ and d) $t=195\,\mbox{ms}$.  To more clearly visualise the instability, the red field lines are  seeded on the equatorial plane close to the neutral line, and the black field lines are seeded on the equatorial plane interior to the neutral line.  The volume rendering is an isopycnic surface at $37\%$ of the central rest-mass density, shown to provide contrast with the field lines.}
\end{figure*}
\end{center}

In figures \ref{3D_non_rotate} we present three-dimensional plots of the magnetic field at various instances throughout the evolution.  The blue volume rendering in these figures is an isopycnic surface of $\rho=0.37\rho_{c}$, where $\rho_{c}$ is the central rest-mass density.  This value was chosen as it lies approximately $50\%$ of the radius of the star, and therefore provides contrast to the magnetic field lines.  Two sets of magnetic field lines are plotted, both of which are seeded on the equatorial plane and traced both in the positive and negative $B$ direction.  The red field lines are seeded close to the neutral line in the equatorial plane, whereas the black field lines have been seeded interior to the neutral line.  It is worth noting that the domain of our Cartesian grid is larger than that plotted in figure \ref{3D_non_rotate} and that the field lines are truncated at the surface of the star for clarity in the images.  A movie of the fiducial simulation lasting $400\,\mbox{ms}$ can be viewed at \url{http://www.tat.physik.uni-tuebingen.de/~tat/grmhd/}

Figure \ref{3D_non_rotate}a shows the initial data as imported from the {\sc lorene} spectral code, as described in section \ref{numerical}.  Figure \ref{3D_non_rotate}b shows the evolution of the system after $t=25\,{\rm ms}=5\,\tau_{A}$.  One can clearly see the onset of the 'sausage' or 'varicose' mode \cite{markey73} in this snapshot.  The sausage mode involves a change in the cross-sectional area of a flux tube around the neutral line.  It is clearly visible that this has developed most strongly in the $m=4$ mode, which is confirmed through a comparison with figure \ref{cmBphi}, which shows this mode clearly dominating over the others in this early phase of the evolution.  

The $m=4$ varicose mode in these non-rotating simulations is the result of a transient excitation attributed to the Cartesian grid.  We have verified that this transient reduces with increased grid resolution, however note that the presence of the varicose mode is an inherent characteristic of the system.  The varicose mode is discussed in more detail in \citet{lasky11} for non-rotating simulations, and is also discussed in greater detail in section \ref{rotation} of the present paper in the case of rotating models.  

The varicose mode visually dominates our simulations for almost ten Alfv\'en crossing times before the 'kink' instability appears and begins to dominate the system.  This is presented in figure \ref{3D_non_rotate}c, which is taken at $t=50\,{\rm ms}=10\,\tau_{A}$.  One can see that the kink instability is acting orthogonal to the gravitational field \cite{markey73}, with the presence of the varicose mode still clearly visible.  This mode will have been excited from the beginning of the simulation (as seen clearly in figure \ref{cmBphi}), however it is only at this point that the exponential growth has reached a point at which it is visually obvious.  This therefore represents the non-linear development of the instability where the change in the field structure is of similar order to the background field.  

\begin{figure}
	\begin{center}
	\includegraphics[angle=0,width=0.95\columnwidth]{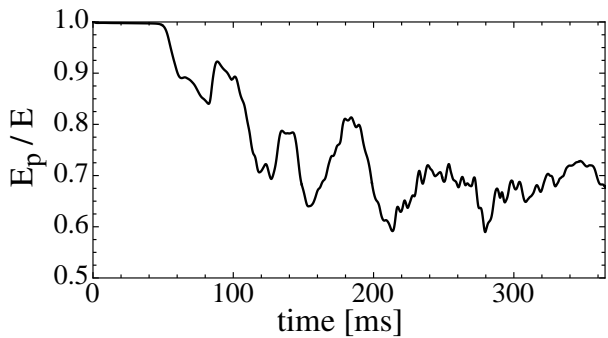}
	\end{center}
	\caption{\label{EpE} Ratio of poloidal-to-total magnetic field energy, $E_{p}/E$, as a function of time for the simulation presented in figures \ref{cmBphi} and \ref{3D_non_rotate}.  The initial configuration is purely poloidal ($E_{p}/E=1$), with the kink instability shown in figure \ref{3D_non_rotate}c causing a large conversion of energy between poloidal and toroidal components.  The instability saturates at approximately $E_{p}/E\sim0.65$.}
\end{figure}

As discussed in \citet{lasky11}, the modal analysis presented in figure \ref{cmBphi} does not distinguish between the varicose and kink modes, implying exponential growth does not indicate an instability in one or the other mode.  In figure \ref{EpE} we plot the ratio of magnetic energy in the poloidal field, $E_{p}$, to the total magnetic energy, $E$, as a function of time for our fiducial simulation.  The initial configuration imported from {\sc lorene} is purely poloidal, i.e. $E_{p}/E=1$.  As discussed, the varicose mode is visually apparent in the simulations for the first $\sim40\,{\rm ms}$, although we can see from figure \ref{EpE} that this mode has no effect on the poloidal energy ratio.  The non-linear development of the kink instability however, causes the $E_{p}/E$ ratio to significantly change.  In other words, it is only the kink mode that introduces a non-zero toroidal component of the field.  This combination of modal analysis and magnetic energy ratios will become important for understanding the evolution of the rotating simulations in section \ref{rotation}.

Finally, in figure \ref{3D_non_rotate}d we show a typical quasi-equilibrium state of the simulation, which, in this case, is $t=195\,{\rm ms}=39\,\tau_{A}$ into the simulation.  We discuss this in more detail in section \ref{endstate}.  

\begin{center}
\begin{figure*}
	\includegraphics[angle=90,width=1.0\textwidth]{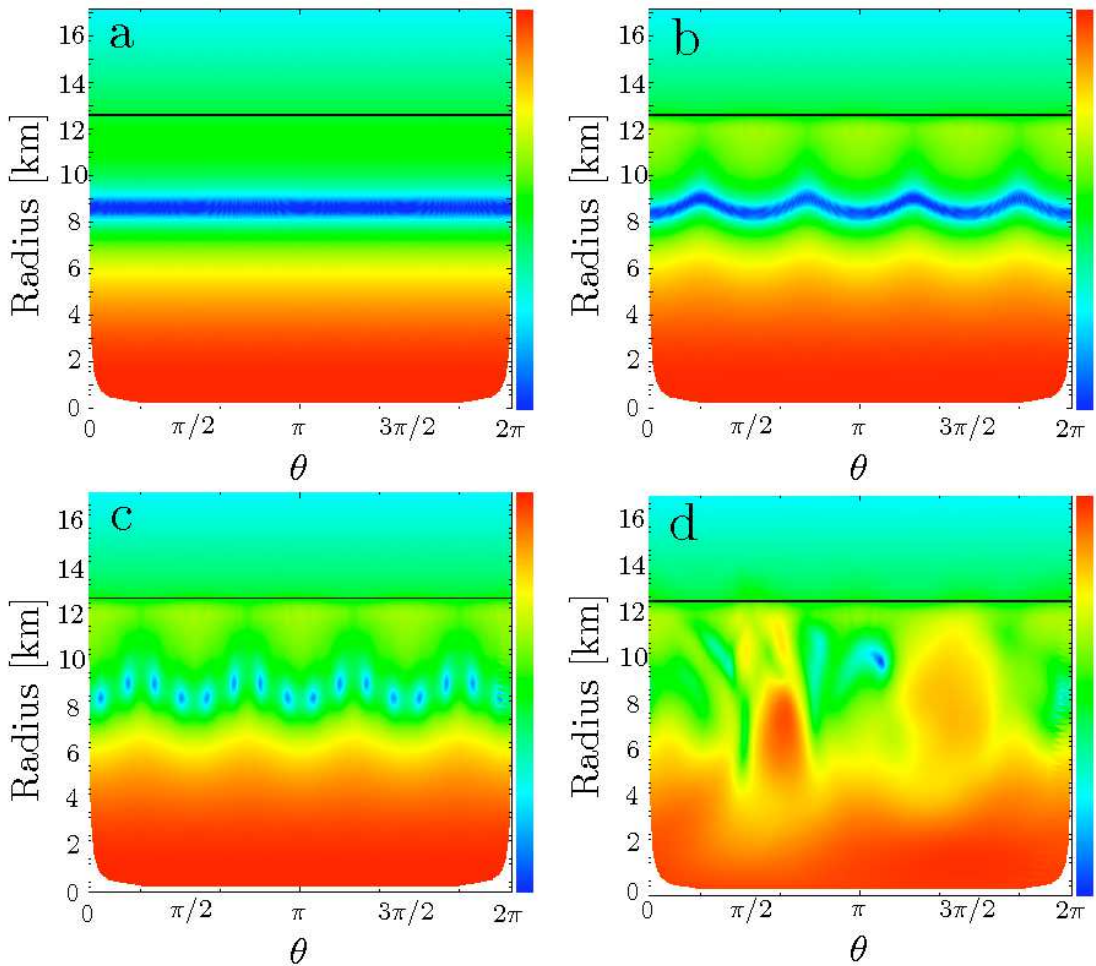}
	\caption{\label{Slice} Equatorial slice of the absolute value of the magnetic field for four different time-points in our fiducial simulation -- i.e. with average magnetic field $\bar{B}_{15}=13$, corresponding to an Alfv\'en wave crossing time of $5\mbox{ms}$.  The magnetic field strength is plotted in a log-scale, with the maximum (shown in red) as $\left|B\right|=4.8\times10^{16}\,{\rm G}$.  Blue represents the minimum value of the magnetic field, and hence the neutral line is clearly visible in figures a, b and c.  The thick black line is the surface of the star.  The figures are a) $t=0\,\mbox{ms}$, b) $t=25\,\mbox{ms}$, c) $t=50\,\mbox{ms}$ and d) $t=195\,\mbox{ms}$.}
\end{figure*}
\end{center}

Figures \ref{3D_non_rotate} can be somewhat misleading in an interpretation of the neutral line of the star.  For this reason, in figures \ref{Slice} we plot the absolute value of the magnetic field on an equatorial slice for the same four timesteps plotted in figure \ref{3D_non_rotate}.  In particular, figure \ref{Slice}a shows the initial conditions where one can see the neutral line at a constant radius of approximately $8.5\,{\rm km}$ (compared to the surface of the star at $12.6\,{\rm km}$).  Figure \ref{Slice}b represents $t=25\,{\rm ms}=5\tau_{A}$ where the $m=4$ varicose mode has come to visually dominate the system.  One can see here that the size of the flux tube around the neutral line has remained constant, however the $m=4$ excitation discussed above has caused the shape of the neutral line to change.  Figure \ref{Slice}c is at $t=50\,{\rm ms}=10\,\tau_{A}$, at which point the kink and varicose modes can both be clearly seen.  In the kink mode, the flux tube around the neutral line has kinked above and below the equatorial plane, which can be seen from the oscillatory nature of the field minima.  Again, figure \ref{Slice}d is after $t=195\,{\rm ms}=39\,\tau_{A}$ and is discussed in more detail in section \ref{endstate}.

\subsection{Gravitational Wave Emission}\label{GWs1}
In \citet{zink12} we utilised these simulations to place estimates on the gravitational wave emission due to an internal rearrangement of a magnetic field.  That is, we used these simulations to mimic the aftermath of a giant magnetar flare that acts to rearrange the internal magnetic field configuration (see also Ref. \cite{ciolfi11}).  We calculate the gravitational wave strain from our simulations utilising the quadrupole formula \cite[e.g.,][]{misner73}
\begin{align}
	h_{ij}=\frac{2\ddot{\rqm}_{ij}}{d},
\end{align}
where an overdot denotes a time derivative, $d$ is the distance between the observer and the source and $\rqm_{ij}$ is the reduced quadrupole moment given by
\begin{align}
	\rqm_{ij}=\int_{V}\rho\left(x_{i}x_{j}-\frac{1}{3}\delta_{ij}x^{2}\right)dV.
\end{align}
Rather than numerically evaluating $\rqm_{ij}$ at each time-step and differentiating twice to get the gravitational wave strain, at each time-step we evaluate $\dot{\rqm}_{ij}$, which can be expressed in terms of spatial quantities utilising the continuity equation.  This significantly decreases differentiation errors due to relatively large time-steps as we are only required to differentiate once with respect to time to find the gravitational wave strain.

\begin{figure}
	\begin{center}
	\includegraphics[angle=0,width=0.95\columnwidth]{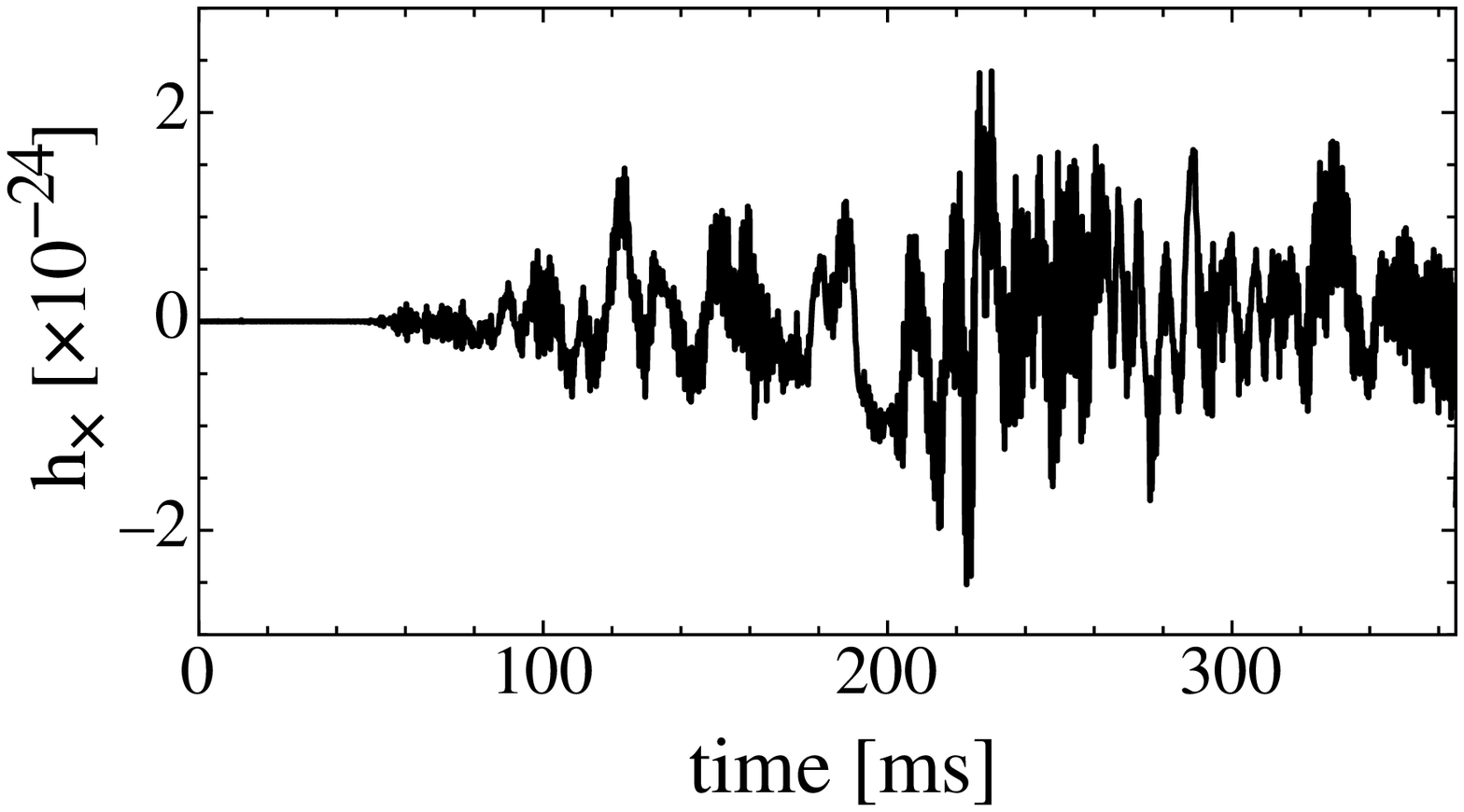}
	\end{center}
	\caption{\label{hecross} Cross-polarisation of the gravitational wave strain, $h_{\times}$, from our fiducial simulation as measured by an observer at $10\,kpc$.  The nonlinear phase of the kink instability induces a prominent gravitational wave signal which is present after approximately $50\,{\rm ms}$.}
\end{figure}

In figure \ref{hecross} we plot the cross-polarisation of the gravitational wave strain as measured by an observer at $d=10\,kpc$.  During the initial phase of the evolution, the strain is of the order of $h_{\times}\sim10^{-27}$, which represents the lower limit of our numerical sensitivity.  The strain amplitude is then excited during the nonlinear phase of the kink instability (i.e. after approximately $50-60\,{\rm ms}$), at which point the strain approaches approximately $h_{\times}\sim10^{-24}$.

\begin{figure}
	\begin{center}
	\includegraphics[angle=0,width=0.95\columnwidth]{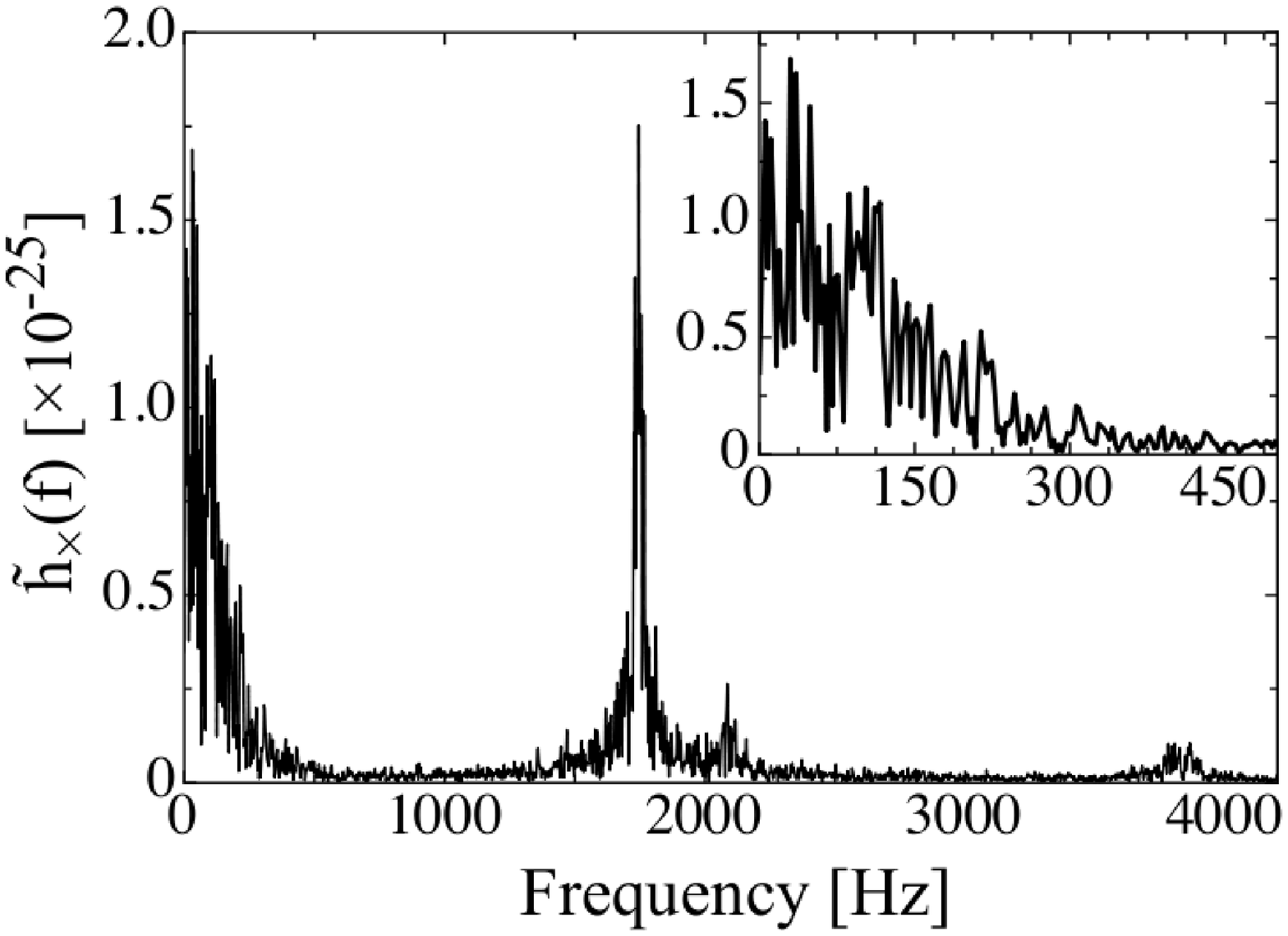}
	\end{center}
	\caption{\label{hecrossFFT} Fourier transform of the cross-polarisation of gravitational wave signal, $\tilde{h}_{\times}(f)$, as a function of oscillation frequency for the fiducial simulation.  The fundamental $f$-mode can be seen as the sharp peak at approximately $1.8\,{\rm kHz}$.  The modes present at less than $300\,{\rm Hz}$ are the lower frequency Alfv\'en modes.  The density of this part of the spectrum can be clearly seen in the plot inset, which shows a zoomed in region between $0$ and $500\,{\rm Hz}$.}
\end{figure}

As expected, the signal in figure \ref{hecross} has the expected $\sim1.8\,{\rm kHz}$ oscillations associated with the $f$-mode.  In figure \ref{hecrossFFT} we plot the Fourier transform of the cross-polarised strain, $\tilde{h}_{\times}\left(f\right)$ for the entirety of the signal shown in figure \ref{hecross} (i.e. from $t=0\,-380\,{\rm ms}$).  The $f$-mode is clearly displayed here as a large peak, with a significantly smaller peak located at $\sim\,3.9\,kHz$.  The signal amplitude present in a gravitational wave detector is given by $\sqrt{T}\left|\tilde{h}_{\times}\left(f\right)\right|$, where $T$ is the damping time of the oscillation.  The dominant mechanism for $f$-mode damping in neutron stars is through gravitational wave emission\footnote{Note that figure \ref{hecross} exhibits no damping of the $f$-mode throughout the long evolution time.  This is because we are working in the Cowling approximation, which implies the system does not lose energy to gravitational radiation.  A more thorough study involving full general relativistic effects is warranted, however we note the extremely large computational costs of simulations evolving the full spacetime render such a task presently intangible.}, which gives a damping timescale of $T\sim100\,-\,300\,{\rm ms}$ \cite[e.g.][]{lindblom83,andersson98}.  For such a situation, this gives a signal amplitude that is well below the detectable limit, even for third generation gravitational wave interferometers such as the proposed Einstein Telescope \cite{sathyaprakash09}.  We discuss this in significantly more detail in section \ref{GWs}.

Figure \ref{hecross} also exhibits large amplitude oscillations at significantly lower frequencies than the $f$-mode.  These can be seen in the temporal evolution of $h_{\times}$ (figure \ref{hecross}), and more clearly in the Fourier transform of this signal (figure \ref{hecrossFFT}).  For our fiducial simulation presented here, the amplitude of the gravitational wave signal in these lower frequency modes is as large, if not larger, than the amplitude of the $f$-mode signal.  The dominant damping mechanism for these modes is largely unknown, although it is generally expected that they will last significantly longer than the $f$-mode signal.  Given that the signal amplitude scales as $\sqrt{T}$, these Alfv\'en modes could be significantly more detectable than the $f$-modes given a large-scale magnetic field rearrangement following a magnetar flare.  This is especially pertinent given that the frequency of these modes coincide with LIGO and VIRGOs most sensitive frequency band.

It is possible that the low frequency Alfv\'en modes seen in the present set of simulations could somehow be associated with the observed quasi-periodic oscillations in the tails of giant flares.  If this were true it would imply a damping time longer than $10$s of seconds, which would significantly enhance the possibility of detection of gravitational waves.  Having said that, we have no direct evidence that the modes we see are in anyway associated with the QPOs, other than them existing in a similar frequency band.  A more rigorous analysis of these modes is required to fully understand their relationship with observations.  However, an explicit analysis of these modes is proving elusive due to the violent dynamics associated with the magnetic field rearrangement.  Moreover, it is expected that the spectrum of these mode will be extremely dense (if not form a mathematical continuum), significantly hampering our efforts to provide a detailed expos\'e on the existence of these modes.  Therefore, for the moment, we shall have to be satisfied with the detailed literature analysing Alfv\'en spectra of magnetars in the linear regime \cite[e.g.][]{levin06,levin07,sotani08a,colaiuda09,lander10,colaiuda11,lander11b,colaiuda12}.

\subsection{Quasi-Equilibria}\label{endstate}
Figures \ref{3D_non_rotate}d and \ref{Slice}d show a typical quasi-equilibrium snapshot at $t=195\,{\rm ms}=39\,\tau_{A}$.  In some ways this snapshot resembles a ``twisted torus'' configuration seen in the nonlinear evolutions of \citet{braithwaite06,braithwaite09}, and in numerous semi-analytic equilibrium derivations that include both poloidal and toroidal field components.  Indeed, the left hand side of the star as seen in figure \ref{3D_non_rotate}d is well approximated by a twisted-torus (i.e. with toroidal components of the field confined to the closed field lines of the poloidal field), and the interior is threaded by a dominantly poloidal field.  However, the remainder of the star exhibits large non-axisymmetric structures with toroidal and poloidal field apparently in equal abundance.  This non-axisymmetric structure can clearly be seen in figure \ref{Slice}d, which also highlights the fact that the neutral line no longer lies on the equatorial plane, and has been severely disturbed by the reconfiguring of the field.

Each of our figures showing time evolution of global quantities (i.e. figures \ref{rhomaxevol}, \ref{cmBphi}, \ref{EpE} and \ref{hecross}) show residual kinetic motion remaining in the star following an evolution lasting almost $80$ Alfv\'en crossing times.  We term the situation we have reached a {\it pseudo}-equilibrium as we can not formally prove whether this state is close to a stable equilibrium or not.  It is clear that the global structure of our pseudo-equilibrium does {\it not} change further with time, in that all global quantities are tightly bounded.  This is a necessary condition for claiming stability, however it is not sufficient.  Given the recent result of \citet{lander12}, who has shown that various twisted-torus configurations originally expected to be stable \cite[e.g.][]{braithwaite08,braithwaite09} actually show instabilities, it is becoming increasingly likely that the gamut of {\it stable} magnetic field equilibria form an extremely complicated set.  

In the exterior region of our star we evolve the magnetic field according to the same prescription as outlined in section \ref{numerical}, however we impose a low density atmosphere ($\rho_{{\rm atm}}\sim10^{-5}\rho_{c}$ where $\rho_{{\rm atm}}$ and $\rho_{c}$ are the atmospheric and central density respectively).  We note that this is not a good model for a magnetosphere as we do not impose a force-free condition.  Moreover, we include boundary conditions that are close to the surface of the star and do not allow significant evolution of the magnetic field (i.e. we use Dirichlet boundary conditions at the outer bounding box of our simulations).  However, we do evolve the magnetic field in the stellar exterior, which allows us to determine the extent to which the external field is altered by the magnetic field instability inside the star.

\begin{figure}
	\begin{center}
	\includegraphics[angle=0,width=0.95\columnwidth]{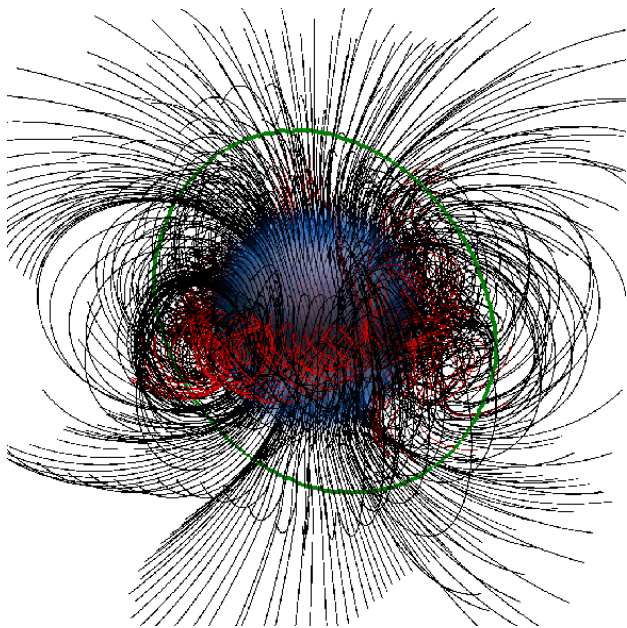}
	\end{center}
	\caption{\label{extfield}Same as figure \ref{3D_non_rotate}d, however now showing the full domain of the grid, and not truncating field lines at the surface of the star, which is shown as the green circle.   Despite a significant rearrangement of the magnetic field inside the star, the exterior has remained significantly unchanged, as evidenced by the fact that it is almost purely poloidal.}
\end{figure}

Figure \ref{extfield} is equivalent to figure \ref{3D_non_rotate}d, however showing the full domain of the Cartesian grid and not truncating the magnetic field lines at the surface of the star.  Despite the large-scale rearrangement of the internal magnetic field topology, one can see from this figure that the exterior magnetic field remains largely unchanged throughout the nonlinear saturation of the instability.  This can be clearly seen by remembering that the initial field is purely poloidal, which is in close approximation to the exterior part of the field shown in figure \ref{extfield}.

\subsection{Effect of Magnetic Field Strength}\label{magfieldstrength}
Until now we have provided a somewhat comprehensive analysis of a single, fiducial simulation with polar surface magnetic field strength of $B_{15}=8.8$, corresponding to an Alfv\'en crossing time of $\tau_{A}=5.0\,{\rm ms}$.  Herein we present the same simulation with differing initial magnetic field strengths between polar surface fields of $B_{15}=6$ and $B_{15}=55$, corresponding to Alfv\'en crossing times of $\tau_{A}=7.3\,{\rm ms}$ and $\tau_{A}=0.9\,{\rm ms}$ respectively.  It is worth noting that our lowest field strength simulation is now only a factor of about two stronger than that observed in magnetars.  

\begin{figure}
	\begin{center}
	\includegraphics[angle=0,width=0.95\columnwidth]{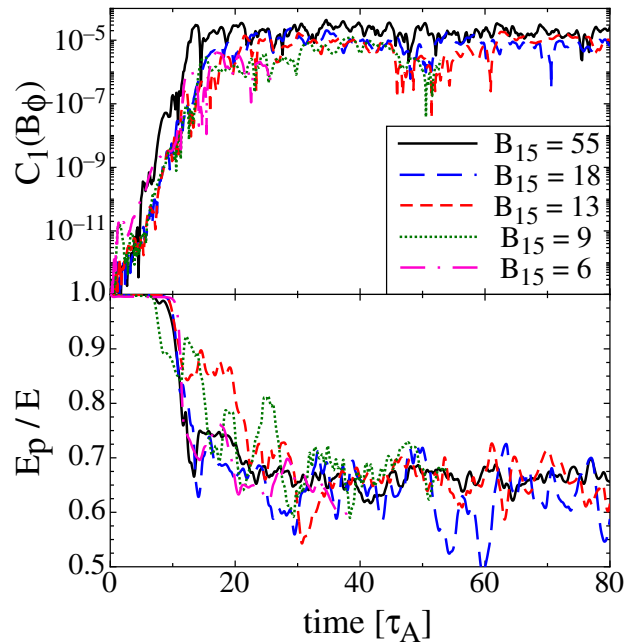}
	\end{center}
	\caption{\label{cmBphimag} (Top panel) Evolution of $C_{m=1}\left(B_{\phi}\right)$ as defined in equation (\ref{cmB}) as a function of time for our fiducial $\Gamma=2$ model with different magnetic field strengths.  The temporal unit is normalised to the characteristic Alfv\'en timescale throughout the star to show invariance of the instability timescale. (Bottom panel) Poloidal-to-total magnetic energy, $E_{p}/E$, as a function of time for our fiducial $\Gamma=2$ model with different magnetic field strengths.  Despite remaining residual motions in the system after up to $100$ Alfv\'en timescales, all models quasi-equilibrium state have $E_{p}/E\sim0.65$.}
\end{figure}

In figure \ref{cmBphimag} we plot the evolution of the $m=1$ component of $C_{m}\left(B_{\phi}\right)$ defined by equation (\ref{cmB}).  We have normalised the temporal axis now to the Alfv\'en crossing time of the system.  One can trivially see an invariance in the growth time of the instability as a function of the Alfv\'en crossing time, and hence magnetic field strength.  

In the bottom panel of figure \ref{cmBphimag} we plot the poloidal-to-total magnetic energy ratio as a function of time, where we have again normalised the temporal unit to the characteristic Alfv\'en crossing time of the system.  While the instability growth timescale of the system can not be as readily established from this plot, one can still see here that this timescale trivially scales with the Alfv\'en crossing time of the system.  Moreover, we see that the poloidal-to-total magnetic energy ratio of the equilibrium state is independent of the magnetic field strength of the simulation.  That is, regardless of the magnetic field strength, the final quasi-equilibrium has $E_{p}/E\sim0.65$.

In \citet{zink12} we established a relationship between the surface magnetic field strength and the gravitational wave emission (both energy and strain).  In the present article we generalise this result to include different polytropic equations of state, and hence stellar radius and mass dependencies.  Rather than review those previous results here, we give a complete analysis in the following section.

\section{Effect of Equation of State}\label{GWs}
The primary purpose of this section is to generalise our gravitational wave results of \citet{zink12} to include the dimensions of the star.  That is, in \citet{zink12} we evaluated the gravitational wave emissions due to a reconfiguring magnetic field, induced by the magnetic field instability.  This is deemed to be a toy model for the aftermath of a magnetar flare, in which the flare causes some form of rearrangement of the magnetic field.  Note that we are not saying that the kink instability causes a magnetar flare, rather that we are simply determining the energy conversion between magnetic field, fluid and spacetime dynamics utilising the kink instability as a mechanism to generate such motions.  

In \citet{zink12} we determined a power-law relationship such that the gravitational wave strain is proportional to the surface magnetic field strength to the power $3.3$.  This only took into account our fiducial $\Gamma=2.0$ polytrope, and therefore was not determined as a function of any other stellar parameters.  In this section we repeat the analysis using a range of soft and hard EoSs to determine a scaling law relating the gravitational wave strain, surface magnetic field strength, stellar radius and mass.  

\begin{center}
\begin{table}
\begin{tabular}{|cccccc|}
	\hline
	Model & $\Gamma$ & $\kappa$ & $M$ & $R$ & Freq.\\
	& & $\left[\frac{\rho_{{\rm n}}c^{2}}{n_{{\rm n}}^{\Gamma}}\right]$ & $\left[M_{\odot}\right]$ & $\left[{\rm km}\right]$ & $\left[kHz\right]$ \\
	\hline\hline
	 A0 & 1.67 & 0.0400 & 1.31 & 19.28 & 1.2 \\
	 B0 & 2.00 & 0.0269 & 1.31 & 12.68 & 1.7\\
	 C0 & 2.34 & 0.0195 & 1.31 & 11.54 & 1.8\\
	 D0 & 2.46 & 0.00936 & 1.31 &  8.47 & 2.5\\
	\hline
\end{tabular}
\caption{\label{EOStable}Equations of State parameters used throughout the article.  Note that model B0 is our non-rotating fiducial model from \cite{lasky11,zink12} and section \ref{nonrotate} and models C0 and D0 are EoSs II and A respectively.  The 0 label represents no rotation, $M$ is the gravitational mass, $R$ the equatorial radius and ${\rm Freq.}$ in the final column is the fundamental $f$-mode frequency of the system giving the characteristic fluid timescale.}
\end{table}
\end{center}

\begin{figure}
	\begin{center}
	\includegraphics[angle=0,width=0.95\columnwidth]{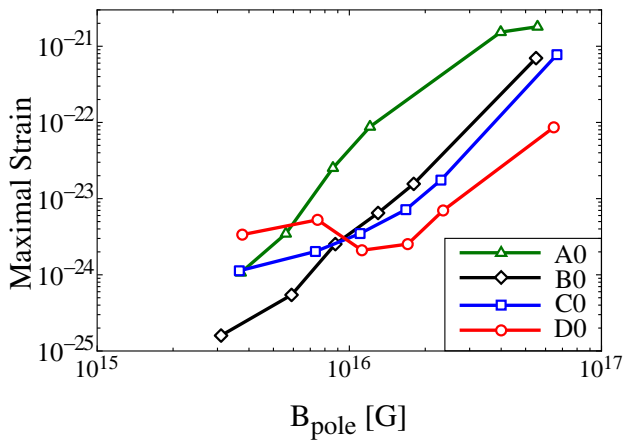}
	\end{center}
	\caption{\label{strainmax}Maximal strain, $h_{\times}^{{\rm max}}$, as a function of polar surface magnetic field strength for non-rotating models presented in table \ref{EOStable}.  These models all have constant stellar radius.}
\end{figure}

Table \ref{EOStable} shows the EoS parameters that we utilise herein.  For each EoS we have constructed a series of models with central magnetic field $1.6\times10^{16}\,{\rm G}\le\left|B_{c}\right|\le2.7\times10^{17}\,{\rm G}$.  Depending on the EoS, these models will have different surface magnetic fields and different Alfv\'en timescales.  For each model we plot the maximum value of the  cross-polarisation of gravitational wave strain, $h_{\times}^{{\rm max}}$, as a function of the surface magnetic field strength at the pole, $B_{{\rm pole}}$, in figure \ref{strainmax}.  As the star becomes more compact (i.e. as $\Gamma$ increases), the strain evaluation for the models with weaker magnetic fields becomes more difficult.  Therefore, for the below analysis, we restrict our attention only to the stronger field strength models for which the power-law relation is valid.  

Figure \ref{strainmax} shows the dependence of strain on the mangetic field and also the equation of state, where explicit dependence is imposed on the radius of the star.  The full functional dependence of this relation will also include the stellar mass (or alternatively the central density).  We have therefore produced extra series' of models for equations of state B0 and C0, allowing the mass to vary.  This allows us to perform a full, least squares fit over the four parameter space $h_{\times}^{{\rm max}}$, $R$, $M$ and $B_{{\rm pole}}$, finding the following relation
\begin{widetext}
\begin{align}
	h_{\times}^{{\rm max}}=8.5\times10^{-28}\left(\frac{10\,{\rm kpc}}{d}\right)\left(\frac{R}{10\,{\rm km}}\right)^{4.8}\left(\frac{M}{M_{\odot}}\right)^{1.8}\left(\frac{B_{{\rm pole}}}{10^{15}\,{\rm G}}\right)^{2.9}\label{GWoutput}
\end{align}
This result is consistent with that presented in \citet{zink12} in that $h_{\times}^{{\rm max}}$ goes approximately to the third power of the magnetic field strength.  The additional factors in this equation are the dependence on the stellar radius and mass, $R$ and $M$, which scale almost to the fifth and second powers respectively.  

As previously discussed, a majority of the energy in the signal associated with $h_{\times}^{{\rm max}}$ is in the $f$-mode.  We can therefore calculate an approximate amount of energy emitted in gravitational radiation by noting \cite{levin11}
\begin{align}
	E_{{\rm GW}}=\frac{2\pi^{2}d^{2}f^{2}c^{3}}{G}\int_{-\infty}^{\infty}\left<h^{2}\right>dt,
\end{align}
where $f$ is the $f$-mode frequency of radiation.  Assuming a gravitational wave damping time of $100\,{\rm ms}$, we find a power-law relation for the total energy emitted in gravitational radiation via the $f$-mode to be\footnote{We note the error in our published article \citet{zink12} where, in equation (2), we have an anomalous dependence on the distance to the source in the relationship for the gravitational wave energy.}
\begin{align}
	E_{{\rm GW}}=1.7\times10^{36}\left(\frac{R}{10\,{\rm km}}\right)^{9.6}\left(\frac{M}{M_{\odot}}\right)^{3.6}\left(\frac{B_{{\rm pole}}}{10^{15\,{\rm G}}}\right)^{5.8}\,{\rm erg}.\label{GWoutputE}
\end{align}
\end{widetext}

This highly non-linear relationship with the radius could have implications for gravitational wave detection.  If nature is kind to us, and we find that the EoS of neutron stars are soft, implying their radii are large with respect to the fiducial model, then gravitational wave detection of $f$-modes from magnetar flares becomes more likely.  

It is worthy of note that, while the lower field values for the D0 model are off-kilter, and hence have been excluded from the above analysis, the three stronger field models are consistent with the above relation.  We have confirmed this by removing all of the D0 models, at which point the least squares analysis gives similar results to those presented above.

In the above we have calculated the maximal strain seen in our simulations rather than a time integrated signal such as a traditional root-sum-square amplitude  $h_{{\rm rss}}^{2}=\int_{-\infty}^{\infty}\left(h_{\times}\right)^{2}dt$ \cite[e.g.][]{ciolfi11}.  The main reason for this is that a time integrated signal would require us to define a time for the onset of the instability.  This is because there is no intrinsic damping in the system (i.e. we are working in the Cowling approximation, implying the dominant $f$-mode damping mechanism is not present), implying we would have to ensure we only integrate for $T$ seconds (where $T$ is the damping time of any particular mode) following the onset of the instability.  Integrating beyond this time would artificially grow the $h_{{\rm rss}}$ signal.  Moreover, were we to integrate from the very beginning of our simulation out to some fixed time, this would introduce a systematic error as a function of the magnetic field strength, due to the onset of the instability being a function of the Alfv\'en timescale of the system.  

\begin{figure*}
	\begin{center}
	\includegraphics[angle=0,width=0.95\textwidth]{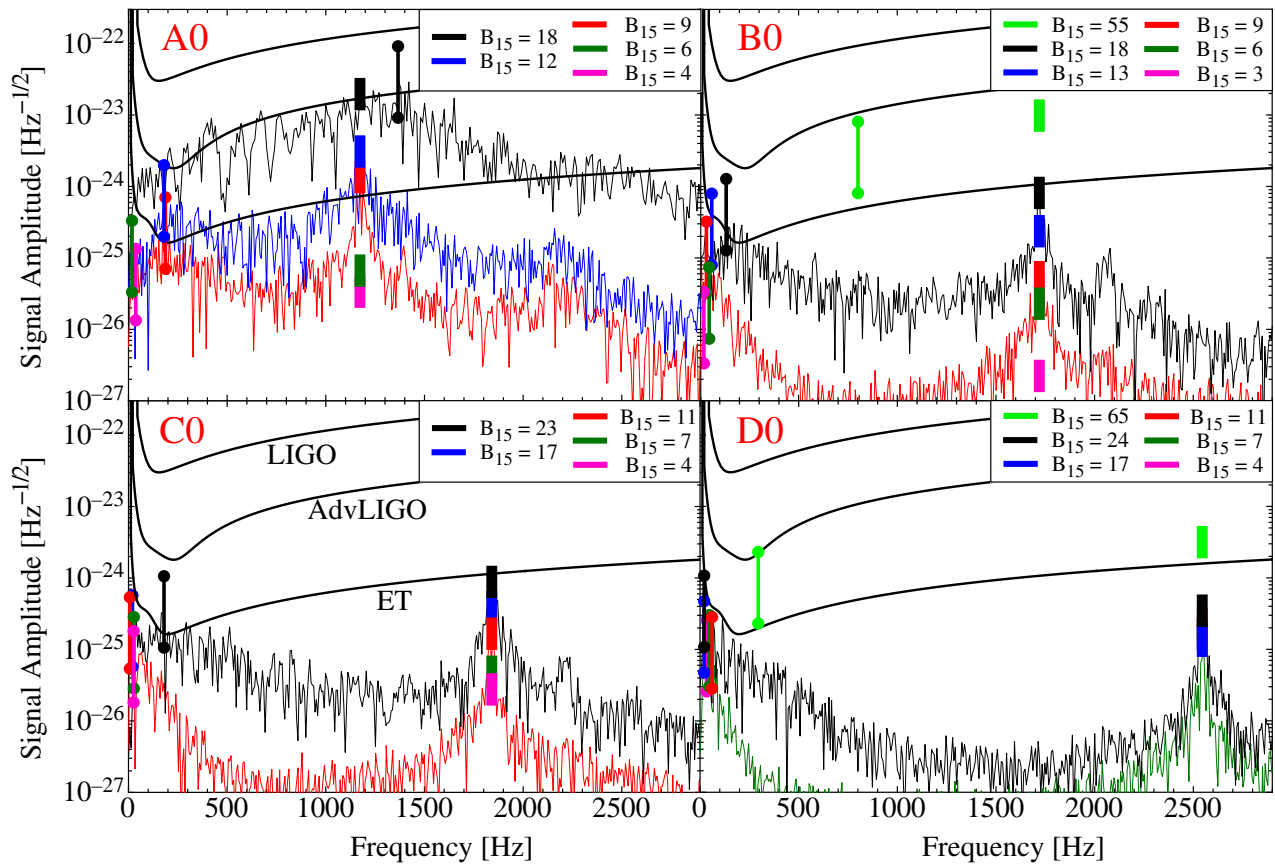}
	\end{center}
	\caption{\label{LIGO} Gravitational wave signal amplitude, $\sqrt{T}\left|\tilde{h}\left(f\right)\right|$, against oscillation frequency for the four equations of state used herein -- A0 (top left panel), B0 (top right), C0 (bottom left) and D0 (bottom right).  For each box model there are several magnetic field strength simulations shown.  The $f$-mode signal amplitude is represented by a coloured box at the appropriate frequency (shown in table \ref{EOStable}), where a source lasting between $50$ and $200\,{\rm ms}$ has been assumed.  The coloured vertical dumbbells (i.e. skinny lines with bulbous ends) represent the maximum Alfv\'en mode seen in the Fourier transform.  These assume a constant source lasting between $10\,{\rm ms}$ and $1\,{\rm s}$.  Both modes scale with $\sqrt{T}$, implying one can easily extrapolate to alternative values of the damping time.  We have further plotted the entire spectrum (assuming a damping time of $T=100\,{\rm ms}$ for numerous models to give an indication of the spectra present.  Finally, the noise power spectral density curves for LIGO, AdvLIGO and ET are shown in each panel.  These have been taken from the review article of \citet{sathyaprakash09}.}
\end{figure*}

Instead of the aforementioned approach, we calculate the signal amplitude, $\sqrt{T}\left|\tilde{h}\left(f\right)\right|$, for our simulations as a function of the frequency, such that this value can be directly compared with the noise power spectral density, $\sqrt{\left|S_{h}\left(f\right)\right|}$, of individual gravitational wave detectors, and the amplitude signal-to-noise ratio, $\sqrt{T}\left|\tilde{h}\left(f\right)\right|/\sqrt{\left|S_{h}\left(f\right)\right|}$, can simply be read of the resultant figure.  To calculate our signal amplitude, we take a Fourier transform of a portion of our derived, $h_{\times}$ that is post kink instability saturation and lasts approximately $150\,{\rm ms}$ (this value ensures we get a reasonable number of oscillations in the lower portion of the spectrum).  We then multiply by the relevant damping times and plot the results in figure \ref{LIGO}.

In particular, each panel of figure \ref{LIGO} represents different EoSs given in table \ref{EOStable}.  For each EoS we have run multiple models with different magnetic field strengths, and for each simulation have located the $f$-mode signal amplitude and the maximal Alfv\'en mode signal amplitude.  The $f$-modes (coloured boxes) are plotted in figure \ref{LIGO} assuming a damping time between $T=50$ and $200\,{\rm ms}$.  One can see that the $f$-mode frequency for the different EoSs ranges from $\sim1.2\,{\rm kHz}$ for models A0 to $\sim\,2.5{\rm kHz}$ for models D0.  Moreover, for all of the models presented herein, only the largest strength magnetic field model (with surface field strength $B_{15}=18$) with the softest EoS is observable by the Advanced LIGO detector.  It is worth noting that the strongest magnetic field observed to date in a magnetar is almost a full order of magnitude {\it less} than that modelled for this particular star.  The detectability situation of $f$-modes could change slightly if, for example, neutron stars have interior toroidal magnetic fields an order of magnitude larger than the observed dipole poloidal field.  Although this situation is unlikely due to stability arguments, a point we discuss in more detail below.  

For each EoS in figure \ref{LIGO} we also plot two or three full spectra with assumed damping time of $T=100\,{\rm ms}$.  While the $f$-mode is abundantly clear in all of these curves (except from model A0 with $B_{15}=18$ -- a point we discuss below), lower frequency Alfv\'en oscillations are also clear in each of these simulations.  As discussed previously, the damping time for these modes is largely unclear, but is likely significantly longer than a millisecond.  Therefore, for each maximal Alfv\'en mode, we have also plotted a vertical dumbbell (i.e. a skinny line with bulbous ends) with damping times between $T=10\,{\rm ms}$ and $1\,{\rm s}$.  For significantly lower values of magnetic field strength, these bars become closer to the lower limit of detection for both the Einstein Telescope and possibly even Advanced LIGO.  This is especially true when one considers that the damping time of these modes could, in fact, be minutes or even significantly longer.  These modes, as with the $f$-modes discussed above, are also better excited for softer EoSs, corresponding to larger neutron stars.  

It is important here to mention the signal amplitude of the A0 model with $B_{15}=18$.  This model has an Alfv\'en crossing time of approximately $\tau_{A}=1.1\,{\rm ms}$, which corresponds to a fundamental Alfv\'en frequency of $f_{A}=910\,{\rm Hz}$.  Moreover, the $f$-mode frequency of the A0 simulations is approximately $1170\,{\rm Hz}$.  One can therefore see from the Fourier transform that the region around the $f$-mode frequency is not a clean peak, but shows a rather dense spectrum consistent with expectations from the lower field simulations.  In fact, while there exists a distinct peak at the $f$-mode frequency, this is not the largest peak present in the Fourier transform.  Rather, the largest peak, less than a factor two larger than the $f-$mode peak, is located at approximately $1.35\,{\rm kHz}$.  We have indicated this on the top left panel of figure \ref{LIGO}.

Our conclusion from figure \ref{LIGO} is the following: Assuming that a giant magnetar flare is somehow related to a catastrophic rearrangement of the core magnetic field, {\it the gravitational wave signal associated with $f$-modes are not observable with present or near-future gravitational wave observatories.  Lower frequency Alfv\'en modes do, however, provide an enticing alternative where efforts could be concentrated.}  Of course, the biggest hurdle to understanding the gravitational wave emission associated with these Alfv\'en modes is understanding their damping time.  If they are sufficiently long-lived, then they {\it may} be detectable in the relatively near future.  

\section{Rotation}\label{rotation}
\citet{frieman60} were the first to show that rigid-body rotation only has a significant effect on hydromagnetic equilibria when the fluid-flow velocity is of the same order as the Alfv\'en velocity.  Certainly, in newly-born neutron stars this is expected to be the case in large regions of the star, even for magnetar field strength stars.  \citet{pitts85} studied rotating, toroidal magnetic fields in cylindrical geometries, showing that the perturbations can be countered by sufficiently large rotational velocities, implying the suppression of certain modes of instabilities.  

The first work on the stability of purely poloidal, rotating fields was from \citet{geppert06}, who performed three-dimensional, non-linear numerical simulations by use of a spectral code.  They found that stars rotating with sufficiently high rotation speeds, quantified as $\Omega_{A}/\Omega\lesssim0.1$, and with roughly aligned magnetic and rotation axes, will have suppressed instabilities.  In contrast, \citet{braithwaite07} also performed nonlinear, global simulations of rotating poloidal magnetic fields, finding that the initial linear phase of instability growth was {\it not} affected by the presence of rotation.  However, while \citet{braithwaite07} found that the nonlinear phase {\it is} affected by rotation, the instability was always present regardless of the rotation speed.  \citet{lander11b} recently performed linear simulations of rotating poloidal fields, showing that some, but not all modes were stabilised by the presence of rotation.  

Herein we attempt to resolve some of the aforementioned contradictions, simultaneously performing the first simulations of rotating poloidal fields in general relativity.  

For our rotating simulations we revert back to the fiducial EoS -- i.e. a $\Gamma=2$ polytrope as described in section \ref{nonrotate}.  We have created a series of rotating models utilising the spectral solver {\sc lorene} with central magnetic field strength of $B_{{\rm c}}=1.0\times10^{17}\,{\rm G}$, corresponding in the non-rotating limit to a surface field strength of $B_{15}=16$.  We summarise the properties of this models in table \ref{Rottable}.

\begin{center}
\begin{table}
\begin{tabular}{|cccccc|}
	\hline
	Model & $B_{{\rm c}}$ & $\Omega$ & $a_{p}/a_{e}$ & $\tau_{\Omega}$ & $\tau_{A}$ \\
	& $\left[10^{15}\,{\rm G}\right]$ & $\left[{\rm Hz}\right]$ & & $\left[{\rm ms}\right]$ & $\left[{\rm ms}\right]$  \\
	\hline\hline
	B0 & 100 & 0 & 0.99 & $\infty$ & 2.4\\
	B100 & 100 & 100 & 0.99 & 10.0 & 2.4 \\
	B200 & 100 & 200 & 0.97 & 5.0 & 2.4 \\
	B300 & 100 & 300 & 0.94 & 3.3 & 2.4\\	
	B400 & 100 & 400 & 0.90 & 2.5 & 2.3 \\	
	 \hline
\end{tabular}
\caption{\label{Rottable}Rotating models.  Our model has equation of state B given in table \ref{EOStable}, with central field of $B_{{\rm c}}=1.0\times10^{17}\,{\rm G}$.  Moreover, $\Omega$ is the rotational frequency, $a_{p}/a_{e}$ is the ratio of equatorial to polar radii and $\tau_{\Omega}$ \& $\tau_{A}$ are the rotational period and Alfv\'en timescale respectively.}
\end{table}
\end{center}

\subsection{Effect of Magnetic Field on Rotation Rate}
Our simulation method is identical to the previous section.  Initial conditions are created on a spectral grid using {\sc lorene}, which is then mapped to our Cartesian grid measuring $120$ grid-points in all directions.  No perturbation is added, and we allow the simulation to evolve.  

\begin{figure}
	\begin{center}
	\includegraphics[angle=0,width=.95\columnwidth]{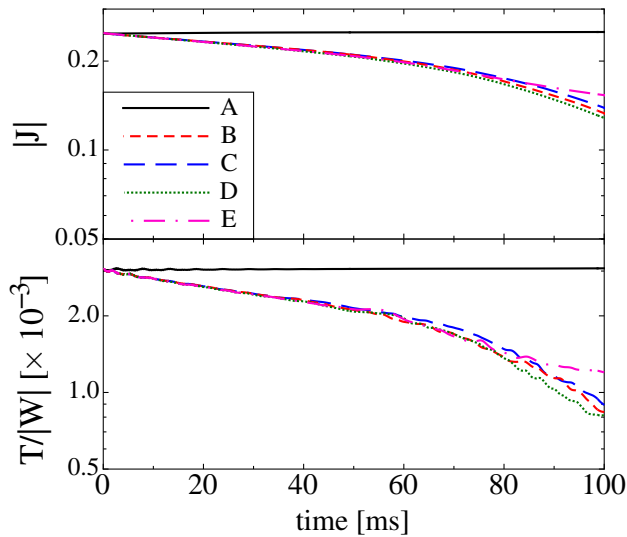}
	\end{center}
	\caption{\label{boundtest}Angular momentum, $J$, and $T/|W|$ for models with initial rotational period of $200\,{\rm Hz}$.  Model A has $120^{3}$ with zero magnetic field.  Models B -- E are all B200; model B has $120^{3}$ grid-points (i.e. is model our canonical simulation set-up), model C has $180^{3}$ grid-points with outer boundaries located further from the star (such that the resolution across the star remains the same as model B).  Model D has $150^{3}$ grid-points with the outer boundary located the same distance as in B, and E has $150^{3}$ grid-points with linear extrapolation for the magnetic field used at the outer boundaries rather than the standard Dirichlet boundary conditions for the other models.}
\end{figure}

It is worth noting from the outset that we see a not-so insignificant spin-down of the rotating star due to the presence of the magnetic field.  This is shown in figure \ref{boundtest}, where we plot measures of the rotational energy in the system for five rotating models, all with initial spin period of  $\tau_{\Omega}=5.0\,{\rm ms}$.  In the top panel of figure \ref{boundtest} we plot the angular momentum, $J$, defined as 
\begin{align}
	J=\int_{\Sigma}n_{a}{T^{a}}_{\phi}\sqrt{\gamma}d^{3}x,
\end{align}
where $n_{\mu}$ is the four-vector that is hypersurface orthogonal to surfaces of constant time, $\gamma_{ij}$ is the three-metric of the spatial hypersurface $\Sigma$ and $T^{\mu\nu}$ is the stress-energy tensor.  In the bottom panel of figure \ref{boundtest} we plot the ratio of the rotational kinetic energy to gravitational binding energy $T/\left|W\right|$.  For definitions of these quantities see \citet{stergioulas03}.  It is worth noting that, for our calculation of the angular momentum, we assume the system remains axisymmetric.  i.e. that $\partial_{\phi}$ is a Killing vector.  While this will introduce an error into the calculation of the angular momentum, we expect this to be minimal due to the fluid remaining almost axisymmetric, with deviations from this only arising due to non-axisymmetries induced by the magnetic field.

Model A in figure \ref{boundtest} has zero magnetic field throughout the star.  The angular momentum and $T/|W|$ are extremely well conserved over the twenty spin periods shown in this figure.  We have evolved such a simulation for almost $700\,{\rm ms}$, showing conservation of angular momentum on the order of $7\%$.   

Models B -- E of figure \ref{boundtest} are all B200 from table \ref{Rottable}; model B has $120^{3}$ grid points with the outer boundary located at approximately $1.4$ times the surface of the star (i.e. this is our canonical simulation set-up that will be used throughout the remainder of the article), model C has $180^{3}$ grid-points, however the outer boundary of the star has been moved to a greater radius such that the grid resolution across the star is the same as in model B.  In model D we have $150^{3}$ grid-points with all other quantities being the same as model B.  Finally, in model E we have used $150^{3}$ grid-points and implemented a linear extrapolation boundary condition for the magnetic field at the outer-boundary of the domain.  This is in contrast to the Dirichlet boundary conditions used for all other simulations.

One can see from figure \ref{boundtest} that the loss of rotational energy is almost independent of our simulation method (in terms of resolution, boundary location and magnetic field boundary conditions).  In fact, the dominant factor in the loss of angular momentum is the strength of the magnetic field, implying this scales almost linearly with the Alfv\'en timescale of the system.  We note here that model E in figure \ref{boundtest} has a slight up-turn in angular momentum and $T/|W|$ after approximately $85\,{\rm ms}$.  This is not a conservation of rotational energy, rather this simulation develops a numerical instability that develops at the boundary of our domain and causes the simulation to crash shortly after $100\,{\rm ms}$.  Higher-order interpolation schemes at the boundary could act to stabilise such a scheme, however figure \ref{boundtest} indicates that this would not change the rate of rotational energy loss from the system.  

\begin{figure}
	\begin{center}
	\includegraphics[angle=0,width=0.95\columnwidth]{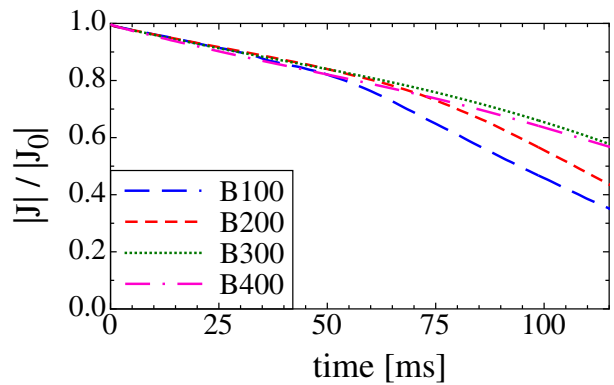}
	\end{center}
	\caption{\label{rotationrate} Normalised absolute value of the angular momentum, $|J|/|J_{0}|$, where $J_{0}=J(t=0)$, as a function of time for our models in table \ref{Rottable}.}
\end{figure}

Because of the loss of angular momentum in our system we are presently restricted to {\it only} discussing the onset and nature of instabilities, and can not provide any insight into the nature of pseudo-equilibria attained after long evolution times.  In figure \ref{rotationrate} we plot the absolute value of the angular momentum, normalised to the initial state, as a function of time for our models presented in table \ref{Rottable}.  One can see here that our simulations retain more than $50\%$ of their angular momentum for approximately $100\,{\rm ms}$, corresponding to more than $40$ Alfv\'en timescales and between $10$ and $40$ initial rotational periods.  As we shall see below, these timescales are long enough to discuss much of the dynamics of the systems in terms of the evolution of the varicose and kink instabilities.  However, we stress that these losses of angular momentum imply that the end states of our simulations are {\it not} necessarily representative of equilibrium states of rotating neutron stars.  We are therefore reticent to discuss such equilibria in the present article, keeping our discussion to that of the dynamics of the instability.

\subsection{Instability}
\begin{figure}
	\begin{center}
	\includegraphics[angle=0,width=0.95\columnwidth]{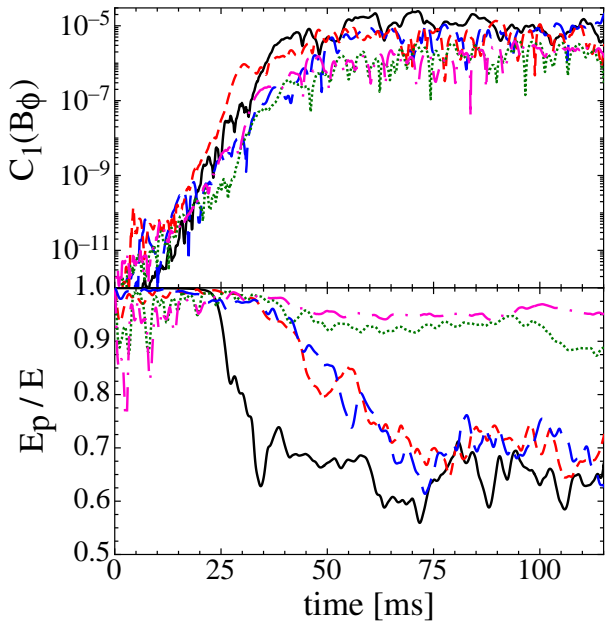}
	\end{center}
	\caption{\label{RotMagEnergy}Magnetic field instability in terms of both $C_{1}\left(B_{\phi}\right)$ (top panel) and poloidal-to-total magnetic energy ratio, $E_{p}/E$, for the models presented in table \ref{Rottable}.  The individual lines are the same as those for figure \ref{rotationrate}, with the black line representing the non-rotating model.}
\end{figure}

In figure \ref{RotMagEnergy} we plot $C_{1}\left(B_{\phi}\right)$ and $E_{p}/E$ (top and bottom panels respectively) for the series of models presented in table \ref{Rottable}, with initial rotational frequencies of between $0$ and $400\,{\rm Hz}$.  

If we focus our attention on the initial first few ${\rm ms}$ of the simulation, particularly in terms of the $E_{p}/E$ quantity (lower panel), we see a large trough reaching as low as $E_{p}/E\sim0.8$ for the $400\,{\rm Hz}$ model, but smaller for lower rotation rates.  This is a result of the mapping between the spectral grid and our Cartesian grid.  The effect is to introduce strongly toroidal components into the magnetic field in the first few milliseconds, particularly around the equatorial region near the neutral line of the field.  As can also be seen in this figure, this non-physical effect also vanishes after the first few Alfv\'en crossing timescales, and the simulation reduces to an almost purely poloidal state.  

The immediately striking part about figure \ref{RotMagEnergy} is that the instability timescale in terms of the $C_{m}\left(B_{\phi}\right)$ quantities is independent of the rotational velocity.  For all initial rotational velocities, we see the $C_{1}\left(B_{\phi}\right)$ grow exponentially by over six orders of magnitude in approximately $50\,{\rm ms}$.  This is the same growth timescale of the instability in terms of the poloidal to total energy ratio for the case with zero rotation rate.  However, this is {\it not} the case for the rotating simulations.  For example, if we focus our attention on model B100 (blue dashed line), one can see that the timescale for the ratio of poloidal to total magnetic energy to reach the canonical ``equilibrium'' value of $E_{p}/E\sim0.65$ is approximately $80\,{\rm ms}$.  In this case, the rotation has slowed the development of the instability in terms of magnetic field reconstructions.  This is more extreme for the $300$ and $400\,{\rm Hz}$ models (dotted green line and dot-dash pink line respectively), which are shown to be stable for the full $115\,{\rm ms}$ evolution shown here.  

The seeming contradiction in growth timescales can be understood by noting that the $C_{m}B_{\phi}$ quantities trace {\it any} change in the structure of the magnetic field around the equator, while $E_{p}/E$ only tracks the growth of toroidal magnetic field.  In terms of the two known instabilities expected to act on such magnetic field configurations, $C_{m}B_{\phi}$ is sensitive to both the varicose {\it and} kink modes, while the poloidal to total magnetic energy ratio is {\it only} sensitive to the kink mode.  This can be understood as the varicose mode only acts to change the cross-sectional areas of flux tubes around the equator, keeping them orthogonal to equipotential surfaces.  The kink instability, on the other hand, acts to directly increase the amount of toroidal magnetic field in the star.

\begin{center}
\begin{figure*}
\includegraphics[width=0.49\textwidth]{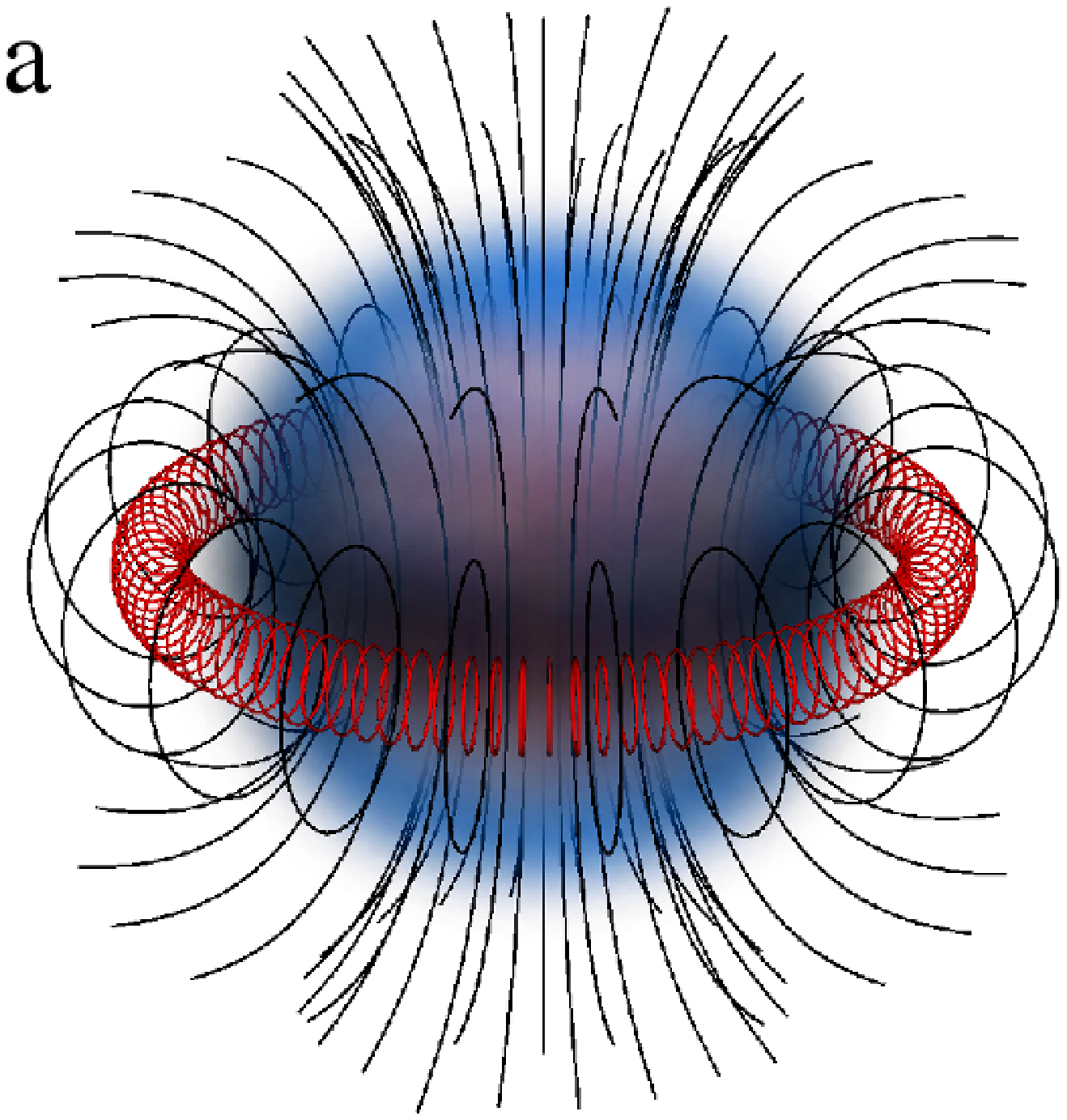}
\includegraphics[width=0.49\textwidth]{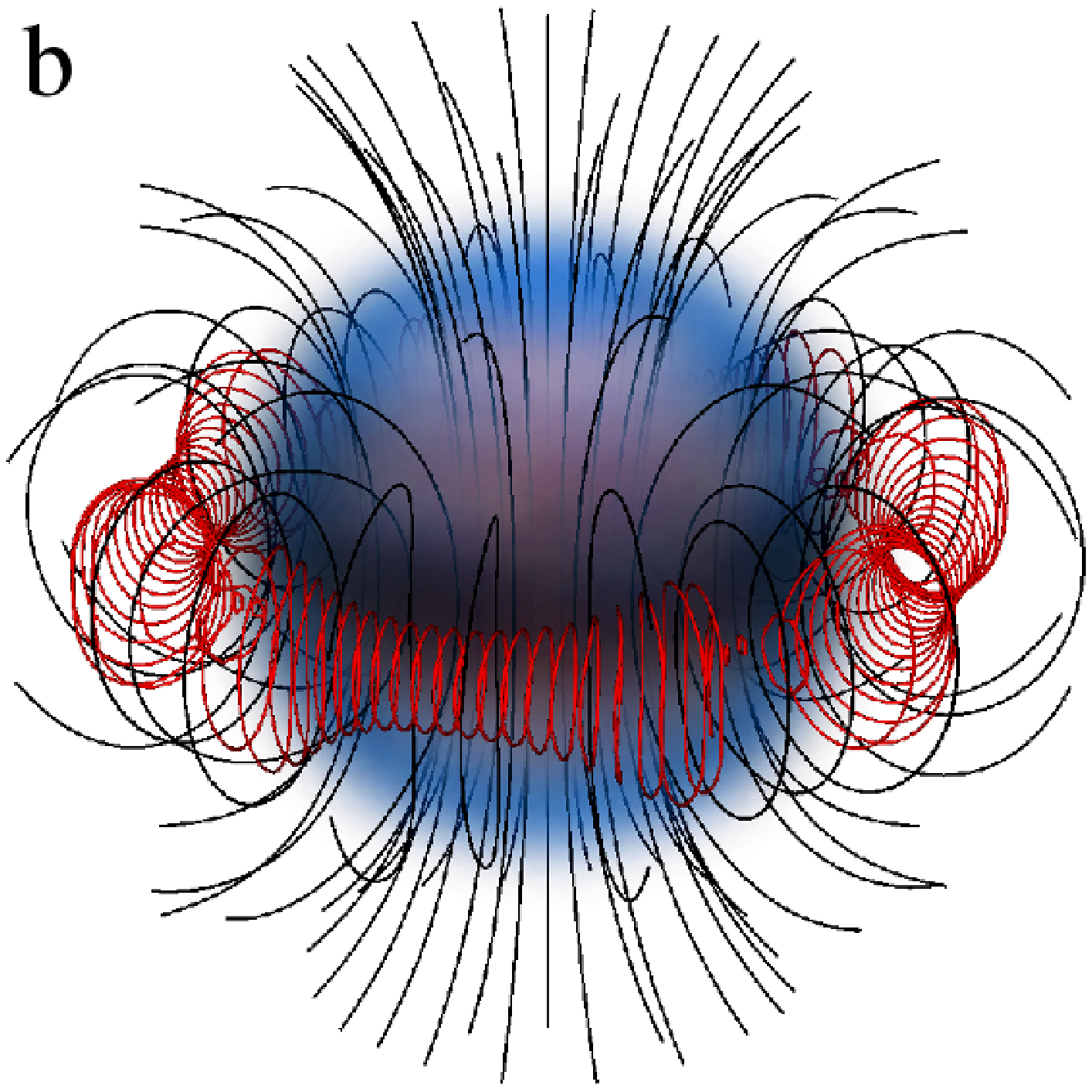}
\includegraphics[width=0.49\textwidth]{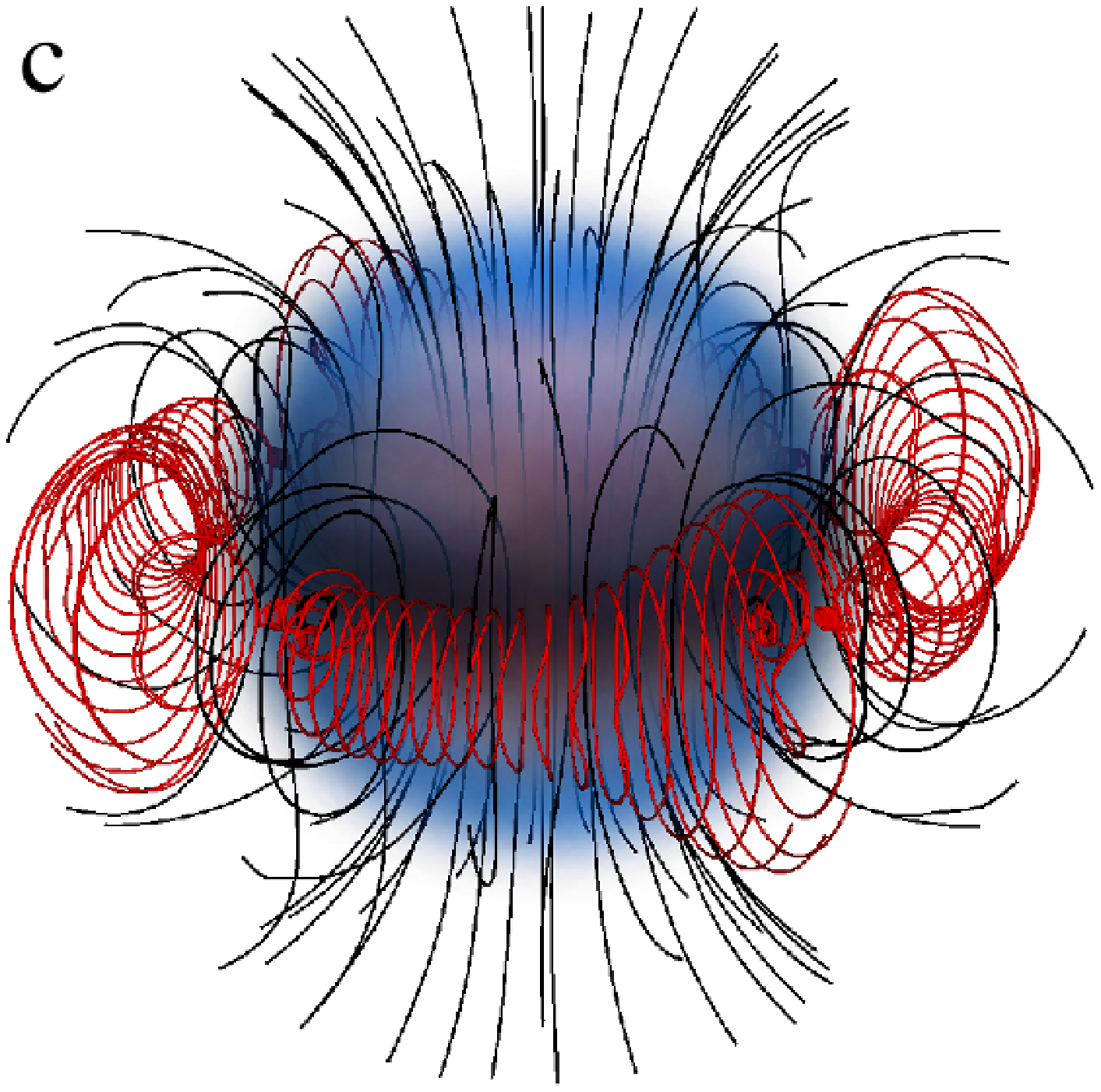}
\includegraphics[width=0.49\textwidth]{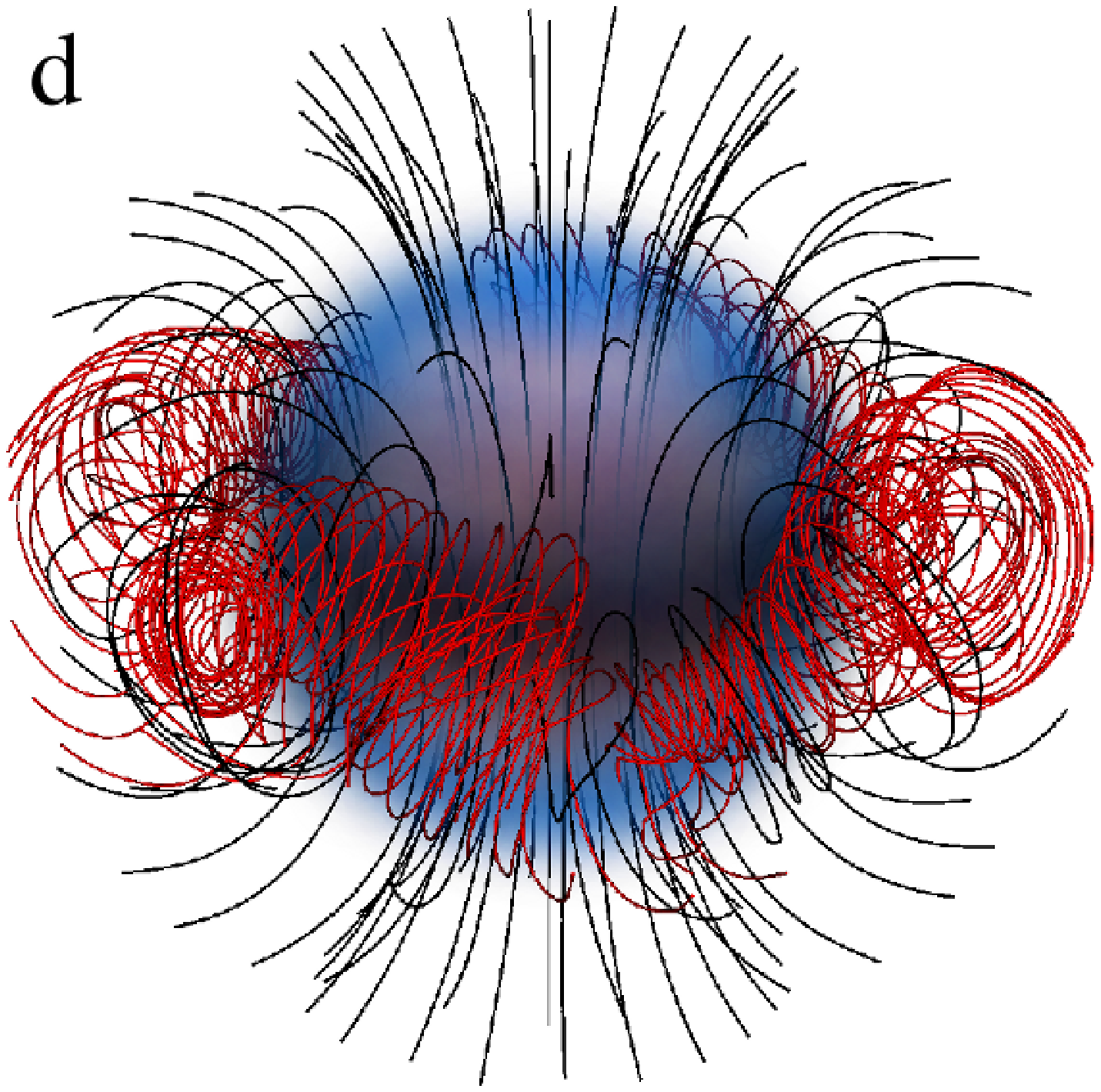}
	\caption{\label{3D_rotate_1.0e17_100Hz}  Time evolution of rotating model B100 defined in table \ref{Rottable} -- i.e. with central magnetic field $\bar{B}_{c}=1.0\times10^{17}\,{\rm G}$, initially rotating at $100\,{\rm Hz}$.  This gives $\tau_{\Omega}/\tau_{A}=4.2$.  The figures are a) $t=0\,\mbox{ms}$, b) $t=17\,\mbox{ms}$, c) $t=27\,\mbox{ms}$ and d) $t=42\,\mbox{ms}$.  A description of the figure is given in the caption to figure \ref{3D_non_rotate}.}
\end{figure*}
\end{center}

To highlight the aforementioned affect we look at three-dimensional plots of the magnetic field lines  to expose the global dynamics of the simulation.  In particular, in figures \ref{3D_rotate_1.0e17_100Hz}, we show three-dimensional snapshots of the B100 simulation (i.e. corresponding to the dashed blue line in figure \ref{RotMagEnergy}).  The four snapshots in this figure are shown after times $t=0\,{\rm ms}$ (figure \ref{3D_rotate_1.0e17_100Hz}a), $t=17\,{\rm ms}$ (figure \ref{3D_rotate_1.0e17_100Hz}b), $t=27\,{\rm ms}$ (figure \ref{3D_rotate_1.0e17_100Hz}c), $t=42\,{\rm ms}$ (figure \ref{3D_rotate_1.0e17_100Hz}d). 

Figure \ref{3D_rotate_1.0e17_100Hz}a represents a similar initial condition to that seen in figure \ref{3D_non_rotate}a.  In figures \ref{3D_rotate_1.0e17_100Hz}b and \ref{3D_rotate_1.0e17_100Hz}c we see the development of a strong varicose mode that has disrupted the magnetic field in a significantly more catastrophic manner than that seen for the non-rotating simulations.  This is especially seen when these figures are compared with figure \ref{3D_non_rotate}b.  If we compare the time of the three-dimensional snapshots seen in figures \ref{3D_rotate_1.0e17_100Hz}b and \ref{3D_rotate_1.0e17_100Hz}c with the instability plots given in figure \ref{RotMagEnergy}, we see that these represent points at which the $C_{m}\left(B_{\phi}\right)$ quantity is growing exponentially, while $E_{p}/E$ remains almost constant.  

As discussed in section \ref{instability}, the varicose mode was excited most strongly in the non-rotating case as an $m=4$ mode.  Although the presence of the varicose mode is a physical effect, the dominance of the $m=4$ mode is attributed to excitations caused by the Cartesian grid.  The present rotating cases however, are moving with respect to the stationary grid, implying this mode is not preferentially excited.  Moreover, we see these modes approximately equally in all azimuthal wavenumbers.  We therefore conclude that the varicose mode seen in our simulations are completely physical.  Moreover, from the top panel of figure \ref{RotMagEnergy}, we conclude that the varicose mode is, in fact, unstable.

Finally, figure \ref{3D_rotate_1.0e17_100Hz}d, shows a strong kink instability developing, corresponding to a transference of poloidal magnetic field energy to toroidal.  As mentioned, this figure is shown after $t=42\,{\rm ms}$ which, in figure \ref{RotMagEnergy} corresponds to $E_{p}/E\sim0.85$.

\begin{center}
\begin{figure*}
\includegraphics[width=0.49\textwidth]{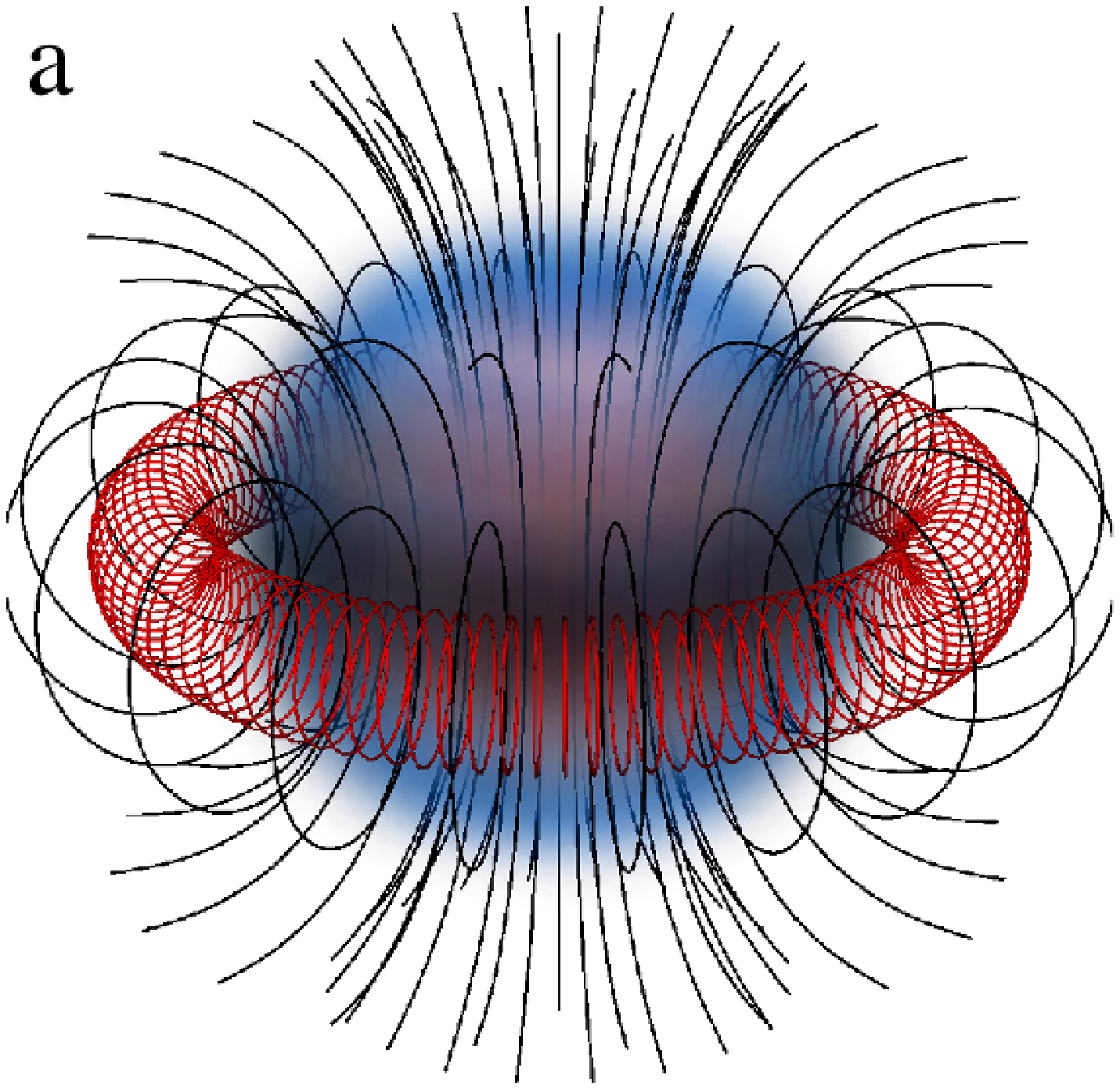}
\includegraphics[width=0.49\textwidth]{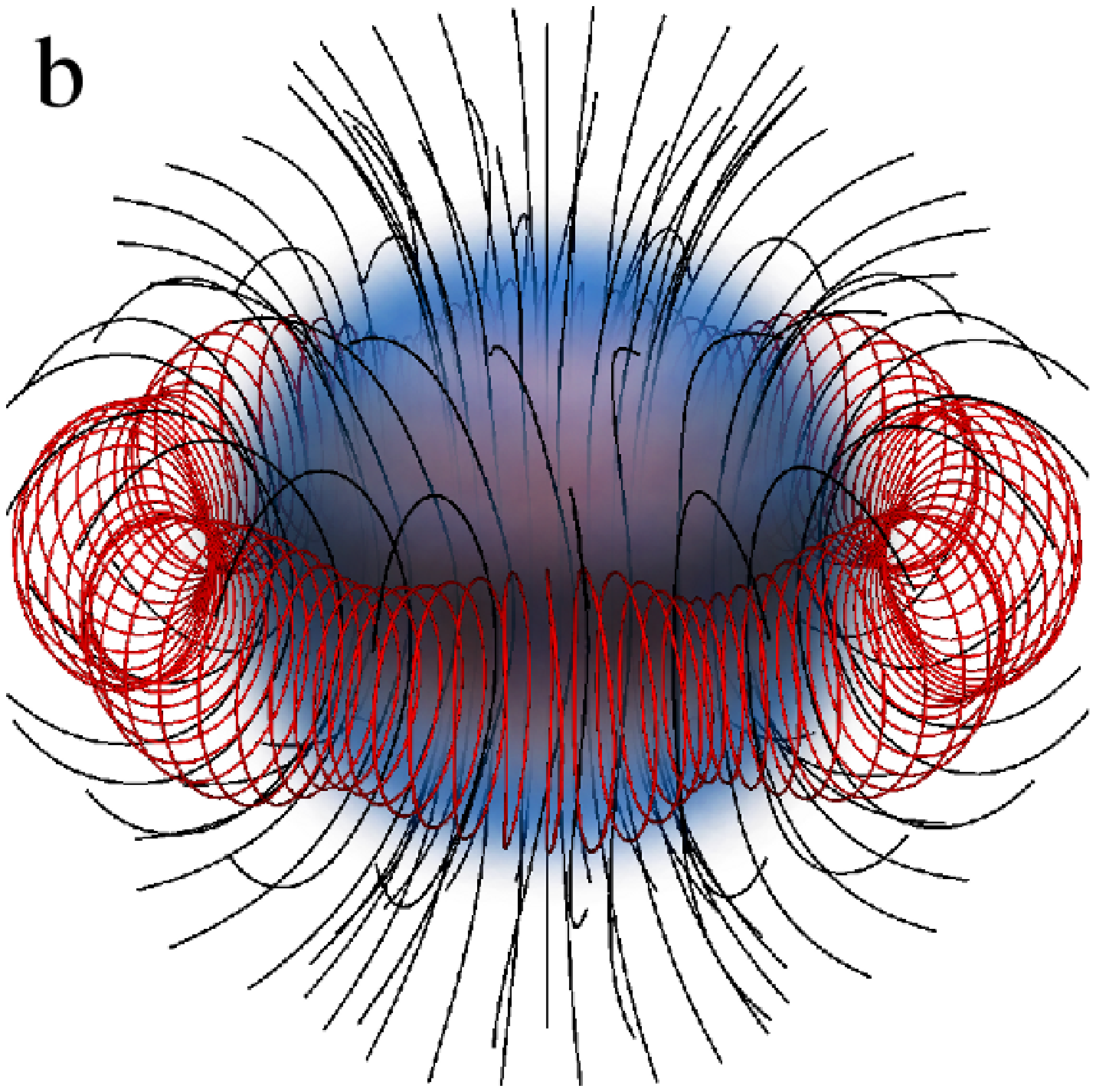}
\includegraphics[width=0.49\textwidth]{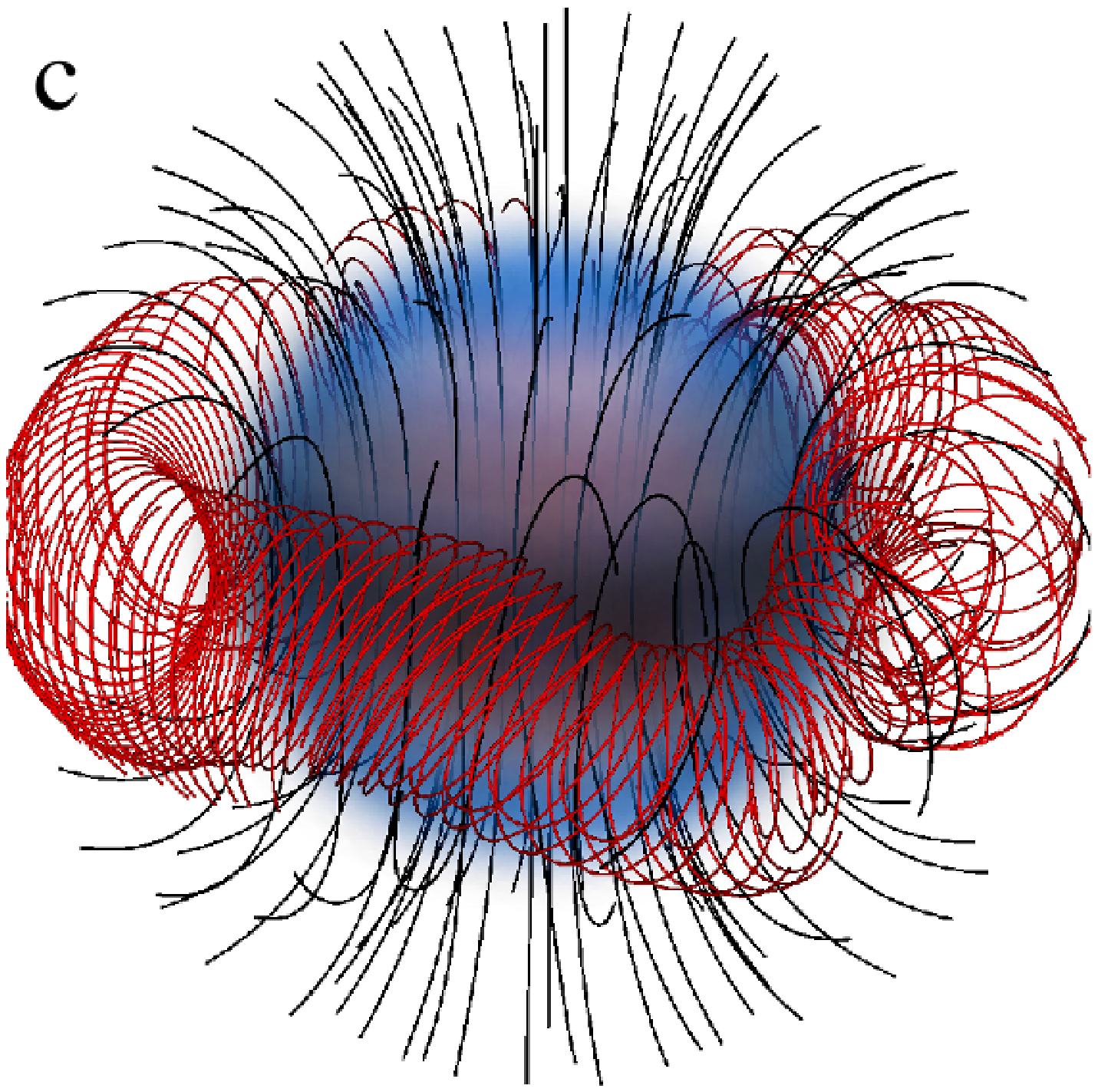}
\includegraphics[width=0.49\textwidth]{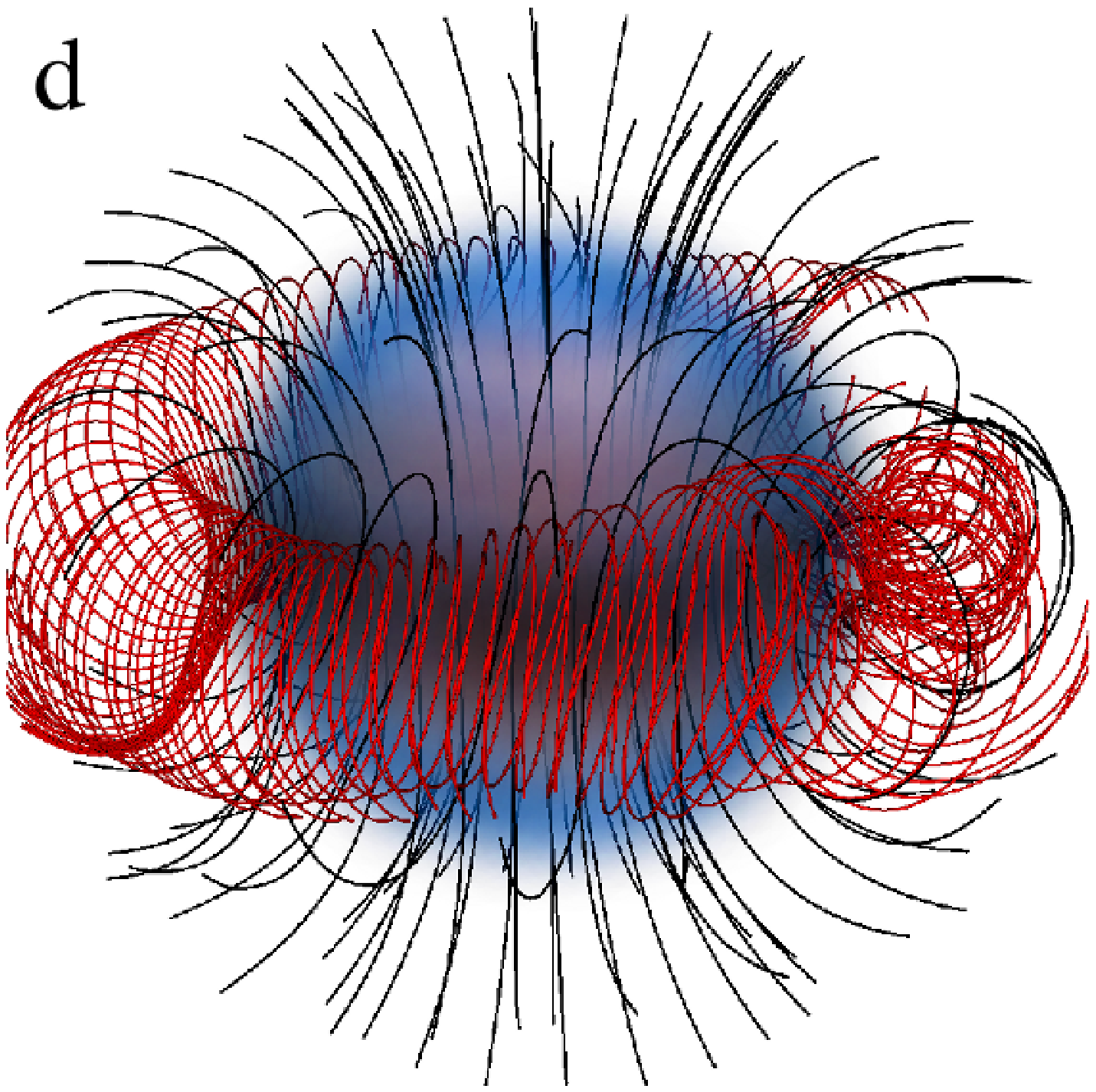}
	\caption{\label{3D_rotate_1.0e17_400Hz}  Time evolution of rotating model B400 defined in table \ref{Rottable} -- i.e. with central magnetic field $\bar{B}_{c}=1.0\times10^{17}\,{\rm G}$, initially rotating at $400\,{\rm Hz}$.  This gives $\tau_{\Omega}/\tau_{A}=1.1$.  The figures are a) $t=0\,\mbox{ms}$, b) $t=27\,\mbox{ms}$, c) $t=73\,\mbox{ms}$ and d) $t=148\,\mbox{ms}$.  A description of the figure is given in the caption to figure \ref{3D_non_rotate}. Note that the field lines are seeded in the same location as figure \ref{3D_non_rotate}, however the fast rotation rate pushes the neutral line further from the center of the star, explaining the larger region covered by the red field lines.}
\end{figure*}
\end{center}

In figures \ref{3D_rotate_1.0e17_400Hz} we again plot three-dimensional snapshots, however this time for the rotating B400 model.  The snapshots shown here are a) $t=0\,\mbox{ms}$, b) $t=27\,\mbox{ms}$, c) $t=73\,\mbox{ms}$ and d) $t=148\,\mbox{ms}$.  In figure \ref{3D_rotate_1.0e17_400Hz}a we see a slight change to the initial conditions to that seen in figure \ref{3D_non_rotate}.  In particular, the red magnetic field lines cover a larger area.  This is due to the initial rearrangement of the magnetic field, which also acts to push the neutral line to a larger equatorial radii.  As the field lines are seeded at a constant radii, the visual effect is that these field lines cover a larger flux surface.  It is worth noting that these field lines are still wholly contained within the star.

Once again, figure \ref{3D_rotate_1.0e17_400Hz}b exhibits a strong varicose mode growing early in the simulation.  This motion remains unchanged for approximately $50$ more ${\rm ms}$.  After approximately $70\,{\rm ms}$ one sees $E_{p}/E\eqsim0.95$.  The corresponding snapshot for this figure exhibits a single kink around the star, which rotates with the star, remaining virtually unchanged for approximately another $100\,{\rm ms}$.  It is extremely tempting to say that this is an equilibrium configuration for a fast rotating (i.e. $\tau_{\Omega}\sim\tau_{A}$), strongly magnetised star, as figure \ref{3D_rotate_1.0e17_400Hz}d, shown after almost $150\,{\rm ms}$, exhibits a very similar structure. Almost $180\,{\rm ms}$ into the simulation one finally sees the kink instability go unstable, and $E_{p}/E$ evolves to the canonical model of $E_{p}/E\sim0.65$.  We have not shown this in figure \ref{RotMagEnergy} as we believe it could possibly be mis-leading due to the substantial rotational energy loss in the star.  It is therefore not clear from this investigation whether the kink instability is only present in these simulations because enough angular momentum has been lost from the simulation such that the condition $\tau_{A}>\tau_{\Omega}$ is once again satisfied, and the kink instability can act.

It is worth pointing out that we have only explored models for which $\tau_{A}\sim\tau_{\Omega}$.  We have not yet broached models whereby $\tau_{\Omega}>>\tau_{A}$, implying the rotational velocity is dominating over the Alfv\'en timescale.  It is in this regime where one expects from linear analysis and Newtonian simulations \cite{geppert06} that the kink instability will be completely suppressed.

\section{Conclusion}\label{conclusion}
We have performed three-dimensional, general relativistic ideal MHD simulations of rotating and non-rotating polytropic neutron stars with initially purely poloidal magnetic field geometries.  This was accomplished by evolving initially self-consistent solutions of the Einstein-Maxwell field equations under the ideal MHD assumption and the Cowling approximation.  Particularly in the non-rotating case, purely poloidal fields are intrinsically unstable to the kink instability, which leeds to a catastrophic rearrangement of the magnetic field.  Such reconfigurations allowed us to explore three separate phenomena:
\begin{enumerate}[label=\roman{*})]
	\item the nature of the instability itself,
	\item the equilibrium configurations derived as steady-state solutions and 
	\item the gravitational wave emission from such magnetic field rearrangements which we interpret as a phenomenological model describing gravitational wave emissions from magnetar flares.  
\end{enumerate}

\subsection{Hydromagnetic Instabilities}
The first part of this work associated with the instability and the pseudo-equilibria configurations represents a significantly more detailed presentation of the first nonlinear GRMHD simulations of such instabilities \cite{lasky11}.  In particular, we showed that the kink instability acts on timescales associated with the Alfv\'en crossing time of the star, $\tau_{A}$, as expected from linear and non-linear analyses alike, saturating after approximately $15\,\tau_{A}$.  The kink instability acts on gravitational equipotential surfaces, and dominates near the neutral line of the poloidal field.  This acts to induce strong toroidal components of the magnetic field, which eventually saturates with a poloidal-to-total magnetic energy ratio of $E_{p}/E\sim0.65$.

We have made a first attempt at including rotation into our simulations, making these the first rotating GRMHD simulations with non-zero magnetic field in the atmosphere of the star.  This provides a mild technological challenge due to the significant numerical difficulties associated with both the stellar boundary and the exterior atmosphere region, particularly with reference to the recovery of primitive variables transformations.  

There has been debate in the literature as to the affect rotation has on the varicose and kink instability.  \citet{geppert06} found that sufficient rotation can act to suppress such instabilities, while \citet{braithwaite07} showed that rotation does {\it not} affect the presence or linear growth rate of such instabilities, although he only performed simulations of rotating toroidal fields.  Our rotating simulations showed that rotation acted to separate the varicose mode from the kink mode, with the former having a growth timescale unaffected by the presence of rotation (i.e. still of order the Alfv\'en crossing time), while the latter being slowed by the rotation.  Despite the slowing of the kink mode, this mode was still present in all of our rotating simulations.  It is worthy of note that our rotating simulations only begin to broach the regime in which $\tau_{\Omega}\sim\tau_{A}$, but does not yet extend into the regime in which $\tau_{\Omega}>>\tau_{A}$.  

\subsection{Equilibria}
A largely unresolved field of research is that of determining the set of stable MHD equilibria possible in barotropic stars.  Our evolutions push further the boundary of understanding.  In particular, our initially axisymmetric systems invariably evolve non-axisymmetric structures early into their evolutions.  All of our subsequent equilibria are therefore likely to retain such non-axisymmetric structures, implying we are not likely to converge on any of the semi-analytic equilibrium models used in the literature \cite[e.g.][]{colaiuda08,haskell08,ciolfi09}, nor are we likely to converge on the axisymmetric numerical results of \citet{braithwaite06b,braithwaite09}.  Despite this, many of the end-states that we derive contain considerable portions of the star that resemble twisted-torii, in that they contain predominantly poloidal regions, where the closed field lines are threaded by toroidal components of the field -- for example figures \ref{3D_non_rotate}d, \ref{3D_rotate_1.0e17_100Hz}d and \ref{3D_rotate_1.0e17_400Hz}d.  In some sense, our equilibria models are roughly a cross between the non-axisymmetric configurations of \citet{braithwaite08}, and the same authors axisymmetric equilibrium configurations \cite{braithwaite09}.  The difference between these two equilibrium derivations was predominantly associated with the position of the neutral line in the star.  

It is worth reiterating that the equilibria derived from our evolutions still show considerable kinetic energy in the system, even after up to a hundred Alfv\'en crossing times.  We describe this situation as a {\it pseudo}-equilibrium as we can not formally prove how close this situation is to a stable, stationary equilibria.  As discussed in the text, it is abundantly clear that the {\it global} structure of our pseudo-equilibria does {\it not} change further in time, and all globally measured quantities are tightly bounded.  For example, the ratio of magnetic energy in the poloidal component of the magnetic field to the total magnetic field is $E_{p}/E\sim0.65$.  

The stability of various twisted-torus configurations has recently been called into question \cite{lander12}.  In that paper, axisymmetric solutions of the baratropic Grad-Shafranov equation with mixed poloidal and toroidal magnetic fields were linearly perturbed, with all resulting equilibria found to be unstable.  The \citet{lander12} 'type-I' solutions are those resembling twisted-torus models in that they have an interior toroidal field threading only the closed field lines of the poloidal field.  Those authors however, could only probe equiblira solutions with $E_{p}/E\gtrsim0.945$\footnote{The quoted maximum value in \citet{lander12} is $E_{T}/E=0.045$, where $E_{T}=E-E_{P}$ is the magnetic energy in the toroidal component, however this has been calculated as a volume integral that includes the exterior region of the star.  The same calculation restricted to the region bounded by the star gives $E_{T}/E=0.055$ \cite{lander12a}.}.  The stability of equilibrium configurations such as those presented herein and in \citet{braithwaite08,braithwaite09} is therefore still an open question -- although, given the bounded nature of any kinetic motion, we expect these pseudo-equilibria to be stable to arbitrary perturbations.

\subsection{Gravitational Waves from Stellar Oscillations}
A further application of the simulations presented herein is to understand the gravitational wave emission from magnetar flares.  We reiterate that we are {\it not} advocating that kink instabilities are {\it responsible} for magnetar flares.  Instead, we are utilising the kink instability to mimic a global reconstruction of the internal magnetic field expected immediately following a magnetar flare.  This is pertinent given the current experimental searches for gravitational waves from magnetar flares from the LIGO Scientific Collaboration \cite{abbott07,abbott08b,abadie11}. 

Our key results in the process of determining overall gravitational wave emission are equations (\ref{GWoutput}) and (\ref{GWoutputE}).  In particular, equation (\ref{GWoutput}) shows the expected gravitational wave strain amplitude given a star of mass $M$, radius $R$, polar magnetic field strength $B_{{\rm pole}}$ and located $d\,{\rm kpc}$ from Earth.  In \citet{zink12} we presented a first self-consistent estimate of this given a single equation of state, finding that the gravitational wave strain scales with approximately the third power of the magnetic field.  The present paper extends this to show that the gravitational wave strain scales with almost the fifth power of the radius and the mass squared.  

To derive the previous result we explored multiple polytropic equations of state, ranging from polytropic index of $\Gamma=1.67$ to $\Gamma=2.46$.  The conclusion for all of these simulations was the same -- {\it the gravitational wave signal associated with $f$-modes is not observable with present or near-future gravitational wave observatories}, including the proposed Einstein Telescope.  Of course this result comes with the caveat that we are not observing extremely bloated stars with larger than expected radii.  

Another factor that could alter our conclusions regarding the $f$-mode excitation is that we have only explored motions due to unstable poloidal fields.  As discussed above, various mixtures of poloidal and toroidal field configurations could exist in the interior of neutron stars, including configurations with a relatively strong toroidal component.  Such configurations would have stronger magnetic fields, hence higher coupling with the $f$-mode and consequently greater gravitational wave luminosities.  Whilst an exploration of these effects is certainly warranted, we do note that such fields are unlikely to provide the orders-of-magnitude difference required in gravitational wave emission to become detectable.  

Although the previous $f$-mode result represents a somewhat disappointing science case for gravitational wave observatories, there is potential for excitement in a somewhat unexpected region of the spectrum.  Our simulations have shown the presence of strong oscillations in the low, i.e. $10$'s to $100$'s of ${\rm Hz}$, region of the spectrum.  These oscillations are consistent with the propagation of Alfv\'en waves in the interior of the star.  Our results only show the indication of these modes, however the detectability requires a better knowledge of the various damping mechanisms relevant for these modes of oscillation.  In figures \ref{LIGO} we have shown these modes assuming a damping time of $10\,{\rm ms}$ to $1\,{\rm s}$.  The gravitational wave signal amplitude, and hence detector sensitivity, scales as $\sqrt{T}$, where $T$ is the length of the signal \cite[e.g.][]{sathyaprakash09}.  Therefore, an increase in damping time from $1\,{\rm s}$ to $1\,{\rm hr}$ increases the signal amplitude by a factor of 60.  To the best of our knowledge, estimates of expected damping times for Alfv\'en modes do not exist in the literature, nor is it entirely clear what the relevant damping mechanism is.  We therefore advocate considerable more research in this area, associated with developing an understanding of the relevant damping mechanism of these modes as well as targeting gravitational wave searches.

\acknowledgments
This work is supported by the Transregio 7 `Gravitational Wave Astronomy', financed by the Deutsche Forschungsgemeinschaft DFG (German Research Foundation) and the Go8-DAAD Australia-Germany Joint Research Co-operation Scheme.  PL was partially supported by the Alexander von Humboldt Foundation, an Australian Research Council Discovery Project (DP110103347) and an internal University of Melbourne Early Career Researcher grant.  Simulations were performed on the Multi-modal Australian ScienceS Imaging and Visualisation Environment (MASSIVE) (www.massive.org.au) through an award under the Merit Allocation Scheme on the NCI National Facility at the ANU and also on the GPU nodes on the nehalem cluster at the High Performance Computing Center Stuttgart (HLRS).  We are extremely grateful to Kostas Glampedakis and Andrew Melatos for invaluable discussions and also to Sam Lander for discussions regarding his instability calculations.

\bibliography{Bib}

\begin{thebibliography}{89}
\expandafter\ifx\csname natexlab\endcsname\relax\def\natexlab#1{#1}\fi
\expandafter\ifx\csname bibnamefont\endcsname\relax
  \def\bibnamefont#1{#1}\fi
\expandafter\ifx\csname bibfnamefont\endcsname\relax
  \def\bibfnamefont#1{#1}\fi
\expandafter\ifx\csname citenamefont\endcsname\relax
  \def\citenamefont#1{#1}\fi
\expandafter\ifx\csname url\endcsname\relax
  \def\url#1{\texttt{#1}}\fi
\expandafter\ifx\csname urlprefix\endcsname\relax\def\urlprefix{URL }\fi
\providecommand{\bibinfo}[2]{#2}
\providecommand{\eprint}[2][]{\url{#2}}

\bibitem[{\citenamefont{Pacini}(1967)}]{pacini67}
\bibinfo{author}{\bibfnamefont{F.}~\bibnamefont{Pacini}},
  \bibinfo{journal}{Nature} \textbf{\bibinfo{volume}{216}},
  \bibinfo{pages}{567} (\bibinfo{year}{1967}).

\bibitem[{\citenamefont{Pacini}(1968)}]{pacini68}
\bibinfo{author}{\bibfnamefont{F.}~\bibnamefont{Pacini}},
  \bibinfo{journal}{Nature} \textbf{\bibinfo{volume}{219}},
  \bibinfo{pages}{145} (\bibinfo{year}{1968}).

\bibitem[{\citenamefont{Eastlund}(1968)}]{eastlund68}
\bibinfo{author}{\bibfnamefont{B.~J.} \bibnamefont{Eastlund}},
  \bibinfo{journal}{Nature} \textbf{\bibinfo{volume}{220}},
  \bibinfo{pages}{1293} (\bibinfo{year}{1968}).

\bibitem[{\citenamefont{Gunn and Ostriker}(1969)}]{gunn69}
\bibinfo{author}{\bibfnamefont{J.~E.} \bibnamefont{Gunn}} \bibnamefont{and}
  \bibinfo{author}{\bibfnamefont{J.~P.} \bibnamefont{Ostriker}},
  \bibinfo{journal}{Nature} \textbf{\bibinfo{volume}{220}},
  \bibinfo{pages}{454} (\bibinfo{year}{1969}).

\bibitem[{\citenamefont{Easson}(1979)}]{easson79}
\bibinfo{author}{\bibfnamefont{I.}~\bibnamefont{Easson}},
  \bibinfo{journal}{Astrophys. J.} \textbf{\bibinfo{volume}{228}},
  \bibinfo{pages}{257} (\bibinfo{year}{1979}).

\bibitem[{\citenamefont{Haskell et~al.}(2008)\citenamefont{Haskell, Samuelsson,
  Glampedakis, and Andersson}}]{haskell08}
\bibinfo{author}{\bibfnamefont{B.}~\bibnamefont{Haskell}},
  \bibinfo{author}{\bibfnamefont{L.}~\bibnamefont{Samuelsson}},
  \bibinfo{author}{\bibfnamefont{K.}~\bibnamefont{Glampedakis}},
  \bibnamefont{and}
  \bibinfo{author}{\bibfnamefont{N.}~\bibnamefont{Andersson}},
  \bibinfo{journal}{Mon. Not. R. Astron. Soc.} \textbf{\bibinfo{volume}{385}},
  \bibinfo{pages}{531} (\bibinfo{year}{2008}).

\bibitem[{\citenamefont{Gualtieri et~al.}(2011)\citenamefont{Gualtieri, Ciolfi,
  and Ferrari}}]{gualtieri11}
\bibinfo{author}{\bibfnamefont{L.}~\bibnamefont{Gualtieri}},
  \bibinfo{author}{\bibfnamefont{R.}~\bibnamefont{Ciolfi}}, \bibnamefont{and}
  \bibinfo{author}{\bibfnamefont{V.}~\bibnamefont{Ferrari}},
  \bibinfo{journal}{Class. Quantum Grav.} \textbf{\bibinfo{volume}{28}},
  \bibinfo{pages}{114014} (\bibinfo{year}{2011}).

\bibitem[{\citenamefont{Mastrano et~al.}(2011)\citenamefont{Mastrano, Melatos,
  Reissenegger, and Akg\"un}}]{mastrano11}
\bibinfo{author}{\bibfnamefont{A.}~\bibnamefont{Mastrano}},
  \bibinfo{author}{\bibfnamefont{A.}~\bibnamefont{Melatos}},
  \bibinfo{author}{\bibfnamefont{A.}~\bibnamefont{Reissenegger}},
  \bibnamefont{and} \bibinfo{author}{\bibfnamefont{T.}~\bibnamefont{Akg\"un}},
  \bibinfo{journal}{Mon. Not. R. Astron. Soc.} \textbf{\bibinfo{volume}{417}},
  \bibinfo{pages}{2288} (\bibinfo{year}{2011}).

\bibitem[{\citenamefont{Payne and Melatos}(2004)}]{payne04}
\bibinfo{author}{\bibfnamefont{D.~J.~B.} \bibnamefont{Payne}} \bibnamefont{and}
  \bibinfo{author}{\bibfnamefont{A.}~\bibnamefont{Melatos}},
  \bibinfo{journal}{Mon. Not. R. Astron. Soc.} \textbf{\bibinfo{volume}{351}},
  \bibinfo{pages}{569} (\bibinfo{year}{2004}).

\bibitem[{\citenamefont{Vigelius and Melatos}(2008)}]{vigelius08}
\bibinfo{author}{\bibfnamefont{M.}~\bibnamefont{Vigelius}} \bibnamefont{and}
  \bibinfo{author}{\bibfnamefont{A.}~\bibnamefont{Melatos}},
  \bibinfo{journal}{Mon. Not. R. Astron. Soc.} \textbf{\bibinfo{volume}{386}},
  \bibinfo{pages}{1294} (\bibinfo{year}{2008}).

\bibitem[{\citenamefont{Thompson and Duncan}(1996)}]{thompson96}
\bibinfo{author}{\bibfnamefont{C.}~\bibnamefont{Thompson}} \bibnamefont{and}
  \bibinfo{author}{\bibfnamefont{R.~C.} \bibnamefont{Duncan}},
  \bibinfo{journal}{Astrophys. J.} \textbf{\bibinfo{volume}{473}},
  \bibinfo{pages}{322} (\bibinfo{year}{1996}).

\bibitem[{\citenamefont{Heyl and Kulkarni}(1998)}]{heyl98}
\bibinfo{author}{\bibfnamefont{J.~S.} \bibnamefont{Heyl}} \bibnamefont{and}
  \bibinfo{author}{\bibfnamefont{S.~R.} \bibnamefont{Kulkarni}},
  \bibinfo{journal}{Astrophys. J.} \textbf{\bibinfo{volume}{506}},
  \bibinfo{pages}{L61} (\bibinfo{year}{1998}).

\bibitem[{\citenamefont{Colpi et~al.}(2000)\citenamefont{Colpi, Geppert, and
  Page}}]{colpi00}
\bibinfo{author}{\bibfnamefont{M.}~\bibnamefont{Colpi}},
  \bibinfo{author}{\bibfnamefont{U.}~\bibnamefont{Geppert}}, \bibnamefont{and}
  \bibinfo{author}{\bibfnamefont{D.}~\bibnamefont{Page}},
  \bibinfo{journal}{Astrophys. J.} \textbf{\bibinfo{volume}{529}},
  \bibinfo{pages}{L29} (\bibinfo{year}{2000}).

\bibitem[{\citenamefont{Ho et~al.}(2011)\citenamefont{Ho, Glampedakis, and
  Andersson}}]{ho11}
\bibinfo{author}{\bibfnamefont{W.~C.~G.} \bibnamefont{Ho}},
  \bibinfo{author}{\bibfnamefont{K.}~\bibnamefont{Glampedakis}},
  \bibnamefont{and} \bibinfo{author}{\bibfnamefont{N.}~\bibnamefont{Andersson}}
  (\bibinfo{year}{2011}), \bibinfo{note}{arXiv:1112.1415}.

\bibitem[{\citenamefont{Duncan and Thompson}(1992)}]{duncan92}
\bibinfo{author}{\bibfnamefont{R.~C.} \bibnamefont{Duncan}} \bibnamefont{and}
  \bibinfo{author}{\bibfnamefont{C.}~\bibnamefont{Thompson}},
  \bibinfo{journal}{Astrophys. J.} \textbf{\bibinfo{volume}{392}},
  \bibinfo{pages}{L9} (\bibinfo{year}{1992}).

\bibitem[{\citenamefont{Thompson and Duncan}(1995)}]{thompson95}
\bibinfo{author}{\bibfnamefont{C.}~\bibnamefont{Thompson}} \bibnamefont{and}
  \bibinfo{author}{\bibfnamefont{R.~C.} \bibnamefont{Duncan}},
  \bibinfo{journal}{Mon. Not. R. Astron. Soc.} \textbf{\bibinfo{volume}{275}},
  \bibinfo{pages}{255} (\bibinfo{year}{1995}).

\bibitem[{\citenamefont{Strohmayer and Watts}(2005)}]{strohmayer05}
\bibinfo{author}{\bibfnamefont{T.~E.} \bibnamefont{Strohmayer}}
  \bibnamefont{and} \bibinfo{author}{\bibfnamefont{A.~L.} \bibnamefont{Watts}},
  \bibinfo{journal}{Astrophys. J.} \textbf{\bibinfo{volume}{632}},
  \bibinfo{pages}{L111} (\bibinfo{year}{2005}).

\bibitem[{\citenamefont{Strohmayer and Watts}(2006)}]{strohmayer06}
\bibinfo{author}{\bibfnamefont{T.~E.} \bibnamefont{Strohmayer}}
  \bibnamefont{and} \bibinfo{author}{\bibfnamefont{A.~L.} \bibnamefont{Watts}},
  \bibinfo{journal}{Astrophys. J.} \textbf{\bibinfo{volume}{653}},
  \bibinfo{pages}{593} (\bibinfo{year}{2006}).

\bibitem[{\citenamefont{Levin}(2006)}]{levin06}
\bibinfo{author}{\bibfnamefont{Y.}~\bibnamefont{Levin}}, \bibinfo{journal}{Mon.
  Not. R. Astron. Soc.} \textbf{\bibinfo{volume}{368}}, \bibinfo{pages}{L35}
  (\bibinfo{year}{2006}).

\bibitem[{\citenamefont{Levin}(2007)}]{levin07}
\bibinfo{author}{\bibfnamefont{Y.}~\bibnamefont{Levin}}, \bibinfo{journal}{Mon.
  Not. R. Astron. Soc.} \textbf{\bibinfo{volume}{377}}, \bibinfo{pages}{159}
  (\bibinfo{year}{2007}).

\bibitem[{\citenamefont{Watts and Strohmayer}(2007)}]{watts07}
\bibinfo{author}{\bibfnamefont{A.~L.} \bibnamefont{Watts}} \bibnamefont{and}
  \bibinfo{author}{\bibfnamefont{T.~E.} \bibnamefont{Strohmayer}},
  \bibinfo{journal}{Adv. Space Res.} \textbf{\bibinfo{volume}{40}},
  \bibinfo{pages}{1446} (\bibinfo{year}{2007}).

\bibitem[{\citenamefont{Sotani et~al.}(2008)\citenamefont{Sotani, Kokkotas, and
  Stergioulas}}]{sotani08a}
\bibinfo{author}{\bibfnamefont{H.}~\bibnamefont{Sotani}},
  \bibinfo{author}{\bibfnamefont{K.~D.} \bibnamefont{Kokkotas}},
  \bibnamefont{and}
  \bibinfo{author}{\bibfnamefont{N.}~\bibnamefont{Stergioulas}},
  \bibinfo{journal}{Mon. Not. R. Astron. Soc.} \textbf{\bibinfo{volume}{385}},
  \bibinfo{pages}{L5} (\bibinfo{year}{2008}).

\bibitem[{\citenamefont{Colaiuda et~al.}(2009)\citenamefont{Colaiuda, Beyer,
  and Kokkotas}}]{colaiuda09}
\bibinfo{author}{\bibfnamefont{A.}~\bibnamefont{Colaiuda}},
  \bibinfo{author}{\bibfnamefont{H.}~\bibnamefont{Beyer}}, \bibnamefont{and}
  \bibinfo{author}{\bibfnamefont{K.~D.} \bibnamefont{Kokkotas}},
  \bibinfo{journal}{Mon. Not. R. Astron. Soc.} \textbf{\bibinfo{volume}{396}},
  \bibinfo{pages}{1441} (\bibinfo{year}{2009}).

\bibitem[{\citenamefont{Cerd\'a-Dur\'an
  et~al.}(2009)\citenamefont{Cerd\'a-Dur\'an, Stergioulas, and
  Font}}]{cerdaduran09}
\bibinfo{author}{\bibfnamefont{P.}~\bibnamefont{Cerd\'a-Dur\'an}},
  \bibinfo{author}{\bibfnamefont{N.}~\bibnamefont{Stergioulas}},
  \bibnamefont{and} \bibinfo{author}{\bibfnamefont{J.~A.} \bibnamefont{Font}},
  \bibinfo{journal}{Mon. Not. R. Astron. Soc.} \textbf{\bibinfo{volume}{397}},
  \bibinfo{pages}{1607} (\bibinfo{year}{2009}).

\bibitem[{\citenamefont{Colaiuda and Kokkotas}(2011)}]{colaiuda11}
\bibinfo{author}{\bibfnamefont{A.}~\bibnamefont{Colaiuda}} \bibnamefont{and}
  \bibinfo{author}{\bibfnamefont{K.~D.} \bibnamefont{Kokkotas}},
  \bibinfo{journal}{Mon. Not. R. Astron. Soc.} \textbf{\bibinfo{volume}{414}},
  \bibinfo{pages}{3014} (\bibinfo{year}{2011}).

\bibitem[{\citenamefont{Gabler et~al.}(2011)\citenamefont{Gabler,
  Cerd\'a-Dur\'an, Font, M\"uller, and Stergioulas}}]{gabler11}
\bibinfo{author}{\bibfnamefont{M.}~\bibnamefont{Gabler}},
  \bibinfo{author}{\bibfnamefont{P.}~\bibnamefont{Cerd\'a-Dur\'an}},
  \bibinfo{author}{\bibfnamefont{J.~A.} \bibnamefont{Font}},
  \bibinfo{author}{\bibfnamefont{E.}~\bibnamefont{M\"uller}}, \bibnamefont{and}
  \bibinfo{author}{\bibfnamefont{N.}~\bibnamefont{Stergioulas}},
  \bibinfo{journal}{Mon. Not. R. Astron. Soc.} \textbf{\bibinfo{volume}{410}},
  \bibinfo{pages}{L37} (\bibinfo{year}{2011}).

\bibitem[{\citenamefont{Colaiuda and Kokkotas}(2012)}]{colaiuda12}
\bibinfo{author}{\bibfnamefont{A.}~\bibnamefont{Colaiuda}} \bibnamefont{and}
  \bibinfo{author}{\bibfnamefont{K.~D.} \bibnamefont{Kokkotas}},
  \bibinfo{journal}{Mon. Not. R. Astron. Soc.}  (\bibinfo{year}{2012}),
  \bibinfo{note}{in press (arXiv:1113.3561)}.

\bibitem[{\citenamefont{Melatos}(1999)}]{melatos99}
\bibinfo{author}{\bibfnamefont{A.}~\bibnamefont{Melatos}},
  \bibinfo{journal}{Astrophys. J.} \textbf{\bibinfo{volume}{519}},
  \bibinfo{pages}{L77} (\bibinfo{year}{1999}).

\bibitem[{\citenamefont{Glampedakis and Jones}(2010)}]{glampedakis10}
\bibinfo{author}{\bibfnamefont{K.}~\bibnamefont{Glampedakis}} \bibnamefont{and}
  \bibinfo{author}{\bibfnamefont{D.~I.} \bibnamefont{Jones}},
  \bibinfo{journal}{Mon. Not. R. Astron. Soc.} \textbf{\bibinfo{volume}{405}},
  \bibinfo{pages}{L6} (\bibinfo{year}{2010}).

\bibitem[{\citenamefont{Zink}(2011)}]{zink11}
\bibinfo{author}{\bibfnamefont{B.}~\bibnamefont{Zink}} (\bibinfo{year}{2011}),
  \bibinfo{note}{arXiv:1102.5202}.

\bibitem[{\citenamefont{Lasky et~al.}(2011)\citenamefont{Lasky, Zink, Kokkotas,
  and Glampedakis}}]{lasky11}
\bibinfo{author}{\bibfnamefont{P.~D.} \bibnamefont{Lasky}},
  \bibinfo{author}{\bibfnamefont{B.}~\bibnamefont{Zink}},
  \bibinfo{author}{\bibfnamefont{K.~D.} \bibnamefont{Kokkotas}},
  \bibnamefont{and}
  \bibinfo{author}{\bibfnamefont{K.}~\bibnamefont{Glampedakis}},
  \bibinfo{journal}{Astrophys. J.} \textbf{\bibinfo{volume}{735}},
  \bibinfo{pages}{L20} (\bibinfo{year}{2011}).

\bibitem[{\citenamefont{Zink et~al.}(2012)\citenamefont{Zink, Lasky, and
  Kokkotas}}]{zink12}
\bibinfo{author}{\bibfnamefont{B.}~\bibnamefont{Zink}},
  \bibinfo{author}{\bibfnamefont{P.~D.} \bibnamefont{Lasky}}, \bibnamefont{and}
  \bibinfo{author}{\bibfnamefont{K.~D.} \bibnamefont{Kokkotas}},
  \bibinfo{journal}{Phys. Rev. D} \textbf{\bibinfo{volume}{85}},
  \bibinfo{pages}{024030} (\bibinfo{year}{2012}),
  \bibinfo{note}{arXiv:1107.1689}.

\bibitem[{\citenamefont{Chandrasekhar and Fermi}(1953)}]{chandrasekhar53}
\bibinfo{author}{\bibfnamefont{S.}~\bibnamefont{Chandrasekhar}}
  \bibnamefont{and} \bibinfo{author}{\bibfnamefont{E.}~\bibnamefont{Fermi}},
  \bibinfo{journal}{Astrophys. J.} \textbf{\bibinfo{volume}{118}},
  \bibinfo{pages}{116} (\bibinfo{year}{1953}).

\bibitem[{\citenamefont{Monaghan}(1965)}]{monaghan65}
\bibinfo{author}{\bibfnamefont{J.~J.} \bibnamefont{Monaghan}},
  \bibinfo{journal}{Mon. Not. R. Astron. Soc.} \textbf{\bibinfo{volume}{131}},
  \bibinfo{pages}{105} (\bibinfo{year}{1965}).

\bibitem[{\citenamefont{Roxburgh}(1966)}]{roxburgh66}
\bibinfo{author}{\bibfnamefont{I.~W.} \bibnamefont{Roxburgh}},
  \bibinfo{journal}{Mon. Not. R. Astron. Soc.} \textbf{\bibinfo{volume}{132}},
  \bibinfo{pages}{347} (\bibinfo{year}{1966}).

\bibitem[{\citenamefont{Parker}(1966)}]{parker66}
\bibinfo{author}{\bibfnamefont{E.~N.} \bibnamefont{Parker}},
  \bibinfo{journal}{Astrophys. J.} \textbf{\bibinfo{volume}{145}},
  \bibinfo{pages}{811} (\bibinfo{year}{1966}).

\bibitem[{\citenamefont{Tayler}(1957)}]{tayler57}
\bibinfo{author}{\bibfnamefont{R.~J.} \bibnamefont{Tayler}},
  \bibinfo{journal}{Proc. Phys. Soc. B} \textbf{\bibinfo{volume}{70}},
  \bibinfo{pages}{31} (\bibinfo{year}{1957}).

\bibitem[{\citenamefont{Tayler}(1973)}]{tayler73}
\bibinfo{author}{\bibfnamefont{R.~J.} \bibnamefont{Tayler}},
  \bibinfo{journal}{Mon. Not. R. Astron. Soc.} \textbf{\bibinfo{volume}{161}},
  \bibinfo{pages}{365} (\bibinfo{year}{1973}).

\bibitem[{\citenamefont{Wright}(1973)}]{wright73}
\bibinfo{author}{\bibfnamefont{G.~A.~E.} \bibnamefont{Wright}},
  \bibinfo{journal}{Mon. Not. R. Astron. Soc.} \textbf{\bibinfo{volume}{162}},
  \bibinfo{pages}{339} (\bibinfo{year}{1973}).

\bibitem[{\citenamefont{Markey and Tayler}(1973)}]{markey73}
\bibinfo{author}{\bibfnamefont{P.}~\bibnamefont{Markey}} \bibnamefont{and}
  \bibinfo{author}{\bibfnamefont{R.~J.} \bibnamefont{Tayler}},
  \bibinfo{journal}{Mon. Not. R. Astron. Soc.} \textbf{\bibinfo{volume}{163}},
  \bibinfo{pages}{77} (\bibinfo{year}{1973}).

\bibitem[{\citenamefont{Markey and Tayler}(1974)}]{markey74}
\bibinfo{author}{\bibfnamefont{P.}~\bibnamefont{Markey}} \bibnamefont{and}
  \bibinfo{author}{\bibfnamefont{R.~J.} \bibnamefont{Tayler}},
  \bibinfo{journal}{Mon. Not. R. Astron. Soc.} \textbf{\bibinfo{volume}{168}},
  \bibinfo{pages}{505} (\bibinfo{year}{1974}).

\bibitem[{\citenamefont{Flowers and Ruderman}(1977)}]{flowers77}
\bibinfo{author}{\bibfnamefont{E.}~\bibnamefont{Flowers}} \bibnamefont{and}
  \bibinfo{author}{\bibfnamefont{M.~A.} \bibnamefont{Ruderman}},
  \bibinfo{journal}{Astrophys. J.} \textbf{\bibinfo{volume}{215}},
  \bibinfo{pages}{302} (\bibinfo{year}{1977}).

\bibitem[{\citenamefont{Tayler}(1980)}]{tayler80}
\bibinfo{author}{\bibfnamefont{R.~J.} \bibnamefont{Tayler}},
  \bibinfo{journal}{Mon. Not. R. Astron. Soc.} \textbf{\bibinfo{volume}{191}},
  \bibinfo{pages}{151} (\bibinfo{year}{1980}).

\bibitem[{\citenamefont{Ioka}(2001)}]{ioka01}
\bibinfo{author}{\bibfnamefont{K.}~\bibnamefont{Ioka}}, \bibinfo{journal}{Mon.
  Not. R. Astron. Soc.} \textbf{\bibinfo{volume}{327}}, \bibinfo{pages}{639}
  (\bibinfo{year}{2001}).

\bibitem[{\citenamefont{Yoshida and Eriguchi}(2006)}]{yoshida06}
\bibinfo{author}{\bibfnamefont{S.}~\bibnamefont{Yoshida}} \bibnamefont{and}
  \bibinfo{author}{\bibfnamefont{Y.}~\bibnamefont{Eriguchi}},
  \bibinfo{journal}{Astrophys. J. S.} \textbf{\bibinfo{volume}{164}},
  \bibinfo{pages}{156} (\bibinfo{year}{2006}).

\bibitem[{\citenamefont{Yoshida et~al.}(2006)\citenamefont{Yoshida, Yoshida,
  and Eriguchi}}]{yoshida06b}
\bibinfo{author}{\bibfnamefont{S.}~\bibnamefont{Yoshida}},
  \bibinfo{author}{\bibfnamefont{S.}~\bibnamefont{Yoshida}}, \bibnamefont{and}
  \bibinfo{author}{\bibfnamefont{Y.}~\bibnamefont{Eriguchi}},
  \bibinfo{journal}{Astrophys. J.} \textbf{\bibinfo{volume}{651}},
  \bibinfo{pages}{462} (\bibinfo{year}{2006}).

\bibitem[{\citenamefont{Ciolfi et~al.}(2009)\citenamefont{Ciolfi, Ferrari,
  Gualtieri, and Pons}}]{ciolfi09}
\bibinfo{author}{\bibfnamefont{R.}~\bibnamefont{Ciolfi}},
  \bibinfo{author}{\bibfnamefont{V.}~\bibnamefont{Ferrari}},
  \bibinfo{author}{\bibfnamefont{L.}~\bibnamefont{Gualtieri}},
  \bibnamefont{and} \bibinfo{author}{\bibfnamefont{J.~A.} \bibnamefont{Pons}},
  \bibinfo{journal}{Mon. Not. R. Astron. Soc.} \textbf{\bibinfo{volume}{397}},
  \bibinfo{pages}{913} (\bibinfo{year}{2009}).

\bibitem[{\citenamefont{Braithwaite and Nordlund}(2006)}]{braithwaite06b}
\bibinfo{author}{\bibfnamefont{J.}~\bibnamefont{Braithwaite}} \bibnamefont{and}
  \bibinfo{author}{\bibfnamefont{A.}~\bibnamefont{Nordlund}},
  \bibinfo{journal}{A\&A} \textbf{\bibinfo{volume}{450}}, \bibinfo{pages}{1077}
  (\bibinfo{year}{2006}).

\bibitem[{\citenamefont{Braithwaite}(2009)}]{braithwaite09}
\bibinfo{author}{\bibfnamefont{J.}~\bibnamefont{Braithwaite}},
  \bibinfo{journal}{Mon. Not. R. Astron. Soc.} \textbf{\bibinfo{volume}{397}},
  \bibinfo{pages}{763} (\bibinfo{year}{2009}).

\bibitem[{\citenamefont{Lander and Jones}(2012)}]{lander12}
\bibinfo{author}{\bibfnamefont{S.~K.} \bibnamefont{Lander}} \bibnamefont{and}
  \bibinfo{author}{\bibfnamefont{D.~I.} \bibnamefont{Jones}}
  (\bibinfo{year}{2012}), \bibinfo{note}{arXiv:1202.2339}.

\bibitem[{\citenamefont{Braithwaite}(2008)}]{braithwaite08}
\bibinfo{author}{\bibfnamefont{J.}~\bibnamefont{Braithwaite}},
  \bibinfo{journal}{Mon. Not. R. Astron. Soc.} \textbf{\bibinfo{volume}{386}},
  \bibinfo{pages}{1947} (\bibinfo{year}{2008}).

\bibitem[{\citenamefont{Frieman and Rotenberg}(1960)}]{frieman60}
\bibinfo{author}{\bibfnamefont{E.}~\bibnamefont{Frieman}} \bibnamefont{and}
  \bibinfo{author}{\bibfnamefont{M.}~\bibnamefont{Rotenberg}},
  \bibinfo{journal}{Rev. Mod. Phys.} \textbf{\bibinfo{volume}{32}},
  \bibinfo{pages}{898} (\bibinfo{year}{1960}).

\bibitem[{\citenamefont{Geppert and Rheinhardt}(2006)}]{geppert06}
\bibinfo{author}{\bibfnamefont{U.}~\bibnamefont{Geppert}} \bibnamefont{and}
  \bibinfo{author}{\bibfnamefont{M.}~\bibnamefont{Rheinhardt}},
  \bibinfo{journal}{A\&A} \textbf{\bibinfo{volume}{456}}, \bibinfo{pages}{639}
  (\bibinfo{year}{2006}).

\bibitem[{\citenamefont{Lander and Jones}(2011)}]{lander11b}
\bibinfo{author}{\bibfnamefont{S.~K.} \bibnamefont{Lander}} \bibnamefont{and}
  \bibinfo{author}{\bibfnamefont{D.~I.} \bibnamefont{Jones}},
  \bibinfo{journal}{Mon. Not. R. Astron. Soc.} \textbf{\bibinfo{volume}{412}},
  \bibinfo{pages}{1730} (\bibinfo{year}{2011}).

\bibitem[{\citenamefont{Gammie et~al.}(2003)\citenamefont{Gammie, McKinney, and
  T\'oth}}]{gammie03}
\bibinfo{author}{\bibfnamefont{C.~F.} \bibnamefont{Gammie}},
  \bibinfo{author}{\bibfnamefont{J.~C.} \bibnamefont{McKinney}},
  \bibnamefont{and} \bibinfo{author}{\bibfnamefont{G.}~\bibnamefont{T\'oth}},
  \bibinfo{journal}{Astrophys. J.} \textbf{\bibinfo{volume}{589}},
  \bibinfo{pages}{444} (\bibinfo{year}{2003}).

\bibitem[{\citenamefont{Zink et~al.}(2008)\citenamefont{Zink, Schnetter, and
  Tiglio}}]{zink08}
\bibinfo{author}{\bibfnamefont{B.}~\bibnamefont{Zink}},
  \bibinfo{author}{\bibfnamefont{E.}~\bibnamefont{Schnetter}},
  \bibnamefont{and} \bibinfo{author}{\bibfnamefont{M.}~\bibnamefont{Tiglio}},
  \bibinfo{journal}{Phys. Rev. D} \textbf{\bibinfo{volume}{77}},
  \bibinfo{pages}{103015} (\bibinfo{year}{2008}).

\bibitem[{\citenamefont{Korobkin et~al.}(2011)\citenamefont{Korobkin,
  Abdikamalov, Schnetter, Stergioulas, and Zink}}]{korobkin11}
\bibinfo{author}{\bibfnamefont{O.}~\bibnamefont{Korobkin}},
  \bibinfo{author}{\bibfnamefont{E.~B.} \bibnamefont{Abdikamalov}},
  \bibinfo{author}{\bibfnamefont{E.}~\bibnamefont{Schnetter}},
  \bibinfo{author}{\bibfnamefont{N.}~\bibnamefont{Stergioulas}},
  \bibnamefont{and} \bibinfo{author}{\bibfnamefont{B.}~\bibnamefont{Zink}},
  \bibinfo{journal}{Phys. Rev. D} \textbf{\bibinfo{volume}{83}},
  \bibinfo{pages}{043007} (\bibinfo{year}{2011}),
  \bibinfo{note}{arXiv:1011.3010}.

\bibitem[{\citenamefont{Bocquet et~al.}(1995)\citenamefont{Bocquet, Bonazzola,
  Gourgoulhon, and Novak}}]{bocquet95}
\bibinfo{author}{\bibfnamefont{M.}~\bibnamefont{Bocquet}},
  \bibinfo{author}{\bibfnamefont{S.}~\bibnamefont{Bonazzola}},
  \bibinfo{author}{\bibfnamefont{E.}~\bibnamefont{Gourgoulhon}},
  \bibnamefont{and} \bibinfo{author}{\bibfnamefont{J.}~\bibnamefont{Novak}},
  \bibinfo{journal}{A\&A} \textbf{\bibinfo{volume}{301}}, \bibinfo{pages}{757}
  (\bibinfo{year}{1995}).

\bibitem[{\citenamefont{Bucciantini and {Del Zanna}}(2011)}]{bucciantini11}
\bibinfo{author}{\bibfnamefont{N.}~\bibnamefont{Bucciantini}} \bibnamefont{and}
  \bibinfo{author}{\bibfnamefont{L.}~\bibnamefont{{Del Zanna}}},
  \bibinfo{journal}{A\&A} \textbf{\bibinfo{volume}{528}}, \bibinfo{pages}{A101}
  (\bibinfo{year}{2011}).

\bibitem[{\citenamefont{Kiuchi et~al.}(2008)\citenamefont{Kiuchi, Shibata, and
  Yoshida}}]{kiuchi08}
\bibinfo{author}{\bibfnamefont{K.}~\bibnamefont{Kiuchi}},
  \bibinfo{author}{\bibfnamefont{M.}~\bibnamefont{Shibata}}, \bibnamefont{and}
  \bibinfo{author}{\bibfnamefont{S.}~\bibnamefont{Yoshida}},
  \bibinfo{journal}{Phys. Rev. D} \textbf{\bibinfo{volume}{78}},
  \bibinfo{pages}{024029} (\bibinfo{year}{2008}).

\bibitem[{\citenamefont{Stergioulas and Friedman}(1995)}]{stergioulas95}
\bibinfo{author}{\bibfnamefont{N.}~\bibnamefont{Stergioulas}} \bibnamefont{and}
  \bibinfo{author}{\bibfnamefont{J.~L.} \bibnamefont{Friedman}},
  \bibinfo{journal}{Astrophys. J.} \textbf{\bibinfo{volume}{444}},
  \bibinfo{pages}{306} (\bibinfo{year}{1995}).

\bibitem[{\citenamefont{Duez et~al.}(2005)\citenamefont{Duez, Liu, Shapiro, and
  Stephens}}]{duez05}
\bibinfo{author}{\bibfnamefont{M.~D.} \bibnamefont{Duez}},
  \bibinfo{author}{\bibfnamefont{Y.~T.} \bibnamefont{Liu}},
  \bibinfo{author}{\bibfnamefont{S.~L.} \bibnamefont{Shapiro}},
  \bibnamefont{and} \bibinfo{author}{\bibfnamefont{B.~C.}
  \bibnamefont{Stephens}}, \bibinfo{journal}{Phys. Rev. D}
  \textbf{\bibinfo{volume}{72}}, \bibinfo{pages}{024028}
  (\bibinfo{year}{2005}).

\bibitem[{\citenamefont{Shibata and Sekiguchi}(2005)}]{shibata05}
\bibinfo{author}{\bibfnamefont{M.}~\bibnamefont{Shibata}} \bibnamefont{and}
  \bibinfo{author}{\bibfnamefont{Y.}~\bibnamefont{Sekiguchi}},
  \bibinfo{journal}{Phys. Rev. D} \textbf{\bibinfo{volume}{72}},
  \bibinfo{pages}{044014} (\bibinfo{year}{2005}).

\bibitem[{\citenamefont{Cerd\'a-Dur\'an
  et~al.}(2008)\citenamefont{Cerd\'a-Dur\'an, Font, Ant{\'o}n, and
  M\"uller}}]{cerdaduran08}
\bibinfo{author}{\bibfnamefont{P.}~\bibnamefont{Cerd\'a-Dur\'an}},
  \bibinfo{author}{\bibfnamefont{J.~A.} \bibnamefont{Font}},
  \bibinfo{author}{\bibfnamefont{L.}~\bibnamefont{Ant{\'o}n}},
  \bibnamefont{and} \bibinfo{author}{\bibfnamefont{E.}~\bibnamefont{M\"uller}}
  (\bibinfo{year}{2008}), \bibinfo{note}{arXiv:0804.4572}.

\bibitem[{\citenamefont{Giacomazzo and Rezzolla}(2007)}]{giacomazzo07}
\bibinfo{author}{\bibfnamefont{B.}~\bibnamefont{Giacomazzo}} \bibnamefont{and}
  \bibinfo{author}{\bibfnamefont{L.}~\bibnamefont{Rezzolla}},
  \bibinfo{journal}{Class. Quantum Grav.} \textbf{\bibinfo{volume}{24}},
  \bibinfo{pages}{S235} (\bibinfo{year}{2007}).

\bibitem[{\citenamefont{Ciolfi et~al.}(2011)\citenamefont{Ciolfi, Lander,
  Manca, and Rezzolla}}]{ciolfi11}
\bibinfo{author}{\bibfnamefont{R.}~\bibnamefont{Ciolfi}},
  \bibinfo{author}{\bibfnamefont{S.~K.} \bibnamefont{Lander}},
  \bibinfo{author}{\bibfnamefont{G.~M.} \bibnamefont{Manca}}, \bibnamefont{and}
  \bibinfo{author}{\bibfnamefont{L.}~\bibnamefont{Rezzolla}},
  \bibinfo{journal}{Astrophys. J.} \textbf{\bibinfo{volume}{736}},
  \bibinfo{pages}{L6} (\bibinfo{year}{2011}), \bibinfo{note}{arXiv:1105.3971}.

\bibitem[{\citenamefont{Detweiler}(1975)}]{detweiler75}
\bibinfo{author}{\bibfnamefont{S.~L.} \bibnamefont{Detweiler}},
  \bibinfo{journal}{Astrophys. J.} \textbf{\bibinfo{volume}{197}},
  \bibinfo{pages}{203} (\bibinfo{year}{1975}).

\bibitem[{\citenamefont{McDermott et~al.}(1988)\citenamefont{McDermott, {van
  Horn}, and Hansen}}]{mcdermott88}
\bibinfo{author}{\bibfnamefont{P.~N.} \bibnamefont{McDermott}},
  \bibinfo{author}{\bibfnamefont{H.~M.} \bibnamefont{{van Horn}}},
  \bibnamefont{and} \bibinfo{author}{\bibfnamefont{C.~J.}
  \bibnamefont{Hansen}}, \bibinfo{journal}{Astrophys. J.}
  \textbf{\bibinfo{volume}{325}}, \bibinfo{pages}{725} (\bibinfo{year}{1988}).

\bibitem[{\citenamefont{Noble et~al.}(2006)\citenamefont{Noble, Gammie, and
  McKinney}}]{noble06}
\bibinfo{author}{\bibfnamefont{S.~C.} \bibnamefont{Noble}},
  \bibinfo{author}{\bibfnamefont{C.~F.} \bibnamefont{Gammie}},
  \bibnamefont{and} \bibinfo{author}{\bibfnamefont{J.~C.}
  \bibnamefont{McKinney}}, \bibinfo{journal}{Astrophys. J.}
  \textbf{\bibinfo{volume}{641}}, \bibinfo{pages}{626} (\bibinfo{year}{2006}).

\bibitem[{\citenamefont{Anderson et~al.}(2006)\citenamefont{Anderson,
  Hirschmann, Liebling, and Neilsen}}]{anderson06}
\bibinfo{author}{\bibfnamefont{M.}~\bibnamefont{Anderson}},
  \bibinfo{author}{\bibfnamefont{E.~W.} \bibnamefont{Hirschmann}},
  \bibinfo{author}{\bibfnamefont{S.~L.} \bibnamefont{Liebling}},
  \bibnamefont{and} \bibinfo{author}{\bibfnamefont{D.}~\bibnamefont{Neilsen}},
  \bibinfo{journal}{Class. Quantum Grav.} \textbf{\bibinfo{volume}{23}},
  \bibinfo{pages}{6503} (\bibinfo{year}{2006}).

\bibitem[{\citenamefont{Sod}(1978)}]{sod78}
\bibinfo{author}{\bibfnamefont{G.~A.} \bibnamefont{Sod}}, \bibinfo{journal}{J.
  Comp. Phys.} \textbf{\bibinfo{volume}{27}}, \bibinfo{pages}{1}
  (\bibinfo{year}{1978}).

\bibitem[{\citenamefont{Balsara}(2001)}]{balsara01}
\bibinfo{author}{\bibfnamefont{D.}~\bibnamefont{Balsara}},
  \bibinfo{journal}{Astrophys. J.} \textbf{\bibinfo{volume}{132}},
  \bibinfo{pages}{83} (\bibinfo{year}{2001}).

\bibitem[{\citenamefont{Gaertig and Kokkotas}(2008)}]{gaertig08}
\bibinfo{author}{\bibfnamefont{E.}~\bibnamefont{Gaertig}} \bibnamefont{and}
  \bibinfo{author}{\bibfnamefont{K.~D.} \bibnamefont{Kokkotas}},
  \bibinfo{journal}{Phys. Rev. D} \textbf{\bibinfo{volume}{78}},
  \bibinfo{pages}{064063} (\bibinfo{year}{2008}).

\bibitem[{\citenamefont{Zink et~al.}(2007)\citenamefont{Zink, Stergioulas,
  Hawke, Ott, Schnetter, and M\"uller}}]{zink07}
\bibinfo{author}{\bibfnamefont{B.}~\bibnamefont{Zink}},
  \bibinfo{author}{\bibfnamefont{N.}~\bibnamefont{Stergioulas}},
  \bibinfo{author}{\bibfnamefont{I.}~\bibnamefont{Hawke}},
  \bibinfo{author}{\bibfnamefont{C.~D.} \bibnamefont{Ott}},
  \bibinfo{author}{\bibfnamefont{E.}~\bibnamefont{Schnetter}},
  \bibnamefont{and} \bibinfo{author}{\bibfnamefont{E.}~\bibnamefont{M\"uller}},
  \bibinfo{journal}{Phys. Rev. D} \textbf{\bibinfo{volume}{76}},
  \bibinfo{pages}{024019} (\bibinfo{year}{2007}).

\bibitem[{\citenamefont{Misner et~al.}(1973)\citenamefont{Misner, Thorne, and
  Wheeler}}]{misner73}
\bibinfo{author}{\bibfnamefont{C.~W.} \bibnamefont{Misner}},
  \bibinfo{author}{\bibfnamefont{K.~S.} \bibnamefont{Thorne}},
  \bibnamefont{and} \bibinfo{author}{\bibfnamefont{J.~A.}
  \bibnamefont{Wheeler}}, \emph{\bibinfo{title}{Gravitation}}
  (\bibinfo{publisher}{Freeman}, \bibinfo{address}{New York},
  \bibinfo{year}{1973}).

\bibitem[{\citenamefont{Lindblom and Detweiler}(1983)}]{lindblom83}
\bibinfo{author}{\bibfnamefont{L.}~\bibnamefont{Lindblom}} \bibnamefont{and}
  \bibinfo{author}{\bibfnamefont{S.~L.} \bibnamefont{Detweiler}},
  \bibinfo{journal}{Astrophys. J. S.} \textbf{\bibinfo{volume}{53}},
  \bibinfo{pages}{73} (\bibinfo{year}{1983}).

\bibitem[{\citenamefont{Andersson and Kokkotas}(1998)}]{andersson98}
\bibinfo{author}{\bibfnamefont{N.}~\bibnamefont{Andersson}} \bibnamefont{and}
  \bibinfo{author}{\bibfnamefont{K.~D.} \bibnamefont{Kokkotas}},
  \bibinfo{journal}{Mon. Not. R. Astron. Soc.} \textbf{\bibinfo{volume}{299}},
  \bibinfo{pages}{1059} (\bibinfo{year}{1998}).

\bibitem[{\citenamefont{Sathyaprakash and Schutz}(2009)}]{sathyaprakash09}
\bibinfo{author}{\bibfnamefont{B.~S.} \bibnamefont{Sathyaprakash}}
  \bibnamefont{and} \bibinfo{author}{\bibfnamefont{B.~F.}
  \bibnamefont{Schutz}}, \bibinfo{journal}{Living Rev. Relativity}
  \textbf{\bibinfo{volume}{12}}, \bibinfo{pages}{2} (\bibinfo{year}{2009}).

\bibitem[{\citenamefont{Lander et~al.}(2010)\citenamefont{Lander, Jones, and
  Passamonti}}]{lander10}
\bibinfo{author}{\bibfnamefont{S.~K.} \bibnamefont{Lander}},
  \bibinfo{author}{\bibfnamefont{D.~I.} \bibnamefont{Jones}}, \bibnamefont{and}
  \bibinfo{author}{\bibfnamefont{A.}~\bibnamefont{Passamonti}},
  \bibinfo{journal}{Mon. Not. R. Astron. Soc.} \textbf{\bibinfo{volume}{405}},
  \bibinfo{pages}{318} (\bibinfo{year}{2010}).

\bibitem[{\citenamefont{Braithwaite and Spruit}(2006)}]{braithwaite06}
\bibinfo{author}{\bibfnamefont{J.}~\bibnamefont{Braithwaite}} \bibnamefont{and}
  \bibinfo{author}{\bibfnamefont{H.~C.} \bibnamefont{Spruit}},
  \bibinfo{journal}{A\&A} \textbf{\bibinfo{volume}{450}}, \bibinfo{pages}{1097}
  (\bibinfo{year}{2006}).

\bibitem[{\citenamefont{Levin and {van Hoven}}(2011)}]{levin11}
\bibinfo{author}{\bibfnamefont{Y.}~\bibnamefont{Levin}} \bibnamefont{and}
  \bibinfo{author}{\bibfnamefont{M.}~\bibnamefont{{van Hoven}}},
  \bibinfo{journal}{Mon. Not. R. Astron. Soc.} \textbf{\bibinfo{volume}{418}},
  \bibinfo{pages}{659} (\bibinfo{year}{2011}).

\bibitem[{\citenamefont{Pitts and Tayler}(1985)}]{pitts85}
\bibinfo{author}{\bibfnamefont{E.}~\bibnamefont{Pitts}} \bibnamefont{and}
  \bibinfo{author}{\bibfnamefont{R.~J.} \bibnamefont{Tayler}},
  \bibinfo{journal}{Mon. Not. R. Astron. Soc.} \textbf{\bibinfo{volume}{216}},
  \bibinfo{pages}{139} (\bibinfo{year}{1985}).

\bibitem[{\citenamefont{Braithwaite}(2007)}]{braithwaite07}
\bibinfo{author}{\bibfnamefont{J.}~\bibnamefont{Braithwaite}},
  \bibinfo{journal}{A\&A} \textbf{\bibinfo{volume}{469}}, \bibinfo{pages}{275}
  (\bibinfo{year}{2007}).

\bibitem[{\citenamefont{Stergioulas}(2003)}]{stergioulas03}
\bibinfo{author}{\bibfnamefont{N.}~\bibnamefont{Stergioulas}},
  \bibinfo{journal}{Living Rev. Relativity} \textbf{\bibinfo{volume}{6}},
  \bibinfo{pages}{3} (\bibinfo{year}{2003}).

\bibitem[{\citenamefont{Colaiuda et~al.}(2008)\citenamefont{Colaiuda, Ferrari,
  Gualtieri, and Pons}}]{colaiuda08}
\bibinfo{author}{\bibfnamefont{A.}~\bibnamefont{Colaiuda}},
  \bibinfo{author}{\bibfnamefont{V.}~\bibnamefont{Ferrari}},
  \bibinfo{author}{\bibfnamefont{L.}~\bibnamefont{Gualtieri}},
  \bibnamefont{and} \bibinfo{author}{\bibfnamefont{J.~A.} \bibnamefont{Pons}},
  \bibinfo{journal}{Mon. Not. R. Astron. Soc.} \textbf{\bibinfo{volume}{385}},
  \bibinfo{pages}{2080} (\bibinfo{year}{2008}).

\bibitem[{\citenamefont{Lander}(2012)}]{lander12a}
\bibinfo{author}{\bibfnamefont{S.~K.} \bibnamefont{Lander}},
  \bibinfo{howpublished}{Private Communication} (\bibinfo{year}{2012}).

\bibitem[{\citenamefont{Abbott and {et al.}}(2007)}]{abbott07}
\bibinfo{author}{\bibfnamefont{B.}~\bibnamefont{Abbott}} \bibnamefont{and}
  \bibinfo{author}{\bibnamefont{{et al.}}}, \bibinfo{journal}{Phys. Rev. D}
  \textbf{\bibinfo{volume}{76}}, \bibinfo{pages}{062003}
  (\bibinfo{year}{2007}).

\bibitem[{\citenamefont{Abbott and {et al.}}(2008)}]{abbott08b}
\bibinfo{author}{\bibfnamefont{B.}~\bibnamefont{Abbott}} \bibnamefont{and}
  \bibinfo{author}{\bibnamefont{{et al.}}}, \bibinfo{journal}{Phys. Rev. Lett.}
  \textbf{\bibinfo{volume}{101}}, \bibinfo{pages}{211102}
  (\bibinfo{year}{2008}).

\bibitem[{\citenamefont{Abadie and {\it et al.}}(2011)}]{abadie11}
\bibinfo{author}{\bibfnamefont{J.}~\bibnamefont{Abadie}} \bibnamefont{and}
  \bibinfo{author}{\bibnamefont{{\it et al.}}}, \bibinfo{journal}{Astrophys.
  J.} \textbf{\bibinfo{volume}{734}}, \bibinfo{pages}{L35}
  (\bibinfo{year}{2011}).

\end{thebibliography}

\end{document}